\documentclass[journal,twocolumn]{IEEEtran}

\usepackage[T1]{fontenc}

\usepackage{microtype}
\usepackage{graphicx}
\usepackage{booktabs} 
\usepackage{enumerate}

\usepackage{hyperref}

\usepackage{amsmath,amssymb}
\usepackage[latin1]{inputenc}
\usepackage{mathabx}
\usepackage{bbm}

\usepackage{pifont}
\usepackage{setspace}
\usepackage{psfrag}
\usepackage{color}
\usepackage{multirow}
\usepackage{wasysym}
\usepackage{times,balance,amsfonts,pgf,tikz,pgffor}
\usepackage{pgfplots}
\usepackage{rotating}
\usepackage{verbatim}
\usetikzlibrary{shadows.blur}
\usetikzlibrary{shapes.symbols}
\usepackage[short]{optidef}
\usepackage[font=small,labelfont=bf]{caption}
\usepackage[caption=false,font=footnotesize]{subfig}
\usepackage{pgfplotstable}

 \usetikzlibrary{positioning}
\usepackage{tabularx}
\usepackage[ruled,linesnumbered]{algorithm2e}

\usepackage[noend]{algpseudocode}
\usetikzlibrary{patterns}

\makeatletter
\def\algbackskip{\hskip-\ALG@thistlm}
\makeatother

\makeatletter
\def\BState{\State\hskip-\ALG@thistlm}
\makeatother

\makeatletter
\newcommand{\multiline}[1]{%
	\begin{tabularx}{\dimexpr\linewidth-\ALG@thistlm}[t]{@{}X@{}}
		#1
	\end{tabularx}
}
\makeatother

\makeatletter
\newcommand{\removelatexerror}{\let\@latex@error\@gobble}
\makeatother

\pgfplotsset{grid style={loosely dashed,gray}}

\usepackage{makecell}

\usepackage{float} 

\definecolor{s1}{rgb}{0.45, 0.66, 0.76}
\definecolor{s2}{rgb}{0.86, 0.44, 0.58}
\definecolor{s3}{rgb}{0.6, 0.47, 0.48}
\definecolor{s4}{rgb}{0.96, 0.64, 0.38}
\definecolor{s5}{RGB}{255, 127, 0}

\newtheorem{theorem}{Theorem}
\newtheorem{definition}{Definition}
\newtheorem{corollary}{Corollary}

\newtheorem{remark}{Remark}
\newtheorem{optproblem}{Optimization Problem}

\newcommand{\volm}[1]{{v}_#1}

\newcommand{\Var}{\operatorname{Var}}

\newcommand{\MMfunc}{$\text{MatMult}$}

\makeatletter
\newcommand*{\rom}[1]{\expandafter{\romannumeral #1\relax}}
\makeatother

\newcommand{\prob}{p}

\newcommand{\success}{q}

\newcommand{\idxset}{\mathcal{I}}

\newcommand{\fMAC}{g}

\newcommand{\ie}{i.e., }
\newcommand{\eg}{e.g., }

\newcommand{\shiftp}{\alpha_{\text{comp}}}
\newcommand{\scalep}{\mu_{\text{comp}}}

\newcommand{\shiftm}{\alpha_{\text{comm}}}
\newcommand{\scalem}{\mu_{\text{comm}}}
\newcommand{\scalee}{\mu}
\newcommand{\shiftt}{\alpha}

\newcommand{\taskbox}{$\text{task block }$}
\newcommand{\taskboxes}{$\text{task blocks }$}

\newcommand{\taskboxno}{$\text{task block}$}

\newcommand{\MAC}{$\text{basic }$}

\newcommand{\infoblock}{$\text{information block }$}

\newcommand{\infoblocks}{$\text{information blocks }$}
\newcommand{\infoblocksno}{$\text{information blocks}$}

\newcommand{\recoverthr}{$\text{recovery threshold }$}

\newcommand{\profile}{$\text{recovery profile }$}
\newcommand{\profileno}{$\text{recovery profile}$}
\newcommand{\Stask}{P}
\newcommand{\stask}{p}

\newcommand{\bicc}{$\text{BICC}$}
\newcommand{\BICC}{$\text{BICC}$}
\newcommand{\mlcc}{$\text{MLCC}$}

\newcommand{\taskmtxs}{$\text{matrices }$}

\newcommand{\enproducts}{$\text{encoded matrix products }$}

\newcommand{\level}{$\text{level }$}

\newcommand{\levelno}{$\text{level}$}
\newcommand{\levels}{$\text{levels }$}

\newcommand{\levelsno}{$\text{levels}$}

\newcommand{\tdec}[1]{T^{\text{dec,#1}}}

\newcommand{\tcombicc}{\bar{T}}
\newcommand{\tcommlcc}{\tilde{T}}

\newcommand{\encc}[1]{C_{#1}^{\text{enc}}}

\newcommand{\commi}[1]{C_{#1}^{\text{in}}}
\newcommand{\commo}[1]{C_{#1}^{\text{out}}}
\newcommand{\comp}[1]{C_{#1}^{\text{comp}}}

\newcommand{\datachunks}{$\text{matrices }$}

\newcommand{\encdatachunks}{encoded submatrices }

\newcommand{\fin}{\tau}
\newcommand{\tcompi}[1]{{T_{#1}^{\text{comp}}}}
\newcommand{\tcomp}{T^{\text{comp}}}
\newcommand{\Fcompi}[1]{{F_{#1}^{\text{comp}}}}

\newcommand{\tcommi}[1]{{T_{#1}^{\text{comm}}}}
\newcommand{\tcomm}{T^{\text{comm}}}
\newcommand{\Fcommi}[1]{{F_{#1}^{\text{comm}}}}

\newcommand{\tcom}{T}

\newcommand{\floor}[1]{\lfloor{#1}\rfloor}
\newcommand{\abs}[1]{\left|#1\right|}

\newcommand{\set}{\mathcal{S}}
\newcommand{\seti}[1]{\mathcal{S}_{#1}}

\newcommand{\setxi}[1]{\mathcal{S}_{x#1}}
\newcommand{\setyi}[1]{\mathcal{S}_{y#1}}
\newcommand{\setzi}[1]{\mathcal{S}_{z#1}}

\newcommand{\inR}[3]{#1 \in \mathbb R^{#2 \times #3}}

\newcommand{\Fs}{q}                   

\newcommand{\Input}[1]{ \textbf{Input}:#1}

\newcommand{\matA}{\mathbf{A}}  
\newcommand{\matB}{\mathbf{B}}
\newcommand{\matC}{\mathbf{C}}

\newcommand{\eA}[2]{\textit a_{#1,#2}}
\newcommand{\eB}[2]{\textit b_{#1,#2}}
\newcommand{\eC}[2]{\textit c_{#1,#2}}
\newcommand{\Xdir}{x}
\newcommand{\Ydir}{y}
\newcommand{\Zdir}{z}
\newcommand{\divx}{M_{x\ly}}
\newcommand{\divy}{M_{y\ly}}
\newcommand{\divz}{M_{z\ly}}
\newcommand{\Divxi}[1]{M_{x#1}}
\newcommand{\Divyi}[1]{M_{y#1}}
\newcommand{\Divzi}[1]{M_{z#1}}
\newcommand{\Divx}{M_{x}}
\newcommand{\Divy}{M_{y}}
\newcommand{\Divz}{M_{z}}
\newcommand{\divxi}{m_x}
\newcommand{\divyi}{m_y}
\newcommand{\divzi}{m_z}

\newcommand{\Aa}{{N_x}}
\newcommand{\AB}{{N_z}}
\newcommand{\Bb}{{N_y}}
\newcommand{\Aai}{i_x}
\newcommand{\ABi}{i_z}
\newcommand{\Bbi}{i_y}
\newcommand{\Ac}[1]{\hat{\matA}(#1)}
\newcommand{\Bc}[1]{\hat{\matB}(#1)}
\newcommand{\Cc}[1]{\hat{\matC}(#1)}
\newcommand{\cA}[2]{\hat{\matA}_{#1}(#2)} 
\newcommand{\cB}[2]{\hat{\matB}_{#1}(#2)} 
\newcommand{\Ai}[1]{\matA_{#1}}           
\newcommand{\Aii}[2]{{\matA_{\tiny{#2}}^{(#1)}}}     
\newcommand{\Bi}[1]{\matB_{#1}}           
\newcommand{\Bii}[2]{{\matB_{\tiny{#2}}^{(#1)}}}     
\newcommand{\LY}{{L}}      
\newcommand{\ly}{l}      
\newcommand{\ND}{{N}}      
\newcommand{\nd}{n}      
\newcommand{\RT}{{R}}      

\newcommand{\rt}[1]{\RT_{#1}}

\newcommand{\DM}{{K}}      
\newcommand{\dm}[1]{\DM_{#1}}
\newcommand{\dmsum}{\DM_{\text{sum}}}

\newcommand{\DMi}{k}
\newcommand{\Xdimi}[1]{N_{x#1}}
\newcommand{\Zdimi}[1]{N_{z#1}}
\newcommand{\Ydimi}[1]{N_{y#1}}

\begin{document}
\title{Hierarchical Coded Matrix Multiplication}

\author{Shahrzad~Kiani,~\IEEEmembership{Graduate~Student~Member,~IEEE,}
        Nuwan~Ferdinand,
        and~Stark~C.~Draper,~\IEEEmembership{Senior~Member,~IEEE}
\thanks{Manuscript received December 23, 2019; revised July 18, 2020; and accepted October 19, 2020. This work was supported in part by the National Science Foundation (NSF) under Grant CCF-1217058, a Discovery Research Grant from the Natural Sciences and Engineering Research Council of Canada (NSERC), an NSERC postdoctoral fellowship, the Ontario Graduate Scholarship, and a grant from Huawei Technologies, Inc. This paper was presented in part in IEEE Int. Symp. Inf. Theory'18~\cite{EXPLOIT:ISIT18}~\cite{HIER:ISIT18}, IEEE Canadian Workshop Inf. Theory'19~\cite{HMM:CWIT19}, and IEEE Int. Conf. Machine Learning (Workshop on Coded Machine Learning)'19~\cite{CUBOID:CODML19}.}
\thanks{S. Kiani and S. Draper are with the Department of Electrical and Computer  Engineering, University of Toronto, Toronto, ON, Canada (Emails: shahrzad.kianidehkordi@mail.utoronto.ca, stark.draper@utoronto.ca).}
\thanks{N. Ferdinand was with the Department of Electrical and Computer  Engineering, University of Toronto, Toronto, ON, Canada (Email: nuwan.ferdinand@utoronto.ca).}
\thanks{Copyright (c) 2017 IEEE. Personal use of this material is permitted.  However, permission to use this material for any other purposes must be obtained from the IEEE by sending a request to pubs-permissions@ieee.org.}
}

\markboth{IEEE Transactions on Information Theory}%
{Shell \MakeLowercase{\textit{et al.}}: Bare Demo of IEEEtran.cls for IEEE Communications Society Journals}

\maketitle

\begin{abstract}
In distributed computing systems slow working nodes, known as stragglers, can greatly extend finishing times. Coded computing is a technique that enables straggler-resistant computation. Most coded computing techniques
presented to date provide robustness by ensuring that the time to
finish depends only on a set of the fastest nodes. However, while
stragglers do compute less work than non-stragglers, in real-world
commercial cloud computing systems (e.g., Amazon's Elastic Compute
Cloud (EC2)) the distinction is often a soft one. In this paper, we
develop \emph{hierarchical} coded computing that exploits the work
completed by all nodes, both fast and slow, automatically integrating
the potential contribution of each. We first present a conceptual
framework to represent the division of work amongst nodes in coded
matrix multiplication as a cuboid partitioning problem. This framework
allows us to unify existing methods and motivates new techniques. We
then develop three methods of hierarchical coded computing that we
term \emph{bit-interleaved} coded computation (BICC), \emph{multilevel} coded
computation (MLCC), and \emph{hybrid} hierarchical coded computation (HHCC). In this paradigm,
each worker is tasked with completing a sequence (a hierarchy) of
ordered subtasks. The sequence of subtasks, and the complexity of each, is designed so that 
partial work completed by stragglers can be used, rather than
ignored. We note that our methods can be used in conjunction with any coded computing method. We illustrate this by showing how we can use our methods to accelerate all previously developed coded computing techniques by enabling them to exploit stragglers. Under a widely studied statistical model of completion time, our approach realizes a $66\%$ improvement in the
expected finishing time. On Amazon EC2, the gain was $27\%$ when stragglers are simulated. 

\end{abstract}

\section{Introduction}
\IEEEPARstart{T}{he} advent of large-scale machine learning algorithms and data analytics has increased the demand for computation. Such computation often cannot be performed in a single computer due to limited processing power and available memory. Parallelization is necessary. In an idealized distributed setting highly parallelizable tasks can be accelerated proportionally to the number of working nodes. However, in many cloud-based systems, slow working nodes, known as \emph{stragglers}, present a bottleneck that can prevent the realization of faster compute times~\cite{ Dean:2012}. Recent studies show that for certain linear algebraic tasks, such as matrix multiplication, the effect of stragglers can be minimized through the use of error correction codes~\cite{SPEEDUP:TIT17, SHORT:NIPS16, PRODUCT:ISIT17, POLY:NIPS17, MATDOT:2018, ENTGL:ISIT18}. This idea, termed \emph{coded computing}, introduces redundant computations so that the completion of any fixed-cardinality subset of tasks suffices to realize the desired solution. Coded computing can greatly accelerate many machine learning algorithms such as those that involve computing gradients, thus accelerating the training of large-scale machine learning applications~\cite{GC:ICML17, GPDOT:ISIT18,COMMCOMP:ICML18,GDREED:ISIT18,Polyregression:2018}. While matrix multiplication and gradient descent are two types of coded computing problems that have been studied, others include coded convolution~\cite{CCONV:ISIT17}, coded approximated computing~\cite{Anytime:16,yang2017coded}, sparse coded matrix multiplication~\cite{Sparse:18}, and heterogeneous coded computing~\cite{HERTO:2019}.

In most of the coded computing work to date a type of {\em erasure} model is assumed. Workers either complete their job or complete no work. While this can be a good model for hardware failure, it is not a perfect model for all decentralized computing systems. In many systems one observes a spectrum of completion speeds. One aim of this paper is to advocate for a more nuanced view of stragglers. A processing node may be slower than the average node but yet may complete some work. On the other hand, it may be faster than the average node, a {\em leader}, and correspondingly able to complete more work. We illustrate one method for maximizing the contributions of both fast and slow workers in the context of accelerating matrix-matrix multiplications. We believe the general idea can be applied quite widely. In the remainder of this section, we discuss the prior work we build on, some contemporary work that thinks about how to exploit stragglers, and then we present our contributions.

\subsection{Background: Stragglers and Coded Computing}
A simple approach to deal with stragglers is to replicate
tasks. This is equivalent to repetition coding. However, if the job can be linearly decomposed, the opportunity arises to introduce
redundancy more efficiently through the use of error-correction
codes. In particular, in~\cite{SPEEDUP:TIT17} maximum distance
separable (MDS) codes are used to develop a straggler-resistant
method of vector-matrix multiplication. This idea is extended to
matrix-matrix multiplication using product codes in~\cite{PRODUCT:ISIT17}. In~\cite{POLY:NIPS17}, coded computation based on polynomial interpolation is introduced. Polynomial codes outperform product codes in terms of their \emph{recovery threshold}. This is the number of workers that are required to complete their tasks for the master to be able to recover the desired calculation. With memory per worker fixed, the recovery threshold of polynomial codes is further improved by MatDot codes~\cite{MATDOT:2018}, albeit at the cost of increased per-worker computation. In addition to MatDot coding,~\cite{MATDOT:2018} introduces polyDot coding as a generalization and unification of polynomial and MatDot codes. PolyDot codes provide a tradeoff between the recovery threshold and the computation load assigned to each worker. In works subsequent to polyDot coding,~\cite{ENTGL:ISIT18} and~\cite{GPDOT:ISIT18} simultaneously arrived at new coding methods that can improve the recovery threshold of polyDot codes. These coding methods are respectively referred to as entangled polynomial and Generalized polyDot codes.

As already mentioned, a drawback to many of the initial coded computing designs is that they rely only on the work completed by a set of the fastest workers. They ignore completely the work done by the slower workers. In the terminology of error correction coding, these slower nodes are modeled as erasures. However, in systems such as the Amazon Elastic Compute Cloud (EC2), we observe \emph{partial} stragglers. Partial stragglers are slower, only able to complete partial tasks by the time at which the faster workers have completed their entire tasks. That said, the amount of work stragglers can complete may be non-negligible. Thus, it can be wasteful to ignore. 

Recent literature (including our work)~\cite{GC:ICML17, EXPLOIT:ISIT18, RATELESS:2018, HIER:ISIT18, HMM:CWIT19, CUBOID:CODML19, CLES:ITW19, UDM:ISIT19, Scheduling:gunduz:19} proposes methods to exploit the work completed by stragglers, rather than ignoring it. The underlying idea is to assign each worker a sequence of multiple small subtasks rather than a single large task. The master is able to complete the job by utilizing the computations completed by all workers, stragglers simply contribute less. The concept of exploiting partial straggler was first studied in~\cite{GC:ICML17} such that each worker is assigned two groups of subtasks: {\em naive} and {\em coded} subtasks. Non-straggler workers process both naive and coded subtasks, while partial stragglers only complete naive tasks. In~\cite{EXPLOIT:ISIT18,RATELESS:2018} each worker is tasked with a fully-coded series of subtasks, where all tasks are coded with respect to a single code. In~\cite{EXPLOIT:ISIT18, RATELESS:2018} the vector-matrix multiplication was first broken into computationally homogeneous subtasks; each subtask was then encoded using a specific code: an MDS code in~\cite{EXPLOIT:ISIT18} and a rateless fountain code in~\cite{RATELESS:2018}. In our paper~\cite{HIER:ISIT18} we further leverage the sequential computing nature of each worker by introducing the concept of hierarchical coding. This concept was explained based on the vector-matrix multiplication problem using MDS codes. In hierarchical coding, workers are all tasked to complete a hierarchy of coded subtasks, each subtask is coded separately. We extended special cases of hierarchical coding in~\cite{HMM:CWIT19, CUBOID:CODML19}. More recent works that aim to exploit stragglers including~\cite{UDM:ISIT19,Scheduling:gunduz:19,CLES:ITW19} complement the idea presented in~\cite{RATELESS:2018} and the idea we present in~\cite{EXPLOIT:ISIT18}. In~\cite{ CLES:ITW19} each worker is tasked with a specified fraction of coded and uncoded computations. In~\cite{UDM:ISIT19} multiple coded subtasks assigned to each worker are generated according to the characteristics of universally decodable matrices. In~\cite{Scheduling:gunduz:19} each worker is tasked with completing a fully-uncoded series of subtasks with respect to a predesigned computation order. 
 
\subsection{Contributions}
In this paper we develop three novel approaches to coded computing. Each leverages the work completed by the average node, exploits the work completed by the stragglers, and enables fast workers to contribute even more to the overall computation. In analogy with some similarly-structured approaches in coded modulation, we term these \emph{bit-interleaved} coded computing (BICC),
\emph{multilevel} coded computing (MLCC), and \emph{hybrid} hierarchical coded computing (HHCC). As in all these three approaches each processing unit is presented with a hierarchy of tasks, we use the unifying term \emph{hierarchical} coding to refer to the methods in aggregate. In our conference publications~\cite{EXPLOIT:ISIT18},~\cite{HIER:ISIT18}, and~\cite{HMM:CWIT19} we presented a few special cases of BICC and MLCC. In this work, we lay out a generalized hierarchical coded computation approach to solve matrix-matrix multiplication that builds upon any classical (non-hierarchical) coding schemes. Two ideas underlie our hierarchical designs: assignment of multiple subtasks to each worker, and recognition of the inherent sequential processing nature of the workers. Before presenting hierarchical coding later in the paper, in Sec.~\ref{SEC:geometric} we first establish an equivalence between task allocation in coded matrix multiplication and a geometric problem of partitioning a rectangular cuboid into subcuboids. The allocation of tasks in the various coded matrix multiplication approaches developed to date correspond to different partitionings of the cuboid. The cuboid visualization will facilitate our extension to hierarchical coding.

We now sketch the observations that lead to our code designs. Our first observation is that by assigning each worker multiple smaller subtasks we can realize a more continuous completion process. This enables the exploitation of a broader spectrum of workers, including stragglers, and motivates the design of BICC. Our second observation is we recognize the inherently sequential processing nature of compute nodes. Once a worker is assigned multiple tasks, and due to sequential processing, more workers will finish their first assigned subtask than their second (or third, fourth, etc.). We exploit this observation in MLCC by grouping subtasks into {\em levels}, one subtask per level. Workers all start with their first level subtask. Since most workers complete their first level subtask, we assign the first level the highest rate; later levels generally receive lower rates. So, while in BICC all subtasks are jointly encoded, in MLCC the encoding of distinct levels is independent. On the one hand, this means that BICC enjoys more flexible recovery requirements than does MLCC, which results in lower computation time. On the other hand, BICC suffers from more complicated decoding and higher communication overhead than MLCC. These characteristics roughly parallel bit-interleaved coded modulation (BICM)~\cite{BICM:1998} and multilevel coding (MLC)~\cite{MLCC:1999}, hence the choice of names. To mitigate the recovery requirements of MLCC, we introduce randomized MLCC (RMLCC) in which each worker randomly picks (without replacement) a level of computation and completes the encoded subtask pertinent to that level. Due to this random permutation of levels across each worker and for large enough number of workers, the same number of workers on average are assigned a different permutation of levels. This makes the recovery mechanism of MLCC more flexible.

 To combine the strengths of BICC and MLCC we introduce HHCC. This hybrid version realizes an achievable design trade-off between computation, decoding, and communication times. We note that our methods can be used in conjunction with any coded computing method. We illustrate this by showing how we can use our methods to accelerate all previously developed coded computing techniques and enabling them to exploit stragglers. 
 
 We prove both theoretically and experimentally that our proposed hierarchical techniques improve the total finishing time when applies to standard coded computing schemes. We numerically show that our method realizes a $66\%$ improvement in the expected finishing time for a widely studied model of shifted exponential completion time of each subtask. On Amazon EC2, the gain was $27\%$ when stragglers are simulated. 

\subsection{Outline}
Other than a brief discussion of notation (next), the rest of the paper is laid out as follows. In Sec.~\ref{SEC:model} we describe a system model that is considered in this paper and discuss the relevant performance measures. In Sec.~\ref{SEC:geometric} we introduce our cuboid partitioning visualization and detail the state-of-the-art coded matrix multiplication approaches through this visualization. In Sec.~\ref{SEC:HMM} we first introduce BICC, MLCC, and HHCC in detail and then compare these hierarchical coding schemes. In Sec.~\ref{CHAPTER:THEORY} we theoretically analyze the finishing time of non-hierarchical and hierarchical designs and provide a parameter design for BICC and MLCC. In Sec.~\ref{sec.sim} we experimentally show that hierarchical coding decreases the total finishing time when compared to non-hierarchical coding. Finally, we conclude our contribution in Sec.~\ref{sec.conclusion}.

\subsection{Notation}
Sets are denoted using calligraphic font, \eg $\mathcal{S}$. The cardinality of a finite set $\set$ is denoted $\abs{\set}$. We use bold upper case, \eg $\matA$, for matrices. The entry in the $i$th row and $j$th column of $\matA$ is denoted as $\eA{i}{j}$. The submatrix of $\matA$ is obtained by selecting out a collection of rows, $\seti{\text{r}}$, and columns, $\seti{\text{c}}$, and is denoted by $\Ai{\set}$ where $\set=\seti{\text{r}} \times \seti{\text{c}}$. To denote the $i$th element of $\seti{\text{r}}$ we write $\seti{\text{r,i}}$, which is a row-index into the $\matA$ matrix. Similarly, $\seti{\text{c,i}}$ is column-index into $\matA$ matrix. We also use the notation $[\Fs]+c=\{1+c,\ldots,\Fs+c\}$ for the shifted index set, where $\Fs,c \in \mathbb{Z}_{+}$; $[\Fs]$ means $c=0$. 

\section{System Model}
\label{SEC:model}
We consider the problem of multiplying two matrices
$\inR{\matA}{\Aa}{\AB}$ and $\inR{\matB}{\AB}{\Bb}$ in a distributed coded system that consists of a central node, referred to as the \emph{master}, and $\ND$ individual nodes, called \emph{workers}. We parallelize the computation of the matrix product $\inR{\matA\matB}{\Aa}{\Bb}$ among the $\ND$ workers by providing each a subset of the data and requesting each to carry out specific computations. In Sec.~\ref{Sec:system} we detail our model of a distributed coded matrix multiplication system. In Sec.~\ref{Sec:metrics} we define the performance metrics we use to compare designs.

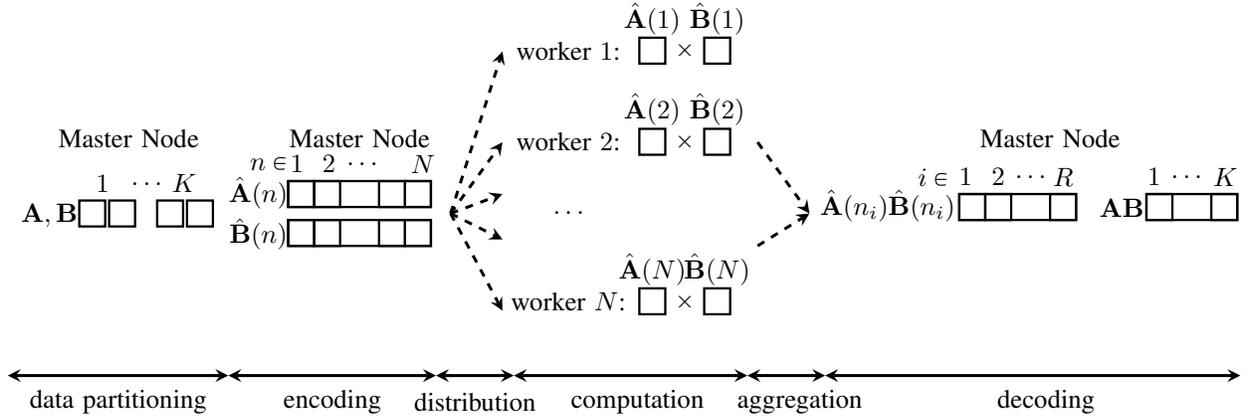
\begin{figure*}[h]
	\centering
	\begin{tikzpicture}
		[scale=0.75,
		fine/.style = {solid,draw=gray},
		shaping/.style = {line width=0.3mm},
        shaping2/.style = {line width=0.25mm},
        shaping3/.style = {line width=0.4mm},
		voronoi/.style = {fill=blue!20!white}]
		\begin{scope}[scale = 2.3]
			\begin{scope}
		 	\node at (-4.65,-1.2){data partitioning};
		 	\node at (-3,-1.2){encoding};
		 	\node at (-1.9,-1.2){distribution};
		 	\node at (-0.65,-1.2){computation};
		 	\node at (0.6,-1.2){aggregation};
		 	\node at (2.5,-1.2){decoding};
\draw[shaping3,<->,>=stealth] (-5.5,-1) -- (-3.8,-1);	
\draw[shaping3,<->,>=stealth] (-3.8,-1) -- (-2.2,-1);
\draw[shaping3,<->,>=stealth] (-2.2,-1) -- (-1.6,-1);
\draw[shaping3,<->,>=stealth] (-1.6,-1) -- (0.2,-1);
\draw[shaping3,<->,>=stealth] (0.2,-1) -- (.8,-1);
\draw[shaping3,<->,>=stealth] (.8,-1) -- (4,-1);
			\end{scope}
		    \begin{scope}[xshift=-130,yshift=18]
		 	\node at (0.0,0.2) {Master Node};
		 	\node at (-0.2,-0.15) {$ 1$};
		 	\node at (0.14,-0.15) {$ \ldots$};
		 	\node at (0.43,-0.15) {$ \DM$}; 
		 	\node at (-0.61,-0.4) {$\matA,\matB$};
		    \end{scope}
		    \begin{scope}[xshift=-130]
		\draw[shaping] (-0.38,.15) rectangle (-0.18,0.35);
		\draw[shaping] (-0.15,.15) rectangle (0.05,0.35);
		\draw[shaping] (0.22,.15)  rectangle (0.42,0.35);		    
		\draw[shaping] (0.45,.15)  rectangle (0.65,0.35);		    
		    
			\end{scope}
		    \begin{scope}[xshift=-95,yshift=18]
		 	\node at (0.55,0.2) {Master Node};
		 	\node at (-0.15,0) {$ \nd \in$};
		 	\node at (0.07,0) {$ 1$};
    	 	\node at (0.315,0) {$ 2$};
		 	\node at (0.57,0) {$ \ldots$};
		 	\node at (1.04,0) {$ \ND$}; 
		 	\node at (-0.23,-0.22) {$\hat{\matA}(n)$};
		 	\node at (-0.23,-0.54) {$\hat{\matB}(n)$};
		    \end{scope}
		    \begin{scope}[xshift=-95]
			\draw[shaping] (-0.0,0) rectangle (1.1,0.2);
			\draw[shaping] (-0.0,0) rectangle (0.2,0.2);
			\draw[shaping] (0.2,0)  rectangle (0.4,0.2);
			\draw[shaping] (0.7,0)  rectangle (0.9,0.2);
			\draw[shaping] (0.9,0)  rectangle (1.1,0.2);
			\draw[shaping] (-0.0,0.3) rectangle (1.1,0.5);
			\draw[shaping] (-0.0,0.3) rectangle (0.2,0.5);
			\draw[shaping] (0.2,0.3) rectangle (0.4,0.5);
			\draw[shaping] (0.7,0.3) rectangle (0.9,0.5);
			\draw[shaping] (0.9,0.3) rectangle (1.1,0.5);
			\end{scope}			
			
			\begin{scope}[xshift=-85]
\draw[shaping3,->,>=stealth,dashed] (0.9,0.25) -- (1.3,1.5);	
\draw[shaping3,->,>=stealth,dashed] (0.9,0.25) -- (1.3,0.8);	
\draw[shaping3,->,>=stealth,dashed] (0.9,0.25) -- (1.3,0.4);	
\draw[shaping3,->,>=stealth,dashed] (0.9,0.25) -- (1.3,0.05);	
\draw[shaping3,->,>=stealth,dashed] (0.9,0.25) -- (1.3,-0.5);	
            \end{scope}
            
            \begin{scope}[xshift=-25,yshift=50]
		 	\node at (-0.3,-0.25) {worker $1$:};
		 	\end{scope}
            \begin{scope}[xshift=-18,yshift=40]
            \node at (0.1,0.35) {$ \hat{\matA}(1)$};
		 	\node at (0.6,0.35) {$ \hat{\matB}(1)$};
			\draw[shaping] (0,0) rectangle (0.2,0.2);
			\draw[shaping] (0.5,0) rectangle (0.7,0.2);
			\node at (0.35,0.1) {$\times$};
            \end{scope}
            
            \begin{scope}[xshift=-25,yshift=30]
      	 	\node at (-0.3,-0.8) {$\ldots$};
		 	\node at (-0.3,-0.25) {worker $2$:};
       
		 	\end{scope}
            \begin{scope}[xshift=-18,yshift=20]
            \node at (0.1,0.35) {$ \hat{\matA}(2)$};
		 	\node at (0.6,0.35) {$ \hat{\matB}(2)$};            
			\draw[shaping] (0,0) rectangle (0.2,0.2);
			\draw[shaping] (0.5,0) rectangle (0.7,0.2);
			\node at (0.35,0.1) {$\times$};
            \end{scope}
            
		    \begin{scope}[xshift=-25,yshift=-5]
		 	\node at (-0.3,-0.25) {worker $\ND$:};

		 	\end{scope}
		    \begin{scope}[xshift=-18,yshift=-15]
		    \node at (0.1,0.35) {$ \hat{\matA}({\ND})$};
		 	\node at (0.6,0.35) {$ \hat{\matB}({\ND})$}; 
			\draw[shaping] (0,0) rectangle (0.2,0.2);
			\draw[shaping] (0.5,0) rectangle (0.7,0.2);
			\node at (0.35,0.1) {$\times$};
            \end{scope}
   			\begin{scope}[xshift=25]
\draw[shaping3,->,>=stealth,dashed] (-0.6,0.8) -- (-0.2,0.25);	
\draw[shaping3,->,>=stealth,dashed] (-0.6,0.) -- (-0.2,0.25);

            \end{scope}            
		    \begin{scope}[xshift=45,yshift=18]
		 	\node at (0.95,0.2) {Master Node};
		 	\node at (0.05,-0.1) {$ i \in$};
		 	\node at (-0.3,-0.32) {$\hat{\matA}(n_i)\hat{\matB}(n_i)$};
		    \end{scope}
		    \begin{scope}[xshift=55]
		    \begin{scope}[yshift=18]
   		 	\node at (-0.05,-0.1) {$ 1$};
    	 	\node at (0.2,-0.1) {$ 2$};
		 	\node at (0.45,-0.1) {$ \ldots$};
		 	\node at (0.7,-0.1) {$ \RT$};
		 	\end{scope} 
			\draw[shaping] (-0.1,0.2) rectangle (0.8,0.4);
			\draw[shaping] (-0.1,0.2) rectangle (0.1,0.4);
			\draw[shaping] (0.1,0.2) rectangle (0.3,0.4);
			\draw[shaping] (0.6,0.2) rectangle (0.8,0.4);
			\end{scope}	
		    \begin{scope}[xshift=102,yshift=18]
		 	\node at (-0.25,-0.1) {$ 1$};
		 	\node at (-0.0,-0.1) {$ \ldots$};
		 	\node at (0.3,-0.1) {$ \DM$}; 
		 	\node at (-0.5,-0.34) {$\matA\matB$};
		    \end{scope}
		    \begin{scope}[xshift=106]
			\draw[shaping] (-0.45,0.2) rectangle (0.25,0.4);
			\draw[shaping] (-0.45,0.2) rectangle (-0.25,0.4);
			\draw[shaping] (0.05,0.2) rectangle (0.25,0.4);
			\end{scope}		
		\end{scope}

\end{tikzpicture} 
 \caption{A master-worker model for distributed coded matrix multiplication problem.} 
 \label{FIG:system}
\end{figure*}

In the following we first explain the system model of distributed coded matrix multiplication through six phases: data partitioning, encoding, distribution, worker computation, result aggregation, and decoding phases. This system model is depicted in Fig.~\ref{FIG:system}. For future reference, we introduce now some new notation used in Fig.~\ref{FIG:system} that will be formally defined in~(\ref{encoding_evaluation}) in Sec.~\ref{Sec:system}. The matrices $\hat{A}(n)$ and $\hat{B}(n)$, which are, respectively, encoded from original matrices $\matA$ and $\matB$, are encoded data distributed to each worker $n\in [N]$.

\subsection{Distributed Coded Matrix Multiplication}
\label{Sec:system}
While we next detail the six phases and define the terminology we use one at a time, the reader may prefer to look forward to Examples~2 and~3 where all the phases are provided in the context of two examples for state-of-the-art coding schemes, polynomial~\cite{POLY:NIPS17} and MatDot~\cite{MATDOT:2018} codes. 

\textbf{Phase 1 (Data partitioning):} We first partition the overall computation of the matrix product $\matA\matB$ into $\DM$ equal-sized\footnote{The load allocation is assumed to be homogeneous; the extension of this assumption to a heterogeneous system is discussed is Sec.~\ref{SEC:Prob}} computations. The master will be able to recover the $\matA\matB$ product if and only if all $\DM$ computations are completed. To accomplish this the master first partitions the data by dividing $\matA$ and (separately) $\matB$ into, respectively, $\Divx\Divz$ and $\Divz\Divy$ equal-sized matrices of dimension $\Aa/\Divx \times \AB/\Divz$ and $\AB/\Divz \times \Bb/\Divy$):
\begin{align} \label{firstphase}
\{\Ai{\seti{\Xdir i}\times\seti{\Zdir j}} &|\, i\in[\Divx],\, j\in[\Divz] \}, \nonumber \\ \{\Bi{\mathcal{{S}}_{\Zdir j}\times \mathcal{{S}}_{\Ydir k}} &| \, j\in[\Divz],\, k\in[\Divy] \},
\end{align} 
where $\DM=\Divx\Divz\Divy$. $\mathcal{{S}}_{\Xdir i}, \mathcal{{S}}_{\Zdir j},$ and $\mathcal{{S}}_{\Ydir k}$ are respectively the subsets of (generally, but not necessarily) consecutive elements of  $[\Aa]$, $[\AB]$, and $[\Bb]$. To ensure that the submatrices $\Ai{\seti{\Xdir i}\times\seti{\Zdir j}} $ (and $\Bi{\mathcal{{S}}_{\Zdir j}\times \mathcal{{S}}_{\Ydir k}}$) partition the entire matrix $\matA$ (and $\matB$), these subsets must satisfy $\bigcup_{i\in[\Divx],\, j\in[\Divz]} \seti{\Xdir i}\times\seti{\Zdir j} = [\Aa] \times [\AB]$ (and $\bigcup_{j\in[\Divz],\, k\in[\Divy] } \mathcal{{S}}_{\Zdir j}\times \mathcal{{S}}_{\Ydir k} = [\AB] \times [\Bb]$). Note that, in general the submatrices $\Ai{\seti{\Xdir i}\times\seti{\Zdir j}} $ (and $\Bi{\mathcal{{S}}_{\Zdir j}\times \mathcal{{S}}_{\Ydir k}}$), $i \in [\Aa], j\in[\AB]$ ($ k\in [\Bb]$), can have overlap; that simply would mean certain elements of the matrix $\matA\matB$ would be computed more than once. For conceptual clarity for the moment, we assume that they are disjoint. 

This partitioning process can be reversed by using a concatenation. For example, let's assume that the matrix $\Ai{\seti{\Xdir i}\times\seti{\Zdir j}} $ contains all elements $\eA{\seti{\Xdir i,i_x}}{\seti{\Zdir j,i_z}}$, where $i_x \in [\floor{\Aa/\Divx}] + (i-1)\floor{\Aa/\Divx}$ and $i_z \in [\floor{\AB/\Divz}] + (j-1)\floor{\AB/\Divz}$. Likewise, the matrix $\Bi{\seti{\Zdir j}\times\seti{\Ydir k}} $ contains all elements $\eB{\seti{\Zdir j,i_z}}{\seti{\Ydir k,i_y}}$, where $i_z \in [\floor{\AB/\Divz}] + (j-1)\floor{\AB/\Divz}$ and $i_y \in [\floor{\Bb/\Divy}] + (k-1)\floor{\Bb/\Divy}$.  We can therefore reverse the $\matA$ and $\matB$ partitioning by concatenating the matrices:
\begin{align} \label{sysmodel_partitioning}
\matA &= 
 \left[ \begin{array}{ccc}
 \Ai{\seti{\Xdir 1}\times\seti{\Zdir 1}} & \ldots & \Ai{\seti{\Xdir 1}\times\seti{\Zdir \Divz}}\\
 \vdots & \ddots & \vdots\\
 \Ai{\seti{\Xdir \Divx}\times\seti{\Zdir 1}} & \ldots & \Ai{\seti{\Xdir \Divx}\times\seti{\Zdir \Divz}}
 \end{array}\right], \nonumber \\
\matB &= 
 \left[ \begin{array}{ccc}
 \Bi{\mathcal{{S}}_{\Zdir 1}\times \mathcal{{S}}_{\Ydir 1}} & \ldots & \Bi{\mathcal{{S}}_{\Zdir 1}\times \mathcal{{S}}_{\Ydir \Divy}}\\
 \vdots & \ddots & \vdots\\
 \Bi{\mathcal{{S}}_{\Zdir \Divz}\times \mathcal{{S}}_{\Ydir 1}} & \ldots & \Bi{\mathcal{{S}}_{\Zdir \Divz}\times \mathcal{{S}}_{\Ydir \Divy}}\\
 \end{array}\right]. 
\end{align}

Once the data $\matA$ and $\matB$ are partitioned, $\DM$ pairs of matrices can be matched up to yield the $\matA\matB$ product, 
\begin{align*}
\{\Ai{\seti{\Xdir i}\times\seti{\Zdir j}}\Bi{\mathcal{{S}}_{\Zdir j}\times \mathcal{{S}}_{\Ydir k}} | \, i\in[\Divx], j\in[\Divz], k\in[\Divy]\}.
\end{align*}

In particular, in Example~1 we will present three particular partitionings of the $\matA\matB$ product, which we refer to as \emph{inner-product}, \emph{outer-product}, and \emph{combinatorial} partitioning.

\textbf{Phase 2 (Encoding):} In this phase redundancy is introduced into computations so that the desired $\matA\matB$ product can be recovered from a subset of completed tasks. To introduce such redundancy, the master encodes the partitioned data from (\ref{firstphase}) to generate $\ND$ pairs of encoded matrices: 
\begin{align}\label{encoding_evaluation}
\{(\Ac{\nd}, \Bc{\nd})| \, \nd \in [\ND]\}.
\end{align}
Typically, each coding scheme corresponds to a specific data partitioning structure. In Examples~2 and~3 we detail the encoding phase of polynomial~\cite{POLY:NIPS17} and MatDot~\cite{MATDOT:2018} codes. In these examples $\Ac{\nd}$ and $\Bc{\nd}$ are the evaluation of the polynomials $\Ac{x}$ and $\Bc{x}$ by substituting $n$ for $x$.

\textbf{Phase 3 (Distribution):} In this phase the master distributes the encoded data, sending $\Ac{\nd} $ and $\Bc{\nd}$ to the $\nd$th worker.

\textbf{Phase 4 (Worker computation):} Worker $\nd$ computes the product $\Ac{\nd}\Bc{\nd}$, and transmits the result to the master as soon as it is completed. 

\textbf{Phase 5 (Results aggregation):} The master aggregates completed results from workers till it receives $\RT$ out of $\ND$ completed matrix products $\Ac{\nd}\Bc{\nd}$. We note that $\RT$ is a function of $\Divx,\Divz,$ and $\Divy$. In Examples~2 and~3 we show that in polynomial codes~\cite{POLY:NIPS17} $\RT_{\text{Poly}}=\Divx\Divy$ and in MatDot codes~\cite{MATDOT:2018} $ \RT_{\text{Mat}}=2\Divz-1.$ As another example, in polyDot coding (cf.~\cite[Theorem V.1]{MATDOT:2018})
\begin{align}\label{EQ:rthr_pdot}
 \RT_{\text{Pdot}}&=2\Divx\Divz\Divy - \Divx\Divy,
\end{align}
and in entangled polynomial coding (cf.~\cite[Eq.~(10)]{ENTGL:ISIT18})
\begin{align}\label{EQ:rthr_ent}
 \RT_{\text{Ent}}&=\Divx\Divz\Divy + \Divz-1.
\end{align}
 
\textbf{Phase 6 (Decoding):} The master recovers the $\matA\matB$ product by implementing the decoding algorithm corresponding to the coding scheme used in the encoding phase. In Examples~2 and~3 we detail the decoding phase of polynomial~\cite{POLY:NIPS17} and MatDot~\cite{MATDOT:2018} codes.

We now highlight the terminology we use to describe the system model of standard coded computing.

\begin{definition} (\textbf{Information dimension}): We use $\DM$ to denote the information dimension of a code. The information dimension refers to the number of useful (non-redundant) computations into which the main computational job is partitioned. In other words, completion of the computational job depends on the completion of at least $\DM$ computations. 
\end{definition}
\begin{definition} (\textbf{Block length}): The block length of a code refers to the total number of coded computational tasks that are encoded from $\DM$ computations. In standard coded computation we use $\ND$ to denote the number of workers which also refers to the block length of a code. 
\end{definition}

Note that in this terminology, $\frac{\DM}{\ND}$ refers as a \emph{rate} of a code. That is, for $\DM$ useful computations, the encoder generates $\ND$ encoded tasks. 
\begin{definition} (\textbf{Recovery threshold}): The recovery threshold per code refers to the minimum number of encoded tasks that are required to be completed so that the main computational job corresponding to that code can be recovered. We use $\RT$ to denote the recovery threshold of a code. (N.B. $\RT$ is reserved for recovery threshold in coded computing, and not rate.)
\end{definition}

We make two comments. First, we note that in the existing literature often only two of the above parameters are specified, e.g., $\ND$ and $\RT$ (and not $\DM$). This is because in many of these previous works there is a functional relationship between $\DM, \ND,$ and $\RT$, e.g., $\DM=\RT_{\text{Poly}}$ in polynomial codes~\cite{POLY:NIPS17}. In the present paper we introduce all three, $\DM, \ND,$ and $\RT$, so that our discussion can encompass all previous coding schemes. Second, in Sec.~\ref{SEC:HMM}, where hierarchical coding is combined with standard coded computing, we will use a separate set of parameters $(\DM,\ND,\RT)$ for each individual code used across the various levels of the hierarchical construction. This means that each level has its own set of parameters rather than a shared set of parameters that pertains to all levels. 

We now exemplify three particular partitionings of the $\matA\matB$ product and then detail the system model of two specific coded matrix multiplication problems.

\textbf{Example 1 (inner-product, outer-product, and combinatorial partitioning)}: In inner-product partitioning the $\matA$ and $\matB$ matrices are respectively partitioned horizontally and vertically. That is $\Divz=1$. The $\matA\matB$ product is thereby divided into $\DM$ inner products, each between a group of rows in $\matA$ and a group of columns in $\matB$. One way to achieve $\DM$ inner products is to divide $\matA$ and $\matB$ into $\sqrt{\DM}$ equal-sized matrices~\footnote{Assume $\sqrt{\DM}\approx \floor{\sqrt{\DM}}$. We ignore the integer effects to clarify the conceptual framework. In our implementation results we address integer effects.}. We partition $\matA$ horizontally into $\Divx=\sqrt{\DM}$ matrices $\Ai{\setxi{i} \times [\AB]}$, where $i \in [\sqrt{\DM}]$ and $ \setxi{i} = [\floor{{\Aa}/{\sqrt{\DM}}}]+ (i-1)\floor{{\Aa}/{\sqrt{\DM}}}$. We partition $\matB$ vertically into $\Divy=\sqrt{\DM}$ matrices $\Bi{[\AB]\times \setyi{i}}$, where $i \in [\sqrt{\DM}]$ and $ \setyi{i}=[\floor{{\Bb}/{\sqrt{\DM}}}]+(i-1)\floor{{\Bb}/{\sqrt{\DM}}}$. The $\matA\matB$ product is thereby divided into $\DM$ computations of matrix products $\Ai{\setxi{i} \times [\AB]}\Bi{[\AB]\times \setyi{j}}$, $i,j \in [\sqrt{\DM}]$.

In contrast to inner-product partitioning, in outer-product partitioning $\matA$ and $\matB$ are, respectively, divided vertically and horizontally into $\Divz=\DM$ sub-matrices. That means that $\matA$ ($\matB$) is divided into $\DM$ matrices $\Ai{[\Aa] \times \setzi{i}}$ ($\Bi{\setzi{i} \times [\Bb] }$), where $i \in [\DM]$ and $ \setzi{i}=[\floor{{\AB}/{\DM}}]+(i-1)\floor{{\AB}/{\DM}}$. Therefore, $\Divx=\Divy=1$. In outer-product partitioning the $\matA\matB$ product is divided into a set of $\DM$ outer products, each between a group of columns in $\matA$ and a group of rows in $\matB$, \ie $\Ai{[\Aa] \times \setzi{i}}\Bi{\setzi{i} \times [\Bb] }$, where $i \in [\DM]$. 

Another option is to combine the inner-product and the outer-product ideas, partitioning $\matA$ and $\matB$ both horizontally and vertically, i.e., $\Divx,\Divz,$ and $\Divy > 1$. We term this \emph{combinatorial} data partitioning.

\textbf{Example 2 (polynomial coded matrix multiplication~\cite{POLY:NIPS17}):} To use polynomial coding, the master uses inner-product data partitioning. Thus, $\Divz=1$. To allocate tasks to $\ND$ workers, in the encoding phase the master applies polynomial codes (separately) to $\Ai{\setxi{i} \times [\AB]}$ and $\Bi{[\AB]\times \mathcal{{S}}_{\Ydir j}}$, where $i,j \in [\sqrt{\DM}]$. This generates $\ND$ \emph{encoded} submatrices $\Ac{\nd}$ and $\Bc{\nd}$, where $\nd\in [\ND]$\footnote{In our setting, the value that the polynomial is evaluated for each worker corresponds to the index of the worker.}. For example, if $\DM=4$ ($\Divx=\Divy=\sqrt{\DM}=2$), 
\begin{align*}
\Ac{\nd} &= \inR{(\Ai{\setxi{1} \times [\AB]} + \Ai{\setxi{2} \times [\AB]} \nd)}{\frac{\Aa}{2}}{\AB}, \\ 
\Bc{\nd} &=\inR{(\Bi{[\AB]\times \mathcal{{S}}_{\Ydir 1}} + \Bi{[\AB]\times \mathcal{{S}}_{\Ydir 2}} \nd^2)}{\AB}{\frac{\Bb}{2}}.
\end{align*}
In the distribution phase the master sends the $\Ac{\nd}$ and $\Bc{\nd}$ matrices to worker $\nd \in [\ND]$ and in the computation phase the $\nd$th worker multiplies the $\Ac{\nd}$ and $\Bc{\nd}$ matrices and sends back the resulting encoded matrix product to the master. 
 For polynomial codes of information dimension $\DM$, the \recoverthr is $\rt{\text{Poly}}=\DM=\Divx\Divy$ (cf.~\cite[Eq.~(3)]{POLY:NIPS17}). For example, if $\DM=4$, in the result aggregation phase the master is required to receive at least $\rt{\text{Poly}}=4$ completed encoded products. For $\DM=4$, the polynomial $\Ac{x}\Bc{x}$ is a polynomial of degree $3$ and therefore can be recovered via polynomial interpolation~\cite{POLY:NIPS17} from any $\rt{\text{Poly}}=4$ distinct values. Once the master receives any $4$ results, in the decoding phase it recovers the $\matA\matB$ matrix via polynomial interpolation.
 
\textbf{Example 3 (MatDot coded matrix multiplication~\cite{MATDOT:2018}):} In contrast to polynomial codes, MatDot codes use outer-product data partitioning. Thus, $\Divx=\Divy=1$. In the encoding phase the master leverages polynomials to generate $\ND$ pairs of \encdatachunks $(\Ac{\nd},\Bc{\nd})$, where $ \nd \in [\ND]$ from the $2\DM$ submatrices $\Ai{[\Aa] \times \setzi{i}}$ and $\Bi{\mathcal{{S}}_{\Zdir i} \times [\Bb] }$, where $i \in {\DM}$. For $\DM=4$ ($\Divz=4$), the \encdatachunks are 
\begin{align*}
\Ac{\nd}&=\inR{(\Ai{[\Aa] \times \setzi{1}} + \Ai{[\Aa] \times \setzi{2}} \nd)}{\Aa}{\frac{\AB}{4}}, \\ \Bc{\nd}&=\inR{(\Bi{\mathcal{{S}}_{\Zdir 1} \times [\Bb] } \nd + \Bi{\mathcal{{S}}_{\Zdir 2} \times [\Bb] })}{\frac{\AB}{4}}{\Bb}. 
\end{align*}
In the distribution phase the master sends the $\Ac{\nd}$ and $\Bc{\nd}$ matrices to the $\nd$th worker and in the computation phase, this worker calculates the $\Ac{\nd}\Bc{\nd}$ matrix product and sends its result to the master.  
For MatDot codes, $\rt{\text{Mat}}=2\DM-1=2\Divz-1$ (cf.~\cite[Theorem III.1]{MATDOT:2018}). For example, if $\DM=4$, then in the result aggregation phase the master needs to receive at least $\rt{\text{Mat}}=7$ completed encoded products. Once the master receives any $\RT_{\text{Mat}}$ results, in the decoding phase it interpolates the polynomial $\Ac{\nd}\Bc{\nd}$ and recovers the $\matA\matB$ product.

\subsection{Performance Measures}
\label{Sec:metrics}
To compare the performance of different coding schemes, we now introduce the measures we use. We start by grouping the measures into four categories: communication, worker computation, encoding, and decoding costs. 

\textbf{Communications cost:}
We define the \emph{input communication load} of the $\nd$th worker as the number of real numbers that the master distributes to the $\nd$th worker in the distribution phase. We use \emph{output communication load} to describe the communication load in the results aggregation phase which is generally not equal to that of the distribution phase. We use $\commi{\nd}$ and $\commo{\nd}$ to denote the respective loads. 

As an illustration, Table~\ref{table:metrics} summarizes $\commi{\nd}$ and $\commo{\nd}$ for the polynomial and MatDot coding examples of~Sec.~\ref{Sec:system}. For polynomial codes, since the input matrices are $\inR{\Ac{\nd}}{\frac{\Aa}{\sqrt{\DM}}}{\AB}$ and $\inR{\Bc{\nd}}{\AB}{\frac{\Bb}{\sqrt{\DM}}}$, the input and output communication load of $\nd$th worker is obtained by $\commi{\nd}={(\Aa\AB + \Bb\AB)}/{\sqrt{\DM}}$ and $\commo{\nd}={\Aa\Bb}/{\DM}$. For MatDot codes, the input matrices are of dimension $\Aa \times \AB/\DM$ and $ \AB/\DM \times \AB$, and thus $\commi{\nd}={(\Aa\AB + \AB\Bb)}/{\DM}$ and $\commo{\nd}=\Aa\Bb$.

In addition to load, the time to communicate is also an important measure. We use $\tcommi{\nd}$ to denote the $\nd$th worker's \emph{communication time}, the time to communicate a scalar real number between master and worker $\nd$ (in either direction).
 We assume a linear relation, $\commi{\nd}\tcommi{\nd}$ to compute the $\nd$th worker's communication time for conveying $\commi{\nd}$ real numbers. In other words, given $\tcommi{\nd}$ it takes the master $\commi{\nd}\tcommi{\nd}$ sec to convey the input data consisting of $\commi{\nd}$ real numbers to the $\nd$th worker. Similarly, it takes the $\nd$th worker $\commo{\nd}\tcommi{\nd}$ sec to convey the output data $\commo{\nd}$ to the master.

\textbf{Worker Computation Cost:}
Similar to communication cost, we define both load and latency measurements for worker computation cost. The \emph{computation load}, denoted by $\comp{\nd}$, measures the number of multiply-and-accumulate operations that worker $\nd$ performs in the computation phase. For a fixed $\DM$, in both examples of Sec.~\ref{Sec:system}, the $\nd$th worker has the same computation load: $\comp{\nd}={\Aa\AB\Bb}/{\DM}$, c.f.~Table~\ref{table:metrics}. 

The latency measurement indicates the amount of time it would take worker $\nd$ to compute a multiply-and-accumulate operation. We call this the \emph{computation time} of the $\nd$th worker and denote it $\tcompi{\nd}$. We assume workers make linear progress. Given $\tcompi{\nd}$, worker $\nd$ would take $\comp{\nd}\tcompi{\nd}$ sec to perform $\comp{\nd}$ multiply-and-accumulate operations. 

\begin{table}
\begin{center}
\begin{tabular}{|l|l|l|l|} \hline
 & $\commi{\nd}$ & $\comp{\nd}$ & $\commo{\nd}$\\ \hline
 Example 2 & $\frac{\Aa\AB}{\sqrt{\DM}} + \frac{\AB\Bb}{\sqrt{\DM}}$ & $\frac{\Aa\AB\Bb}{\DM}$& $\frac{\Aa\Bb}{\DM}$\\ \hline
 Example 3 & $\frac{\Aa\AB}{\DM} + \frac{\AB\Bb}{\DM}$ & $\frac{\Aa\AB\Bb}{\DM}$ & $\Aa\Bb$\\ \hline
\end{tabular}
\end{center}
 \caption{Comparison of polynomial codes and MatDot codes across load measurements.}
 \label{table:metrics}
\end{table}

\textbf{Encoding Cost:}
\label{encc}
The encoding cost of a scheme counts the number of multiply-and-accumulate operations required. We use $\encc{\nd}$ to denote the encoding cost of worker $\nd$. In the polynomial code example, to generate $\Ac{\nd}$ we first multiply size $\floor{\frac{\Aa}{\sqrt{\DM}}}\times \AB$ matrices by a scalar. We then sum $\sqrt{\DM}$ matrices of dimensions $\floor{\frac{\Aa}{\sqrt{\DM}}}\times \AB$. This requires $\Aa\AB$ multiply-and-accumulate operations. Similarly, to generate $\Bc{\nd}$, $\AB\Bb$ multiply-and-accumulate operations are required. Therefore, the encoding cost of worker $\nd$ in the example is $\encc{\nd}=\Aa\AB + \AB\Bb$. This is also equal to $\encc{\nd}$ for the MatDot codes example. We note that, in the case of dealing with very large matrices, where $\DM \ll \min(\Aa,\AB,\Bb)$, the encoding cost is negligible in comparison to the computation load. In particular, $\Aa\AB+\AB\Bb \ll {\Aa\AB\Bb}/{\DM}$ in the two examples. Therefore, we ignore the encoding time when calculating the finishing time in Sec.~\ref{CHAPTER:THEORY}.

\textbf{Decoding Cost:}
Decoding complexity depends on many factors such as the hardware specification and the decoding algorithm implementation. For instance, in both Examples~2 and~3 decoding complexity is governed by the complexity of interpolating a degree $\RT-1$ polynomial, which is order $\mathcal{O}((\RT-1) \log^2 (\RT-1))$~\cite{POLY:NIPS17}.

\section{A Unified Geometric Model}
\label{SEC:geometric}
We now present a conceptual framework wherein the decomposition of a matrix multiplication task into smaller computations can geometrically be visualized as the partitioning of a three-dimensional cuboid. This visualization would be used in the data partitioning phase of Sec.~\ref{Sec:system}. In this section we show that the data partitioning phase in the various prior coded matrix multiplication approaches, e.g., polynomial~\cite{POLY:NIPS17}, MatDot~\cite{MATDOT:2018} codes, and others, corresponds to different partitions of the cuboid. Starting from this geometric perspective, in Sec.~\ref{SEC:HMM} we design hierarchical coding in such a way that this idea can immediately be combined with all coded matrix multiplication schemes. 

\subsection{$3$D Visualization: Standard Matrix Multiplication} 
\label{SEC:3dmm}
Standard (non-coded) matrix multiplication techniques for computing the product $\matA\matB$, where $\inR{\matA}{\Aa}{\AB}$ and $\inR{\matB}{\AB}{\Bb}$, require $\Aa\AB\Bb$ \emph{\MAC}operations, each of which is a multiply-and-accumulate. This \MAC operation $\fMAC: \mathbb R \times \mathbb R \times \mathbb R \rightarrow \mathbb R$ is defined pointwise as $\fMAC(a,b,c)=ab+c$. One method to compute each entry of $\matA\matB$ is iteratively to apply the \MAC operation $\AB$ times to calculate an inner product (cf., Alg.~\ref{Alg:MM}). 
\begin{figure}[ht]
  \centering
  \begin{minipage}{.7\linewidth}
\removelatexerror
    \begin{algorithm}[H]\label{Alg:MM}
    \caption{[$\matC$] = \MMfunc ($\matA,\matB$) }
\textbf{Input}:$\inR{\matA}{\Aa}{\AB}$, $\inR{\matB}{\AB}{\Bb}$, \\
\textbf{Output}:$\inR{\matC}{\Aa}{\Bb}$
\begin{algorithmic}[1]
\State	\emph{\textbf{forall} $\Aai \in [\Aa], \Bbi \in [\Bb]$}: 
\State $\,\,$ $\eC{\Aai}{\Bbi}=0$
\State $\,\,$ \emph{\textbf{for} $\ABi \in [\AB]$}: 
\State $\,\,\,\,$ $\eC{\Aai}{\Bbi} = \fMAC(\eA{\Aai}{\ABi},\eB{\ABi}{\Bbi},\eC{\Aai}{\Bbi})$ 
\State $\,\,$ \emph{\textbf{end for}}
\BState \emph{\textbf{end forall}}
\end{algorithmic}
\end{algorithm}
  \end{minipage}
\end{figure}

Each \MAC operation is indexed by a positive integer triple $(\Aai, \ABi, \Bbi) \in \idxset = [\Aa] \times [\AB] \times [\Bb]$ such that the pairs $(\Aai, \ABi)$ and $(\ABi,\Bbi)$ index the entries of $\matA$ and $\matB$ that serve as the $a$ and $b$ entries in $\fMAC(a,b,c)$. In $3$D space each integer triple $(\Aai, \ABi, \Bbi)$ can be visualized as indexing a unit cube situated within a rectangular cuboid of integer edge lengths $(\Aa, \AB, \Bb)$ (cf., Fig.~\ref{FIG:dmm}). The unit square in either the $\Xdir\Zdir$ or $ \Zdir \Ydir$ plane corresponds to the index pair $(\Aai,\ABi)$ or $(\ABi,\Bbi)$ and geometrically specifies the $\eA{\Aai}{\ABi}$ or $\eB{\ABi}{\Bbi}$ element of the $\matA$ or $\matB$ matrix. Each unit square in the $\Xdir\Ydir$ plane represents an entry of the desired matrix product $\matA\matB$. 

\subsection{$3$D Visualization: Data Partitioning in Coded Computing}
\label{SEC:3dcoded}
We now employ the 3D cuboid visualization to understand the data partitioning phase of coded matrix multiplication. The partitioning of the $\matA\matB$ product into $\DM=\Divx\Divz\Divy$ computations can be visualized as a partitioning of the 3D cuboid. This partitioning \emph{slices} the cuboid into $\DM$ equal-sized subcuboids by making $\Divx-1$ parallel cuts along the $\Xdir$-axis, $\Divz-1$ parallel cuts along the $\Zdir$-axis, and $\Divy-1$ parallel cuts along the $\Ydir$-axis. By specifying the axes that are sliced, we can categorize all possible cuboid partitionings into eight possible categories. Each of the eight partitioning categories is defined by slicing the $3$D cuboid along a subset of directions $\{\Xdir,\Zdir,\Ydir\}$. There are eight possibilities --- depending on whether or not a slice is made along each axis. All eight possible partitionings are represented by a tree in Fig.~\ref{FIG:cluster}. Slicing or not slicing along each axis is illustrated by labeling the respective edge of the tree with a ``1'' or a ``0''. 

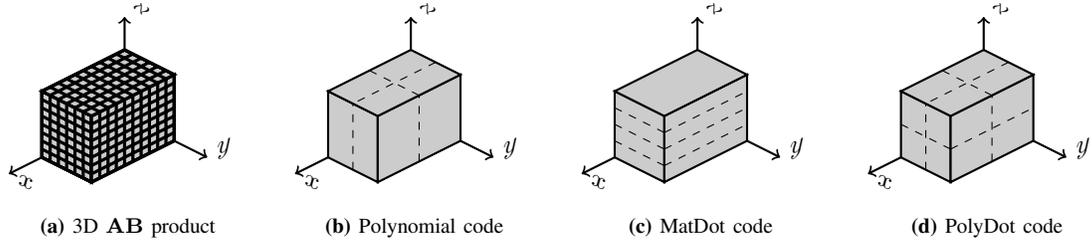
\begin{figure*}[h]
\centering 
		\subfloat[3D $\matA\matB$ product]{\tikzset{every mark/.append style={scale=0.8}}	
\begin{tikzpicture} 
        [cube/.style={very thick,black},
            grid/.style={very thin,gray},
            axis/.style={->,black,thick}]

 \begin{scope}[scale=0.22,yshift=160,xshift=75]
 \begin{scope}[every node/.append style={yslant=-0.5},yslant=-0.5]
 [cube/.style={very thick,black},
            axis/.style={->,blue,thick}]
   \draw[axis] (5,5,0) -- (-4,-4,0) node[anchor=west]{$\Xdir$};
   \draw [step=0.5cm,very thick,fill=black!20!white](-2,-2) grid +(3,4) rectangle (-2,-2);
 \end{scope}

 \begin{scope}[every node/.append style={yslant=0.5},yslant=0.5]
 \draw[axis] (3,0,0) -- (8,-5,0) node[anchor=west]{$\Ydir$};

   \draw[step=0.5cm,very thick,fill=black!20!white] (1,-3) grid +(5,4) rectangle (1,-3);

 \end{scope}
 \begin{scope}[every node/.append style={
     yslant=0.5,xslant=-1},yslant=0.5,xslant=-1
   ]
   \draw[axis] (7,4,0) -- (9,6,0) node[anchor=west]{$\Zdir$};
   \draw[step=0.5cm,very thick,fill=black!20!white] (7,4) grid +(-5,-3)  rectangle (7,4);
 \end{scope}
 \end{scope}
 \end{tikzpicture}
    \label{FIG:dmm}}         
		\quad
		\subfloat[Polynomial code]{\tikzset{every mark/.append style={scale=0.8}}	
	 \begin{tikzpicture}
        [cube/.style={very thick,black},
            grid/.style={very thin,gray},
            axis/.style={->,black,thick}]

 \begin{scope}[scale=0.22,yshift=160,xshift=75]
 \begin{scope}[every node/.append style={yslant=-0.5},yslant=-0.5]
 [cube/.style={very thick,black},
            axis/.style={->,blue,thick}]
   \draw[axis] (5,5,0) -- (-4,-4,0) node[anchor=west]{$\Xdir$};
   \draw [step=0.5cm,thick,fill=black!20!white](-2,-2) rectangle +(3,4);
   \draw [dashed] (-0.5,-2) -- (-0.5,2);
 \end{scope}

 \begin{scope}[every node/.append style={yslant=0.5},yslant=0.5]
 \draw[axis] (3,0,0) -- (8,-5,0) node[anchor=west]{$\Ydir$};

   \draw[step=0.5cm,thick,fill=black!20!white] (1,-3) rectangle +(5,4);
   \draw [dashed] (3.5,-3) -- (3.5,1);
 \end{scope}

 \begin{scope}[every node/.append style={
     yslant=0.5,xslant=-1},yslant=0.5,xslant=-1
   ]
   \draw[axis] (7,4,0) -- (9,6,0) node[anchor=west]{$\Zdir$};
   \draw[step=0.5cm,thick,fill=black!20!white] (7,4) rectangle +(-5,-3);
   \draw [dashed] (7,2.5) -- (2,2.5);
   \draw [dashed] (4.5,4) -- (4.5,1);  
 \end{scope}
 \end{scope}
 \end{tikzpicture}        
       \label{FIG:xy} }
       	\quad	
      \subfloat[MatDot code]{\tikzset{every mark/.append style={scale=0.8}}	
\begin{tikzpicture}
        [cube/.style={very thick,black},
            grid/.style={very thin,gray},
            axis/.style={->,black,thick}]

 \begin{scope}[scale=0.22,yshift=160,xshift=75]
 \begin{scope}[every node/.append style={yslant=-0.5},yslant=-0.5]
 [cube/.style={very thick,black},
            axis/.style={->,blue,thick}]
   \draw[axis] (5,5,0) -- (-4,-4,0) node[anchor=west]{$\Xdir$};
   \draw [step=0.5cm,thick,fill=black!20!white](-2,-2) rectangle +(3,4);
   \draw [dashed] (-2,-1) -- (1,-1);
   \draw [dashed](-2,0) -- (1,0);
   \draw [dashed](-2,1) -- (1,1);
 \end{scope}

 \begin{scope}[every node/.append style={yslant=0.5},yslant=0.5]
 \draw[axis] (3,0,0) -- (8,-5,0) node[anchor=west]{$\Ydir$};
   \draw[step=0.5cm,thick,fill=black!20!white] (1,-3) rectangle +(5,4);
   \draw [dashed](1,-2) -- (6,-2);
   \draw [dashed](1,-1) -- (6,-1);
   \draw [dashed](1,0) -- (6,0);
 \end{scope}

 \begin{scope}[every node/.append style={
     yslant=0.5,xslant=-1},yslant=0.5,xslant=-1
   ]
   \draw[axis] (7,4,0) -- (9,6,0) node[anchor=west]{$\Zdir$};
   \draw[step=0.5cm,thick,fill=black!20!white] (7,4) rectangle +(-5,-3);
 \end{scope}
 \end{scope}
 \end{tikzpicture}       
       \label{FIG:z} }
       \quad
       \subfloat[PolyDot code]{\tikzset{every mark/.append style={scale=0.8}}
	   \begin{tikzpicture}
        [cube/.style={very thick,black},
            grid/.style={very thin,gray},
            axis/.style={->,black,thick}]

 \begin{scope}[scale=0.22,yshift=160,xshift=75]
 \begin{scope}[every node/.append style={yslant=-0.5},yslant=-0.5]
 [cube/.style={very thick,black},
            axis/.style={->,blue,thick}]
   \draw[axis] (5,5,0) -- (-4,-4,0) node[anchor=west]{$\Xdir$};
   \draw [step=0.5cm,thick,fill=black!20!white](-2,-2) rectangle +(3,4);
   \draw [dashed] (-0.5,-2) -- (-0.5,2);
      \draw [dashed](-2,0) -- (1,0);
 \end{scope}

 \begin{scope}[every node/.append style={yslant=0.5},yslant=0.5]
 \draw[axis] (3,0,0) -- (8,-5,0) node[anchor=west]{$\Ydir$};

   \draw[step=0.5cm,thick,fill=black!20!white] (1,-3) rectangle +(5,4);
   \draw [dashed] (3.5,-3) -- (3.5,1);
      \draw [dashed](1,-1) -- (6,-1);
 \end{scope}

 \begin{scope}[every node/.append style={
     yslant=0.5,xslant=-1},yslant=0.5,xslant=-1
   ]
   \draw[axis] (7,4,0) -- (9,6,0) node[anchor=west]{$\Zdir$};
   \draw[step=0.5cm,thick,fill=black!20!white] (7,4) rectangle +(-5,-3);
   \draw [dashed] (7,2.5) -- (2,2.5);
   \draw [dashed] (4.5,4) -- (4.5,1);  
 \end{scope}
 \end{scope}
 \end{tikzpicture}         \label{FIG:xyz}}
	\quad	
       \caption{(a) $3$D visualization of the basic operations involved in matrix multiplication where $\Aa=10, \Bb=6,$ and $\AB=8$. (b)-(d) Cuboid partitioning structure for: (b) Polynomial codes where $\Divx=\Divy=2,\DM=4,\RT=4,$ and $\ND\geq 4$; (c) MatDot codes where $ \Divz=4, \DM=4, \RT=7,$ and $\ND\geq 7$; (d) PolyDot codes, where $\Divx=\Divy=\Divz=2,\DM=8, \RT=12,$ and $\ND\geq 12$.}
\end{figure*}

\begin{figure*}[h]
	\centering 
	 \includegraphics[width=0.95\textwidth]{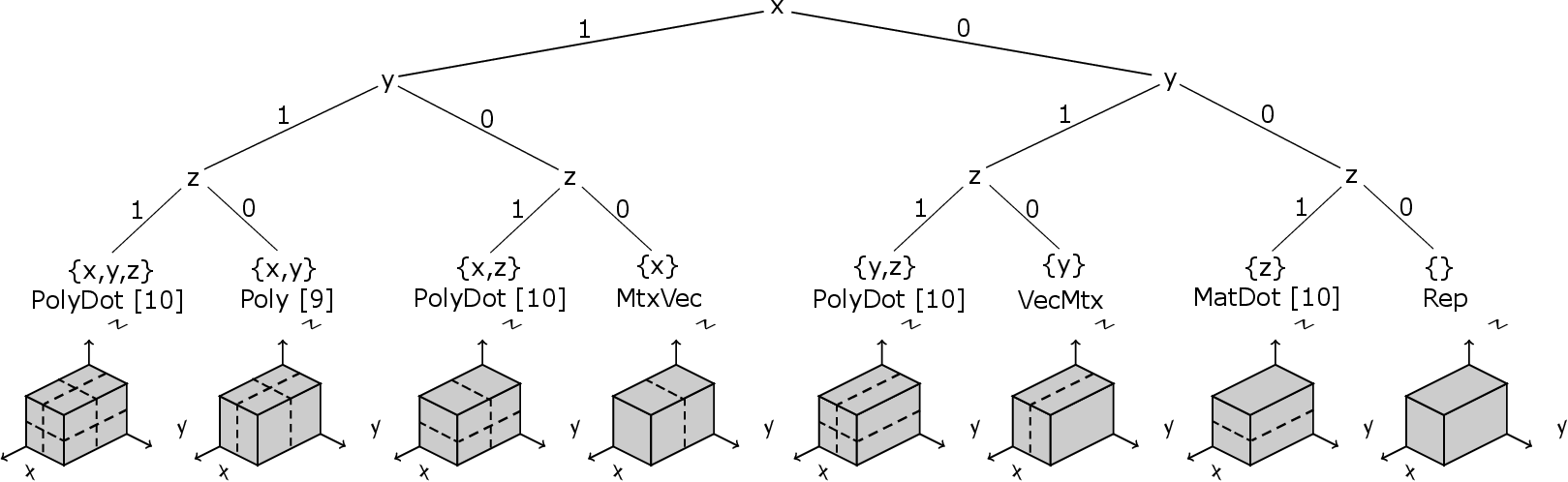}
    \caption{Decision tree for the 3D cuboid partitioning structure, where ``1'' and ``0'' on the edges of the tree identify the slicing and not slicing along each axis, respectively. The $x$-axis is sliced in the top branch, the $y$-axis in the middle branches, and the $z$-axis in the lowest branches. We use ``Rep'', ``MtxVec'', and ``VecMtx'' as the respective abbreviation for repetition codes, matrix-vector multiplication, and vector-matrix multiplication. }
    \label{FIG:cluster}
     \vspace*{-4ex}
\end{figure*}

Each partitioning structure corresponds to a distinct approach to coded matrix multiplication. For instance, product codes~\cite{ PRODUCT:ISIT17} and polynomial codes~\cite{POLY:NIPS17} correspond to slicing along the $\Xdir$- and $\Ydir$-axes. This is because both of these coding schemes use inner-product partitioning in their data partitioning phase. Recall from Sec.~\ref{SEC:model} that in inner-product partitioning with information dimension $\DM=\Divx\Divy$ the $\matA$ matrix is divided horizontally into $\Divx$ submatrices and the $\matB$ matrix is divided vertically into $\Divy$ submatrices. Such a decomposition slices the 3D cuboid into $\DM$ equal-sized subcuboids by making $\Divx-1$ parallel cuts along the $\Xdir$-axis and $\Divy-1$ parallel cuts along the $\Ydir$-axis. This partitioning is depicted in Fig.~\ref{FIG:xy} for $\Divx=\Divy=2$. MatDot codes~\cite{MATDOT:2018} slice along the $\Zdir$-axis, resulting in outer-product data partitioning. Recall that in outer-product partitioning to encode the $\matA\matB$ product, the $\matA$ matrix is divided vertically into $\Divz$ submatrices and the $\matB$ matrix is divided horizontally into $\Divz$ submatrices. As is illustrated in Fig.~\ref{FIG:z}, the 3D cuboid partitioning for MatDot codes involves $\DM-1=3$ slices along the $\Zdir$-axis. 

 PolyDot codes~\cite{MATDOT:2018} and entangled polynomial codes~\cite{ENTGL:ISIT18} slice the 3D cuboid along all ($\Xdir$-, $\Ydir$-, $\Zdir$-) axes. This is what we termed combinatorial partitioning (cf., Fig.~\ref{FIG:xyz}). Slicing along either ($\Xdir$-,$\Zdir$-) or ($\Ydir$-,$\Zdir$-) axes are special cases of slicing along the ($\Xdir$-, $\Ydir$-, $\Zdir$-) axes, where only one matrix is split both horizontally and vertically and the other matrix is split in only one direction. Slicing along $\Xdir$- or $\Ydir$-axis specifically models coded vector-matrix or coded matrix-vector multiplication. It also, more generally, models coded matrix multiplication in which only one matrix is split. The last category, denoted by the empty set, represents an $(\ND,\ND)$ repetition code in which each of the $\ND$ workers is tasked with completing the entire $\matA\matB$ product.
	
The $\DM$ equal-sized subcuboids into which the 3D cuboid is partitioned correspond to $\DM$ distinct computations. We use the term \emph{\infoblock}to refer to each of these subcuboids. Note that the number of information blocks is equal to the information dimension $\DM$ of the error correction code that we introduced earlier.

\section{Hierarchical Coded Matrix Multiplication}
\label{SEC:HMM}
Building from our cuboid perspective, we now introduce hierarchical coded computing. The intuition underlying hierarchical coding is to subdivide the single task that each worker is assigned with in (non-hierarchical) coded computing into a number of smaller subtasks. The division into a number of subtasks allows each worker to make progress through its assigned tasks in a more continuous, incremental manner. Since each worker progresses through a sequence or {\em hierarchy} of subtasks, we term these hierarchical methods. Once we break each (original) task into the smaller subtasks we have flexibility in how to apply coding across the subtasks. We present three alternatives: BICC (bit-interleaved coded computing), MLCC (multilevel coded computing), and HHCC (hybrid hierarchical coded computing). Each has its respective strengths.

\begin{figure*}[h] 
 \centering
\subfloat[BICC]{ \includegraphics[keepaspectratio,height=0.4\columnwidth, width=0.6\columnwidth]{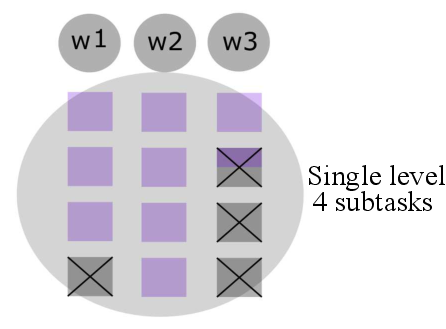}\label{toy_bicc}}
	\quad 
\subfloat[MLCC]{ \includegraphics[keepaspectratio,height=0.4\columnwidth, width=0.6\columnwidth]{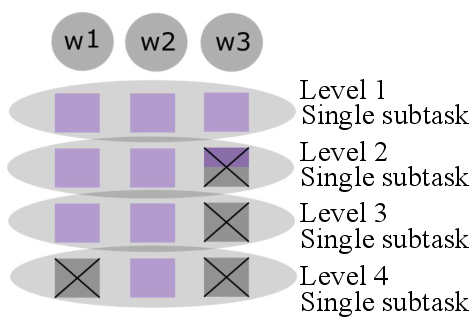}\label{toy_mlcc}}
	\quad 
\subfloat[HHCC]{ \includegraphics[keepaspectratio,height=0.4\columnwidth, width=0.6\columnwidth]{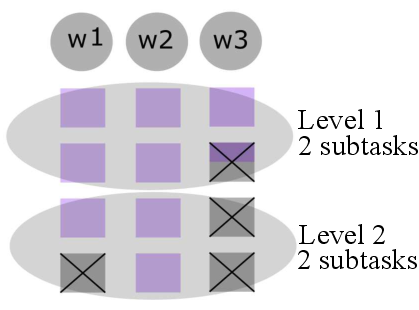}\label{toy_hhcc}}
	\quad 
  \caption{Toy examples of non-hierarchical and hierarchical coded computation, where we need 2 out of 3 computations to be completed and recovery thresholds are (a) $\RT_{\text{BICC}}=8$, (b) $\RT_{\ly} \in \{3,2,2,1\}$, and (c) $\RT_{\ly,\text{HHCC}}\in \{5,3\}$.}\label{toy}
\end{figure*}

In Fig.~\ref{toy} we illustrate a toy example for each scheme. Figure~\ref{toy_bicc} corresponds to BICC, Fig.~\ref{toy_mlcc} to MLCC, and Fig.~\ref{toy_hhcc} to HHCC. We assume that we have $\ND=3$ workers, and we need two third of computations to be completed in order to complete the main computational job. In all three figures each worker is provided $\Stask=4$ encoded subtasks. In traditional non-hierarchical coded computing $P_{\text{Non-h}}=1$, and there would be a single task for each worker. As Fig.~\ref{toy_bicc} illustrates, one hierarchical approach is BICC that applies a single code across all the subtasks, such that the block length of the code, which equals the number of encoded subtasks, is $\ND\Stask = 12$. Two third of these subtasks, which results in the recovery threshold $\RT_{\text{\bicc}}=8$, are required to be completed. 

 A second hierarchical approach is MLCC that groups the subtasks into $\LY=4$ levels of computation. Each level contains a single subtask for each worker, hence $\Stask_l=1$ for all $\ly \in [4]$ (cf. Fig.~\ref{toy_mlcc}). In MLCC the master collects the eight completed subtasks through four smaller codes each of which has a block length $\ND=3$. The $8$ completed subtasks should be split into per-level recovery thresholds $\rt{\ly}\in \{3,2,2,1 \}$, that satisfy $\sum_{\ly\in[4]} \rt{\ly} =8$. We note that since all workers start with their first (encoded) subtask and sequentially work their way through their $\LY$ subtasks, more workers will complete their first subtask than their $4$-th. Therefore, we (in general) use a higher-rate code, i.e., a larger recover threshold $\RT_{\ly}$, for $\ly=1$ than for $\ly=4$. 
 
 Finally, as is illustrated in Fig.~\ref{toy_hhcc}, in HHCC we combine these two ideas. In this example, there are $\LY_{\text{HHCC}}=2$ levels of computation but each worker also has $\Stask_{\ly,\text{HHCC}}=2$ subtasks per level $, \ly\in[2]$. The block length of both levels is thus $\Stask_{\ly,\text{HHCC}}\ND=6$, while it is $12$ in BICC and 3 in MLCC. The master collects eight required subtasks through two codes with recovery thresholds $\rt{1,\text{HHCC}}=5$ and $\rt{2,\text{HHCC}}=3$, so that $\sum_{\ly} \rt{\ly,\text{HHCC}} = 8$.
 
One way of conceiving of the difference between BICC and MLCC is by analogy with the two coded modulation techniques: BICM (bit-interleaved coded modulation)~\cite{BICM:1998} and MLC (multilevel coding)~\cite{MLCC:1999}. In BICM a vector of encoded bits is first generated from information bits using a binary encoder and then is permuted using a bit-level interleaver. This single interleaved vector is then passed to the mapper. In contrast, an alternative technique is MLC in which the information bits are first separated into multiple smaller message vectors. Next, each message vector is independently encoded and is passed to the mapper. The mapper receives multiple encoded bits in MLC, while in BICM a single vector of interleaved bits is passed to the mapper. This distinction is analogous to that of BICC and MLCC. We use the term bit-interleaved and multilevel for our proposed coded computation schemes to highlight such parallels.

The categorization of traditional non-hierarchical coded computing, BICC, MLCC, and HHCC is presented in Table~\ref{table:payoff}. There are two design axes. The first is the number of codes used. A single code, which corresponds to a single level of the hierarchy (non-hierarchical and BICC) or multiple codes, i.e., multiple levels (MLCC and HHCC).  The other design axis is the number of encoded subtasks each worker is assigned per level. Either a single subtask (non-hierarchical and MLCC) or multiple subtasks (BICC and HHCC) per level. In Table~\ref{table:terminology} we summarize the notation we use for both hierarchical and non-hierarchical schemes.

 \begin{table*}[h]
  	\centering
    \setlength{\extrarowheight}{2pt}
    \begin{tabular}{*{4}{c|}}
      \multicolumn{2}{c}{} & \multicolumn{2}{c}{}\\\cline{3-4}
      \multicolumn{1}{c}{} &  & Single subtask per level & Multi-subtask per level \\\cline{2-4}
      \multirow{2}*{}  & Single level & \makecell{Non-h\\($\LY_{\text{Non-h}}=1,\Stask_{\text{Non-h}}=1$)} &  \makecell{BICC\\($\LY_{\text{\BICC}}=1,\Stask>1$)} \\\cline{2-4}
      & Multi-level & \makecell{ MLCC\\($\LY>1,\Stask_{\ly}=1$)} & \makecell{HHCC\\($\LY_{\text{HHCC}}>1,\Stask_{\ly,\text{HHCC}}>1$)} \\\cline{2-4}
    \end{tabular}
     \caption{Parameter decision in different methods of hierarchical coding and non-hierarchical schemes.}
 \label{table:payoff}
  \end{table*}
  
\begin{table*}[h]
\center
\begin{tabular}{|l|l|l|l|l|} \hline
 & \makecell{Non-hier} & \makecell{BICC} & \makecell{MLCC} & \makecell{HHCC} \\ \hline
 Block length & $\ND$ & $\ND\Stask$& $\ND$& $\{\ND\Stask_{\ly}\}_{\ly \in [\LY_{\text{HHCC}}]}$ \\ \hline
 \makecell{Information \\ dimension} & $\DM$& $\DM_{\text{\bicc}}=\DM\Stask$& $\DM_{\ly}, \sum\limits_{\ly=1}^{{\Stask}} \DM_{\ly} = \dmsum = \DM \Stask$ & $\DM_{\ly,{\text{HHCC}}}, \sum\limits_{\ly=1}^{\LY_{\text{HHCC}}} \DM_{\ly,\text{HHCC}} = \DM \Stask$ \\ \hline
 \makecell{Recovery \\ threshold} &$\RT$& $\RT_{\text{\bicc}}\approx \RT\Stask$& $\RT_{\ly}, \sum\limits_{\ly=1}^{{\Stask}} \RT_{\ly} \approx \RT\Stask$ & $\RT_{\ly,{\text{HHCC}}}, \sum\limits_{\ly=1}^{\LY_{\text{HHCC}}} \RT_{\ly,\text{HHCC}} \approx \RT\Stask$ \\ \hline
 \makecell{$\#$ Levels} & 1 & $\LY_{\text{\bicc}}=1$ & $\LY=\Stask$& $\LY_{\text{HHCC}} \in [ \Stask]$\\ \hline
 	\makecell{\makecell{$\#$ Subtasks}\\ per level} & 1 & $\Stask$ & $\Stask_l = 1, \sum\limits_{\ly=1}^{{\Stask}} \Stask_{\ly} = \Stask$ & $\Stask_{\ly,\text{HHCC}} \in [\Stask], \sum\limits_{\ly=1}^{\LY_{\text{HHCC}}} \Stask_{\ly,\text{HHCC}} = \Stask$ \\ \hline 
 \end{tabular}
 \caption{Summary of parameters used in this paper.}
 \label{table:terminology}
\end{table*}

\subsection{Bit-Interleaved Coded Matrix Multiplication}
\label{BICC}

In the following We detail the six phases of the system model for BICC with parameters $(\LY_{\text{\BICC}}=1,\DM_{\text{\BICC}},\Stask,\ND,\RT_{\text{\BICC}})$. 

\textbf{Data partitioning phase and cuboid visualization:} 
Similar to non-hierarchical coding, in BICC $\LY_{\text{\BICC}}=1$. The master directly decomposes the single-level computational job into $\DM_{\text{\BICC}}$ equal-sized sub-computations. From the cuboid perspective, to realize the information dimension $\DM_{\text{\BICC}}$ we first assume that $\DM_{\text{\BICC}}=\Divxi{,\text{\BICC}}\Divzi{,\text{\BICC}}\Divyi{,\text{\BICC}}$. Based on this assumption, we partition the cuboid into $\DM_{\text{\BICC}}$ equal-sized information blocks\footnote{As before, we comment that these information blocks can be allowed to overlap. For simplicity of presentation, we assume they are disjoint.} with $\Divxi{,\text{\BICC}}-1$ slices along the $\Xdir$-axis and  $\Divzi{,\text{\BICC}}-1$ and $\Divyi{,\text{\BICC}}-1$ slices along the $\Zdir$- and $\Ydir$-axes, respectively. Such a partitioning is equivalent to partitioning the matrix $\matA$ into $\Divxi{,\text{\BICC}}\Divzi{,\text{\BICC}}$ equal-sized matrices $\Ai{\seti{\Xdir i}\times\seti{\Zdir j}}$, where $i \in [\Divxi{,\text{\BICC}}]$ and $j \in [\Divzi{,\text{\BICC}}]$. Likewise, the master partitions the matrix $\matB$ into $\Divzi{,\text{\BICC}}\Divyi{,\text{\BICC}}$ equal-sized matrices $\Bi{\seti{\Zdir j}\times\seti{\Ydir k}}$, where $j \in [\Divzi{,\text{\BICC}}]$ and $k \in [\Divyi{,\text{\BICC}}]$. The matrix $\Ai{\seti{\Xdir i}\times\seti{\Zdir j}}$ is of dimension ${\Aa/\Divxi{,\text{\BICC}}}\times{\AB/\Divzi{,\text{\BICC}}}$, and $\Bi{\seti{\Zdir j}\times\seti{\Ydir k}}$ is of dimension ${\AB/\Divzi{,\text{\BICC}}}\times{\Bb/\Divyi{,\text{\BICC}}}$ for all $i \in [\Divxi{,\text{\BICC}}], j \in [\Divzi{,\text{\BICC}}],$ and $k \in [\Divyi{,\text{\BICC}}]$. Note that given $\DM_{\text{\BICC}}$ we have freedom to choose the design parameters $\Divxi{,\text{\BICC}}, \Divzi{,\text{\BICC}}$, and $\Divyi{,\text{\BICC}}$. Different choices yield different performance in terms of the input and output communication time. In next sections we provide and numerically test a specific choice that makes analyses easier. 

\textbf{Encoding phase:} The master encodes $2\DM_{\text{\BICC}}$ matrices $\Ai{\seti{\DMi}}$ and (separately) $\Bi{\mathcal{\bar{S}}_{\DMi}}$, $\DMi \in [\DM_{\text{\BICC}}]$, to generate $\Stask\ND$ pairs of encoded matrices $(\Ac{i}, \Bc{i})$, where $i \in [\Stask\ND]$. For instance, if we follow~\cite{POLY:NIPS17}, polynomials are used to generate encoded matrices. To generate encoded matrices for worker $\nd \in [\ND]$, the master evaluates polynomials $\Ac{x}$ and $\Bc{x}$ at $\Stask$ distinct points. We use $ x \in \{(\nd-1)\Stask+1,\ldots,(\nd-1)\Stask+\Stask \}$. 

\textbf{Distribution phase:} The master transmits $\Stask$ distinct pairs of encoded matrices $(\Ac{(\nd-1)\Stask+\stask},\Bc{(\nd-1)\Stask+\stask})$, where $\stask \in [\Stask]$ to worker $\nd \in [\ND]$. All matrices are distributed to distinct workers, \ie no single matrix is given to two workers. We note that $\Ac{\nd} $ and $\Bc{\nd}$ are, respectively, of dimensions $\Aa/\Divxi{,\text{\BICC}} \times \AB/\Divzi{,\text{\BICC}}$ and $\AB/\Divzi{,\text{\BICC}} \times \Bb/\Divyi{,\text{\BICC}}$.
 
\textbf{Worker computation phase:} The $\nd$-th worker 
first computes $\Cc{(\nd-1)\Stask+1}=\Ac{(\nd-1)\Stask+1}\Bc{(\nd-1)\Stask+1}$ and transmits the result $\Cc{(\nd-1)\Stask+1}$ back to the master. That same worker next computes $\Cc{(\nd-1)\Stask+2}=\Ac{(\nd-1)\Stask+2}\Bc{(\nd-1)\Stask+2}$ and sends the result to the master. Likewise, it sequentially completes subtasks up to $\Stask$ subtasks, transmitting each result to the master. The transmission of partial (per-sub-task) results is a novel aspect of BICC and is an essential aspect required to exploit the work performed by all workers.

\textbf{Result aggregation phase:} The master receives $\Aa/\Divxi{,\text{\BICC}} \times \Bb/\Divyi{,\text{\BICC}}$ matrices sequentially from all workers till it receives at least $\RT_{\text{\BICC}}$ distinct completed subtasks. 

\textbf{Decoding phase:} Once the master collects $\RT_{\text{\BICC}}$ completed subtasks, it starts decoding. The decoding algorithm depends on the code. If the master uses polynomial or MatDot codes in the encoding phase, the master can use a polynomial interpolation algorithm or, equivalently, a Reed-Solomon decoder~\cite{REED:2009}. 

\begin{remark}\label{remark_bicc}
In contrast to the information dimension $\DM$ in non-hierarchical coding, in BICC the information dimension is $\DM_{\text{\BICC}}=\Stask\DM$. That means in BICC the number of equal-sized sub-computations into which the matrix product $\matA\matB$ is partitioned is $\Stask$ times larger than that of non-hierarchical coding. This makes each worker's computational load equal for a fair comparison. Also, the recovery threshold of BICC must be approximately $\Stask$ times larger than the recovery threshold of non-hierarchical coding, i.e., $\RT_{\text{\BICC}}\approx\Stask\RT$. Through this assumption, the same amount of computation is required to be completed in both BICC and non-hierarchical coding.
\end{remark}

\subsection{Multilevel Coded Matrix Multiplication}
\label{SEC:MLCC}
We now introduce the system model for MLCC of parameters $(\LY,\DM_{\ly},\Stask_{\ly}=1,\ND,\RT_{\ly})$. 

\textbf{Data partitioning and cuboid visualization} In contrast to the one-phase cuboid partitioning of previous (bit-interleaved or non-hierarchical) coded matrix multiplication schemes, in MLCC the master partitions the cuboid in two steps. It first partitions the cuboid into $\LY$ subcuboids each of which we think of as a \emph{\level}of computation. We use the term \emph{\taskbox}for each of these ($\LY$) first-stage subcuboids. The \taskbox $\ly\in[\LY]$ can be visualized as a cuboid of dimensions $\Xdimi{\ly} \times \Ydimi{\ly} \times \Zdimi{\ly}$, where 
\begin{align} \label{fixed_volume}
\sum_{l \in [\LY]} \Xdimi{\ly} \Ydimi{\ly} \Zdimi{\ly}   =  \Aa\AB\Bb. 
\end{align}

The above equality holds by the assumption of disjoint task blocks. Partitioning the cuboid into task blocks is equivalent to dividing the matrix $\matA$ into $\LY$ \taskmtxs $\{\Ai{\setxi{\ly} \times\setzi{\ly}}|\, \ly \in [\LY]\}$ and dividing the matrix $\matB$ into $\LY$ \taskmtxs $\{\Bi{\setzi{\ly} \times\setyi{\ly}}|\, \ly \in [\LY]\}$, where $\Xdimi{\ly} = \abs{\setxi{\ly}}, \Zdimi{\ly} = \abs{\setzi{\ly}},$ and $\Ydimi{\ly} = \abs{\setyi{\ly}}$. Through this first step of partitioning, the master decomposes the $\matA\matB$ product into $\LY$ computations $\Ai{\setxi{\ly} \times\setzi{\ly}}\Bi{\setzi{\ly} \times\setyi{\ly}}$.

In the second partitioning step the master subdivides the \taskbox $\ly \in [\LY]$ into $\dm{\ly}=\divx\divz\divy$ equal-sized information blocks each of dimensions ${\Xdimi{\ly}}/{\divx} \times {\Zdimi{\ly}}/{\divz} \times {\Ydimi{\ly}}/{\divy}$. Through this partitioning phase, the master partitions $\Ai{\setxi{\ly}\times\setzi{\ly}}$ into $\divx \divz$ equal-sized matrices denoted by $\Aii{\ly}{\divxi,\divzi}$, where $(\divxi,\divzi) \in [\divx] \times [\divz]$ and $\ly \in [\LY]$. Likewise the matrices $\Bi{\setzi{\ly}\times\setyi{\ly}}$ are partitioned into $\divz \divy$ equal-sized matrices $\Bii{\ly}{\divzi,\divyi}$, where $(\divzi,\divyi) \in [\divz] \times [\divy]$ and $\ly \in [\LY]$. Note that the matrix product $\Ai{\setxi{\ly}\times\setzi{\ly}}\Bi{\setzi{\ly}\times\setyi{\ly}}$ pertinent to the $\ly$ task block is partitioned into $\DM_{\ly}$ matrix multiplication $\Aii{\ly}{\divxi,\divzi}\Bii{\ly}{\divzi,\divyi}$. 

Two comments are in order. First, we assume that $\Xdimi{\ly}, \Zdimi{\ly}$ and $\Ydimi{\ly}$ are, respectively, much larger than $\divx,\divz$ and $\divy$. This assumption allows us to ignore integer effects. Second, if $\volm{\ly}$ denotes the integer volume of the $\ly$th \taskboxno, \ie $\volm{\ly} = \Xdimi{\ly}\Ydimi{\ly}\Zdimi{\ly}$, we choose $\divx,\divz$ and $\divy$ so that $\volm{\ly}/(\divx\divz\divy)$ is (approximately) constant for all $\ly \in [\LY]$. While we need not make this choice, we make it to keep the quanta of computation constant across levels. The implication is that \infoblocks will be of (approximately) constant volume. In particular, we choose there to be $\dmsum=\sum_{\ly \in [\LY]} \dm{\ly}$ \infoblocks each of (approximate) volume $\Aa\AB\Bb/\dmsum$. This assumption will prove useful when computing the response times of workers and when comparing to the results of previous papers. We note that the assumption that we keep 
\begin{align}
\volm{\ly}/(\divx\divz\divy) \approx \Aa\AB\Bb/\dmsum
\label{assumption}
\end{align}
constant does {\em not} mean that the height, width, and depth of \infoblocks must be the same across different \levelsno. Only the volume of information blocks is kept constant. The volume corresponds to the number of basic operations in each computation and thus each subtask is an equal amount of work. 

\textbf{Encoding phase:} For each level $\ly \in [\LY]$, the master generates $\ND$ pairs of encoded matrices $(\cA{\ly}{\nd},\cB{\ly}{\nd})$, $ \nd \in [\ND]$, from the $2\dm{\ly}$ matrices $\Aii{\ly}{\divxi,\divzi}$ and $\Bii{\ly}{\divzi,\divyi}$, where $(\divxi, \divzi, \divyi) \in [\divx] \times[\divz] \times [\divy]$. To do this, the master applies coding across the \datachunks $\Aii{\ly}{\divxi,\divzi}$ and (separately)  $\Bii{\ly}{\divzi,\divyi}$. To generate $\LY$ encoded matrices for worker $\nd \in [\ND]$, $\LY$ distinct codes are used. For instance, by using polynomial codes~\cite{POLY:NIPS17}, the master evaluates $\LY$ pairs of polynomials $(\cA{\ly}{x},\cB{\ly}{x})$, $\ly \in [\LY]$, at $x=n$ to generates encoded matrices for worker $\nd \in [\ND]$. 
  
\textbf{Distribution phase:} The master sends $\LY$ pairs of encoded matrices to each worker. For the $\nd$th worker this is the set $\{(\cA{\ly}{\nd}, \cB{\ly}{\nd}) \, | \, \ly \in [\LY]\}$. Note that, for $\ly \in [\LY]$, the encoded matrices $\cA{\ly}{\nd}$ and $\cB{\ly}{\nd}$ are, respectively, of dimensions ${\Xdimi{\ly}}/{\divx} \times {\Zdimi{\ly}}/{\divz} $ and ${\Zdimi{\ly}}/{\divz} \times {\Ydimi{\ly}}/{\divy}$. 

\textbf{Worker computation phase:} The worker sequentially computes its $\LY$ levels of subtasks, $\cA{1}{\nd}\cB{1}{\nd}$ through $\cA{\LY}{\nd}\cB{\LY}{\nd}$. Note that for the level $\ly\in[\LY]$, worker $\nd$ is tasked with exactly a single task of $\cA{\ly}{\nd}\cB{\ly}{\nd}$ product, i.e., the number of tasks pertinent to level $\ly$ is $\Stask_{\ly}=1$. Each worker sends each completed subtask to the master as soon as it is finished. 

\textbf{Result aggregation phase:} To recover all the \infoblocks that make up the $\ly$th \level of computation (and thus to recover the $\ly$th \taskboxno), the master must receive at least $\rt{\ly}$ jobs from the $\ND$ workers. Any subset of cardinality at least $\rt{\ly}$ of $\{\cA{\ly}{\nd}\cB{\ly}{\nd} \, | \, \nd \in [\ND]\}$ will suffice. We term the choice of per-level recovery thresholds $\{\rt{\ly}\}_{\ly \in [\LY]}$ the \emph{\profileno}. The \profile is something that we optimize in Sec.~\ref{CHAPTER:THEORY}, based on the statistics of the distribution of processing times. 

\textbf{Decoding phase:} 
Due to the independence encoding of each level, the decoding phase of MLCC can be carried out either in a serial, a parallel, or a streaming manner across levels. In serial and parallel decoding the master starts decoding when it receives enough completed jobs from all levels. In serial decoding the master pipelines the decoding. In the parallel decoder, decoding of all levels is run in parallel after that the computation of all levels is finished. In streaming decoding the master starts decoding each level once it receives enough completed jobs of that specific level. The master doesn't need to wait for all levels to be finished.

\begin{remark}\label{remark_mlcc1}
We note that uniform randomly {\em shuffling} the order of levels which workers work through can mitigate the need to design an optimal recovery profile in MLCC. This is indeed a special case of MLCC in which workers randomly pick an encoded subtask (without replacement) and sends its result to the master as soon as it is completed. We term this approach {\em randomized} MLCC (RMLCC). In RMLCC, we set the recovery profile to be $\rt{\ly}=\RT$ for all $\ly \in [\LY]$, thus mitigating the need to profile design. The intuition behind this setting is because of the fact that for $\ND\gg\LY$, on average an equal-sized group of workers would try different permutations of orders. Therefore, each level would get the same attention. This makes the expected finishing time of levels to be approximately the same without a need to design the recovery profile.     
\end{remark}

\begin{remark}\label{remark_mlcc2}
For a fair comparison, in MLCC each worker is tasked with the same amount of computation to that of BICC and non-hierarchical coding. Therefore, we set $\LY=\Stask$ and $\dmsum = \DM \Stask$. Furthermore, in MLCC the same amount of computation compared to BICC is required to be completed, \ie $\sum_{\ly=1}^{\LY} \RT_{\ly}\approx\RT\Stask$.
\end{remark}

\subsection{Hybrid Hierarchical Coded Matrix Multiplication}
We now introduce HHCC as a general hierarchical coding approach, with BICC and MLCC as two special cases. Given ($\LY_{\text{HHCC}},\DM_{\ly,\text{HHCC}},\Stask_{\ly,\text{HHCC}},\ND,\rt{\ly,\text{HHCC}})$, we next detail HHCC.

\textbf{Data partitioning phase and cuboid visualization:} In HHCC the master partitions the cuboid in two steps. In the first step the master partitions the cuboid into $\LY_{\text{HHCC}}$ heterogeneously sized task blocks, the $\ly$th of which is of dimensions $\Xdimi{\ly,\text{HHCC}} \times \Zdimi{\ly,\text{HHCC}} \times \Ydimi{\ly,\text{HHCC}}$. The parameter $\LY_{\text{HHCC}}$ is larger than the number of levels in BICC, which is $\LY_{\text{\BICC}}=1$, and smaller than the number of levels in MLCC, which is $\LY=\Stask$. Therefore, the number of levels in HHCC is somewhere between those of MLCC and BICC, \ie $\LY_{\text{HHCC}} \in [\Stask]$. Assuming that task blocks are disjoint yields a similar constraint to (\ref{fixed_volume}), \ie
\begin{align}
\sum_{\ly \in [\LY_{\text{HHCC}}]} \Xdimi{\ly,\text{HHCC}}\Zdimi{\ly,\text{HHCC}}\Ydimi{\ly,\text{HHCC}} = \Aa\AB\Bb.
\end{align}
Through the first step of partitioning the matrix product $\matA\matB$ is divided into $\LY_{\text{HHCC}}$ computations $\Ai{\setxi{\ly} \times\setzi{\ly}}\Bi{\setzi{\ly} \times\setyi{\ly}}$, where $\ly \in [\LY_{\text{HHCC}}]$, $\abs{\setxi{\ly}}=\Xdimi{\ly,\text{HHCC}}, \abs{\setzi{\ly}}=\Zdimi{\ly,\text{HHCC}},$ and $\abs{\setyi{\ly}}=\Ydimi{\ly,\text{HHCC}}$. 

In the second step the master subdivides the $\ly$th task block into $\DM_{\ly,\text{HHCC}}=\Divxi{\ly,\text{HHCC}}\Divzi{\ly,\text{HHCC}}\Divyi{\ly,\text{HHCC}}$ equal-sized information blocks, where $\ly \in [\LY_{\text{HHCC}}]$. To do this, the master cuts the cuboid with $\Divxi{\ly,\text{HHCC}}-1$ slices along the $\Xdir$-axis, and $\Divzi{\ly,\text{HHCC}}-1$ and $\Divyi{\ly,\text{HHCC}}-1$ slices along $\Zdir$- and $\Ydir$-axes, respectively. Through the second step of partitioning the matrix $\Ai{\setxi{\ly} \times\setzi{\ly}}$ is subdivided into $\Divxi{\ly,\text{HHCC}}\Divzi{\ly,\text{HHCC}}$ equal-sized matrices $\Aii{\ly}{\divxi,\divzi}$, where $(\divxi,\divzi) \in [\Divxi{\ly,\text{HHCC}}] \times [\Divzi{\ly,\text{HHCC}}]$ and $\ly \in [\LY_{\text{HHCC}}]$. Likewise, each of $\Bi{\setzi{\ly} \times\setyi{\ly}}$ matrices, $\ly \in [\LY_{\text{HHCC}}]$, is subdivided into $\Divzi{\ly,\text{HHCC}}\Divyi{\ly,\text{HHCC}}$ equal-sized matrices $\Bii{\ly}{\divzi,\divyi}$, where $(\divzi,\divyi) \in [\Divzi{\ly,\text{HHCC}}] \times [\Divyi{\ly,\text{HHCC}}]$. At the end of the partitioning phase, the matrix product $\Ai{\setxi{\ly} \times\setzi{\ly}}\Bi{\setzi{\ly} \times\setyi{\ly}}$ is partitioned into $\DM_{\ly,\text{HHCC}}$ multiplications of $\Aii{\ly}{\divxi,\divzi}$ and $\Bii{\ly}{\divzi,\divyi}$ matrices, where $(\divxi,\divzi,\divyi) \in [\Divxi{\ly,\text{HHCC}}] \times [\Divzi{\ly,\text{HHCC}}] \times [\Divyi{\ly,\text{HHCC}}]$. Note that the $\Aii{\ly}{\divxi,\divzi}$ and $\Bii{\ly}{\divzi,\divyi}$ matrices are, respectively, of dimensions $\Xdimi{\ly,\text{HHCC}}/\Divxi{\ly,\text{HHCC}} \times \Zdimi{\ly,\text{HHCC}}/\Divzi{\ly,\text{HHCC}}$ and $\Zdimi{\ly,\text{HHCC}}/\Divzi{\ly,\text{HHCC}} \times \Ydimi{\ly,\text{HHCC}}/\Divyi{\ly,\text{HHCC}}$. 

\textbf{Encoding phase:} For each $\ly \in [\LY_{\text{HHCC}}]$, the master encodes $2\DM_{\ly,\text{HHCC}}$ matrices $\Aii{\ly}{\divxi,\divzi}$ and $\Bii{\ly}{\divzi,\divyi}$ to generate $\Stask_{\ly,\text{HHCC}}\ND$ pairs of encoded matrices $(\cA{\ly}{i},\cB{\ly}{i})$, $ i \in [\Stask_{\ly,\text{HHCC}}\ND]$. We use $\Stask_{\ly,\text{HHCC}}$ to denote the number of subtasks each worker is assigned to contribute to the completion of the $\ly$th level of computation. In MLCC we have $\Stask_{\ly}=1$ for all $\ly \in [\LY]$. On the other hands, in BICC we have only one level ($\LY_{\text{\BICC}}=1$); thus, $\Stask_{1}=\Stask$. If we again follow~\cite{POLY:NIPS17}, the master uses polynomial codes to generate encoded matrices. For level $\ly \in [\LY_{\text{HHCC}}]$, the master evaluates polynomials $\cA{\ly}{x}$ and $\cB{\ly}{x}$ at $\Stask_{\ly,\text{HHCC}}$ distinct points $x \in \{(\nd-1)\Stask_{\ly,\text{HHCC}}+1,\ldots,(\nd-1)\Stask_{\ly,\text{HHCC}}+\Stask_{\ly,\text{HHCC}} \}$ to generate encoded matrices for worker $\nd \in [\ND]$. 

\textbf{Distribution phase:} The master next sends $\Stask_{\ly,\text{HHCC}}$ pairs of encoded submatrices to each worker, one pair per level, $\ly \in [\LY_{\text{HHCC}}]$. Worker $\nd \in [\ND]$ requires in total $\sum_{\ly \in [\LY_{\text{HHCC}}]} \Stask_{\ly,\text{HHCC}}$ encoded matrices $\cA{\ly}{(\nd-1)\Stask_{\ly,\text{HHCC}}+\stask}$ and $\cB{\ly}{(\nd-1)\Stask_{\ly,\text{HHCC}}+\stask}$, for all $\ly \in [\LY_{\text{HHCC}}]$ and $\stask \in [\Stask_{\ly,\text{HHCC}}]$. We note that the $\cA{\ly}{(\nd-1)\Stask_{\ly,\text{HHCC}}+\stask}$ and $\cB{\ly}{(\nd-1)\Stask_{\ly,\text{HHCC}}+\stask}$ matrices are, respectively, of dimensions $\Xdimi{\ly,\text{HHCC}}/\Divxi{\ly,\text{HHCC}} \times \Zdimi{\ly,\text{HHCC}}/\Divzi{\ly,\text{HHCC}}$ and $\Zdimi{\ly,\text{HHCC}}/\Divzi{\ly,\text{HHCC}} \times \Ydimi{\ly,\text{HHCC}}/\Divyi{\ly,\text{HHCC}}$. 

\textbf{Worker computation phase:} Similar to MLCC and BICC, in HHCC each worker sequentially completes its $\sum_{\ly \in [\LY_{\text{HHCC}}]} \Stask_{\ly,{\text{HHCC}}}$ subtasks and transmits each result to the master as soon as each is completed. This means that, worker $\nd \in [\ND]$ first completes a sequence of $\Stask_{1,\text{HHCC}}$ matrix multiplications $\{\cA{1}{(\nd-1)\Stask_{1,\text{HHCC}}+1}\cB{1}{(\nd-1)\Stask_{1,\text{HHCC}}+1}, \ldots,\cA{1}{(\nd-1)\Stask_{1,\text{HHCC}}+\Stask_{1,\text{HHCC}}}\cB{1}{(\nd-1)\Stask_{1,\text{HHCC}}+\Stask_{1,\text{HHCC}}}\}$. It then performs $\Stask_{2,\text{HHCC}}$ matrix multiplications pertinent to the second level, \ie $\cA{2}{(\nd-1)\Stask_{2,\text{HHCC}}+\stask}\cB{2}{(\nd-1)\Stask_{2,\text{HHCC}}+\stask}$, where $\stask \in [\Stask_{2,\text{HHCC}}]$. It continues to sequentially multiply matrices up to completion of $\cA{\LY_{\text{HHCC}}}{(\nd-1)\Stask_{\LY_{\text{HHCC}},\text{HHCC}}+1}\cB{\LY_{\text{HHCC}}}{(\nd-1)\Stask_{\LY_{\text{HHCC}},\text{HHCC}}+1}$ through $\cA{\LY_{\text{HHCC}}}{(\nd-1)\Stask_{\LY_{\text{HHCC}},\text{HHCC}}+\Stask_{\LY_{\text{HHCC}},\text{HHCC}}}\cB{\LY_{\text{HHCC}}}{(\nd-1)\Stask_{\LY_{\text{HHCC}},\text{HHCC}}+\Stask_{\LY_{\text{HHCC}},\text{HHCC}}}$.

\textbf{Result aggregation phase:} The master receives subtasks sequentially from all workers till it receives $\RT_{\ly,\text{HHCC}}$ completed subtasks for the $\ly$th level. For each $\ly \in [\LY_{\text{HHCC}}]$, any subset of cardinality at least $\RT_{\ly,\text{HHCC}}$ of $\{\cA{\ly}{i}\cB{\ly}{i} |i \in [\Stask_{\ly,\text{HHCC}}\ND]\}$ is required. 

\textbf{Decoding phase:} Once the master collects enough completed subtasks for level $\ly$, it starts decoding all subtasks pertinent to that level. In MLCC decoding can be parallelized, while in BICC the subtasks of all workers are decoded jointly as part of a single code. The decoding algorithm the master uses in this phase depends on the coding approach the master used in the encoding phase. For instance, in the case of using polynomial or MatDot codes, the master uses a polynomial interpolation decoder, while in the case of product codes, the master uses a simple peeling decoder.

\begin{remark}\label{remark_hhcc1}
To make the amount of computation in HHCC equal to that of BICC and MLCC, the master sends in total $\Stask$ pairs of encoded submatrices to each worker, i.e., $ \sum_{\ly \in [\LY_{\text{HHCC}}]} \Stask_{\ly,\text{HHCC}}=\Stask$. Additionally, the total amount of sub-computations that $\matA\matB$ product is partitioned into should be equal to that of previous hierarchical coding schemes, i.e., $\sum_{\ly=1}^{\LY_{\text{HHCC}}}  \DM_{\ly,\text{HHCC}} = \Stask\DM$. Furthermore, in  HHCC we consider $\sum_{\ly=1}^{\LY_{\text{HHCC}}}  \RT_{\ly,\text{HHCC}} \approx \Stask\RT$ for a fair comparison. 
\end{remark}

\begin{remark}\label{remark_hhcc2}
In this paper we focus on a master-worker model where both workers and the master have only a single core. However, in practice cloud-based systems often provide multicore processing nodes. We next explain a simple and direct extension of the proposed hierarchical coding to a multicore model, where each worker is assumed to have $c>1$ cores. While our comments here are brief, we hope they will help initiate further research in this matter. Note that in a multicore model, all $c$ cores of a single worker typically operate at the same speed and straggling behavior is mainly caused by external delays, such as communication bottlenecks or contention for shared resources~\cite{lee2017coded}. Therefore, to apply hierarchical coding into a multicore model, we can subdivide each subcomputation into $c$ parts, and assign each part to a core. For instance, in the level $\ly$ of HHCC, each worker will be tasked with $\Stask_{\ly,\text{HHCC}}$ subtasks of $\Xdimi{\ly,\text{HHCC}}\Zdimi{\ly,\text{HHCC}} \Ydimi{\ly,\text{HHCC}}/(\Divxi{\ly,\text{HHCC}}\Divzi{\ly,\text{HHCC}}\Divyi{\ly,\text{HHCC}})$ computation. In the worker computation phase, each multicore worker parallelizes its subtask $\stask\in[\Stask_{\ly,\text{HHCC}}]$ in level $\ly$ across its $c$ cores, such that each core is tasked with $\Xdimi{\ly,\text{HHCC}}\Zdimi{\ly,\text{HHCC}} \Ydimi{\ly,\text{HHCC}}/(c\Divxi{\ly,\text{HHCC}}\Divzi{\ly,\text{HHCC}}\Divyi{\ly,\text{HHCC}})$ computation. Due to parallelization, each computational task will be finished $c$ times faster than before. Other phases will be as before.
\end{remark}

\subsection{Examples}
\label{exp}
To illustrate our results we use polynomial codes~\cite{POLY:NIPS17} to present different hierarchical coded computing approaches. In the following four examples we first detail the usage of polynomial codes~\cite{POLY:NIPS17} in BICC. We then present in detail MLCC when polynomial coding is used for each level. For illustrative reasons, in both of these examples we assume that $\Stask=4$ subtasks are assigned to each worker. The computation load of the $\nd$th worker is kept constant at $\comp{\nd}=\frac{\Aa\AB\Bb}{4}$ to enable a fair comparison. In the last two examples we use polynomial codes in HHCC. In Example~6 we use HHCC with $\LY_{\text{HHCC}}=2$, while in Example~7 we use $\LY_{\text{HHCC}}=3$.

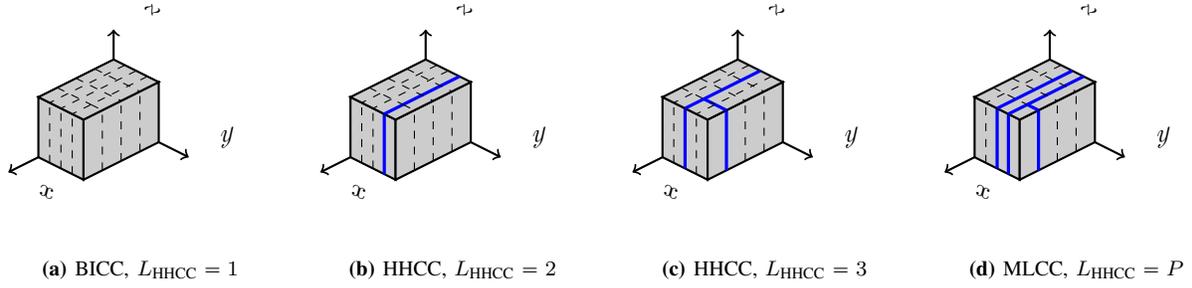
\begin{figure*}[h]
\centering 
	\subfloat[BICC, $\LY_{\text{HHCC}}=1$]{\tikzset{every mark/.append style={scale=0.8}}	
\begin{tikzpicture}
        [cube/.style={very thick,black},
            grid/.style={very thin,gray},
            axis/.style={->,black,thick},every node/.style={minimum size=1cm},on grid]

 \begin{scope}[scale=0.2,yshift=0,xshift=75]
 \begin{scope}[every node/.append style={yslant=-0.5},yslant=-0.5]
 [cube/.style={very thick,black},
            axis/.style={->,blue,thick}]
   \draw[axis] (5,5,0) -- (-4,-4,0) node[anchor=west]{$\Xdir$};
   \draw [step=0.5cm,thick,fill=black!20!white](-2,-2) rectangle +(3,4);
   \draw [dashed](-1.25,-2) -- (-1.25,2);
   \draw[dashed] (0.25,-2) -- (0.25,2);
   \draw [dashed](-0.5,-2) -- (-0.5,2);
 \end{scope}

 \begin{scope}[every node/.append style={yslant=0.5},yslant=0.5]
 \draw[axis] (3,0,0) -- (8,-5,0) node[anchor=west]{$\Ydir$};

   \draw[step=0.5cm,thick,fill=black!20!white] (1,-3) rectangle +(5,4);
   \draw [dashed] (4.75,-3) -- (4.75,1);
   \draw [dashed](3.5,-3) -- (3.5,1);
   \draw[dashed] (2.25,-3) -- (2.25,1);

 \end{scope}

 \begin{scope}[every node/.append style={
     yslant=0.5,xslant=-1},yslant=0.5,xslant=-1
   ]
   \draw[axis] (7,4,0) -- (9,6,0) node[anchor=west]{$\Zdir$};
   \draw[step=0.5cm,thick,fill=black!20!white] (7,4) rectangle +(-5,-3);
   \draw [dashed](7,3.25) -- (2,3.25);
   \draw  [dashed] (7,1.75) -- (2,1.75);
   \draw [dashed](7,2.5) -- (2,2.5);
   
   \draw[ dashed] (5.75,4) -- (5.75,1);
   \draw[dashed] (4.5,4) -- (4.5,1);
   \draw [dashed](3.25,1.75) -- (3.25,1);
   \draw [ dashed] (3.25,4) -- (3.25,1.75);
 \end{scope}
 \end{scope}
 \end{tikzpicture}        
       \label{FIG:HL1} }
       	\quad
	\subfloat[HHCC, $\LY_{\text{HHCC}}=2$]{\tikzset{every mark/.append style={scale=0.8}}	
\begin{tikzpicture}
        [cube/.style={very thick,black},
            grid/.style={very thin,gray},
            axis/.style={->,black,thick},every node/.style={minimum size=1cm},on grid]

 \begin{scope}[scale=0.2,yshift=0,xshift=75]
 \begin{scope}[every node/.append style={yslant=-0.5},yslant=-0.5]
 [cube/.style={very thick,black},
            axis/.style={->,blue,thick}]
   \draw[axis] (5,5,0) -- (-4,-4,0) node[anchor=west]{$\Xdir$};
   \draw [step=0.5cm,thick,fill=black!20!white](-2,-2) rectangle +(3,4);
   \draw [dashed](-1.25,-2) -- (-1.25,2);
   \draw[blue,very thick] (0.25,-2) -- (0.25,2);
   \draw [dashed](-0.5,-2) -- (-0.5,2);
 \end{scope}

 \begin{scope}[every node/.append style={yslant=0.5},yslant=0.5]
 \draw[axis] (3,0,0) -- (8,-5,0) node[anchor=west]{$\Ydir$};

   \draw[step=0.5cm,thick,fill=black!20!white] (1,-3) rectangle +(5,4);
   \draw [dashed] (4.75,-3) -- (4.75,1);
   \draw [dashed](3.5,-3) -- (3.5,1);
   \draw[dashed] (2.25,-3) -- (2.25,1);

 \end{scope}

 \begin{scope}[every node/.append style={
     yslant=0.5,xslant=-1},yslant=0.5,xslant=-1
   ]
   \draw[axis] (7,4,0) -- (9,6,0) node[anchor=west]{$\Zdir$};
   \draw[step=0.5cm,thick,fill=black!20!white] (7,4) rectangle +(-5,-3);
   \draw [dashed](7,3.25) -- (2,3.25);
   \draw  [blue,very thick] (7,1.75) -- (2,1.75);
   \draw [dashed](7,2.5) -- (2,2.5);
   
   \draw[ dashed] (5.75,4) -- (5.75,1);
   \draw[dashed] (4.5,4) -- (4.5,1);
   \draw [dashed](3.25,1.75) -- (3.25,1);
   \draw [ dashed] (3.25,4) -- (3.25,1.75);
 \end{scope}
 \end{scope}
 \end{tikzpicture}        
       \label{FIG:HL2} }
       	\quad
       	\subfloat[HHCC, $\LY_{\text{HHCC}}=3$]{\tikzset{every mark/.append style={scale=0.8}}	
\begin{tikzpicture}
        [cube/.style={very thick,black},
            grid/.style={very thin,gray},
            axis/.style={->,black,thick},every node/.style={minimum size=1cm},on grid]

 \begin{scope}[scale=0.2,yshift=0,xshift=75]
 \begin{scope}[every node/.append style={yslant=-0.5},yslant=-0.5]
 [cube/.style={very thick,black},
            axis/.style={->,blue,thick}]
   \draw[axis] (5,5,0) -- (-4,-4,0) node[anchor=west]{$\Xdir$};
   \draw [step=0.5cm,thick,fill=black!20!white](-2,-2) rectangle +(3,4);
   \draw [dashed](-1.25,-2) -- (-1.25,2);
   \draw[dashed] (0.25,-2) -- (0.25,2);
   \draw [blue,very thick](-0.5,-2) -- (-0.5,2);
 \end{scope}

 \begin{scope}[every node/.append style={yslant=0.5},yslant=0.5]
 \draw[axis] (3,0,0) -- (8,-5,0) node[anchor=west]{$\Ydir$};

   \draw[step=0.5cm,thick,fill=black!20!white] (1,-3) rectangle +(5,4);
   \draw [dashed] (4.75,-3) -- (4.75,1);
   \draw [dashed](3.5,-3) -- (3.5,1);
   \draw[blue,very thick] (2.25,-3) -- (2.25,1);

 \end{scope}

 \begin{scope}[every node/.append style={
     yslant=0.5,xslant=-1},yslant=0.5,xslant=-1
   ]
   \draw[axis] (7,4,0) -- (9,6,0) node[anchor=west]{$\Zdir$};
   \draw[step=0.5cm,thick,fill=black!20!white] (7,4) rectangle +(-5,-3);
   \draw [dashed](7,3.25) -- (2,3.25);
   \draw  [dashed] (7,1.75) -- (2,1.75);
   \draw [blue,very thick](7,2.5) -- (2,2.5);
   
   \draw[ dashed] (5.75,4) -- (5.75,1);
   \draw[dashed] (4.5,4) -- (4.5,1);
   \draw [blue,very thick](3.25,2.5) -- (3.25,1);
   \draw [ dashed] (3.25,4) -- (3.25,1.75);
 \end{scope}
 \end{scope}
 \end{tikzpicture}        
       \label{FIG:HL3} }
       	\quad
       	\subfloat[MLCC, $\LY_{\text{HHCC}}=\Stask$]{\tikzset{every mark/.append style={scale=0.8}}	
\begin{tikzpicture}
        [cube/.style={very thick,black},
            grid/.style={very thin,gray},
            axis/.style={->,black,thick},every node/.style={minimum size=1cm},on grid]

 \begin{scope}[scale=0.2,yshift=0,xshift=75]
 \begin{scope}[every node/.append style={yslant=-0.5},yslant=-0.5]
 [cube/.style={very thick,black},
            axis/.style={->,blue,thick}]
   \draw[axis] (5,5,0) -- (-4,-4,0) node[anchor=west]{$\Xdir$};
   \draw [step=0.5cm,thick,fill=black!20!white](-2,-2) rectangle +(3,4);
   \draw [dashed](-1.25,-2) -- (-1.25,2);
   \draw[blue,very thick] (0.25,-2) -- (0.25,2);
   \draw [blue,very thick](-0.5,-2) -- (-0.5,2);
 \end{scope}

 \begin{scope}[every node/.append style={yslant=0.5},yslant=0.5]
 \draw[axis] (3,0,0) -- (8,-5,0) node[anchor=west]{$\Ydir$};

   \draw[step=0.5cm,thick,fill=black!20!white] (1,-3) rectangle +(5,4);
   \draw [dashed] (4.75,-3) -- (4.75,1);
   \draw [dashed](3.5,-3) -- (3.5,1);
   \draw[blue,very thick] (2.25,-3) -- (2.25,1);

 \end{scope}

 \begin{scope}[every node/.append style={
     yslant=0.5,xslant=-1},yslant=0.5,xslant=-1
   ]
   \draw[axis] (7,4,0) -- (9,6,0) node[anchor=west]{$\Zdir$};
   \draw[step=0.5cm,thick,fill=black!20!white] (7,4) rectangle +(-5,-3);
   \draw [dashed](7,3.25) -- (2,3.25);
   \draw  [blue,very thick] (7,1.75) -- (2,1.75);
   \draw [blue,very thick](7,2.5) -- (2,2.5);
   
   \draw[ dashed] (5.75,4) -- (5.75,1);
   \draw[dashed] (4.5,4) -- (4.5,1);
   \draw [blue,very thick](3.25,1.75) -- (3.25,1);
   \draw [ dashed] (3.25,4) -- (3.25,1.75);
 \end{scope}
 \end{scope}
 \end{tikzpicture}        
       \label{FIG:HL4} }
       	\quad
       \caption{Cuboid partitioning structure for hierarchical coding schemes, where $\Stask=4$: (a) BICC, a hierarchical approach with $\LY_{\text{HHCC}}=1,\RT_{\text{\BICC}}=16$; (b) HHCC, a hierarchical approach with $\LY_{\text{HHCC}}=2,\RT_{\ly,\text{HHCC}} \in \{12,4\}$; (c) HHCC, a hierarchical approach with $\LY_{\text{HHCC}}=3,\RT_{\ly,\text{HHCC}} \in \{8,6,2\}$; (d) MLCC, a hierarchical approach with $\LY_{\text{HHCC}}=4,\RT_{\ly} \in \{8,4,3,1\}$.} \label{FIG:HL_cuboid}
\end{figure*}

\textbf{Example~4 (Bit-interleaved polynomial coding):} 
 Similar to polynomial coding in Example~2, bit-interleaved polynomial coding has a one-step cuboid partitioning. This means that the master partitions the cuboid into $\DM_{\text{\BICC}}=16$ equal-sized information blocks. The partitioning used in this example is depicted in Fig.~\ref{FIG:HL1}. Note that the information dimension $\DM_{\text{\BICC}}$ of this code is $\Stask=4$ times larger than that of the polynomial codes (which had $\DM_{\text{Poly}}=4$). In other words, $\DM_{\text{\BICC}}=\Stask\DM_{\text{Poly}}$. To achieve the information dimension of $\DM_{\text{\BICC}}=16$, the master divides each of the $\matA$ and $\matB$ matrices into four equal-sized submatrices.  Respectively, these are 
\begin{align*}
\matA^T&=[{\tilde{\matA}}_1^T, {\tilde{\matA}}_2^T, {\tilde{\matA}}_3^T, {\tilde{\matA}}_4^T], \nonumber\\
\matB&=[{\tilde{\matB}}_1, {\tilde{\matB}}_2, {\tilde{\matB}}_3, {\tilde{\matB}}_4],
\end{align*}
where $\tilde{\matA}_i \in \mathbb{R}^{\frac{\Aa}{4} \times \AB}$ and $\tilde{\matB}_i \in \mathbb{R}^{\AB \times \frac{\Bb}{4}}$ for $i \in [4]$. 

After partitioning, the master encodes the $\tilde{\matA}_i$ and $\tilde{\matB}_i$ separately using polynomial codes, to generate encoded submatrices 
\begin{align*}
\hat{\matA}(x) &= \tilde{\matA}_1 + \tilde{\matA}_2 x + \tilde{\matA}_3 x^2 + \tilde{\matA}_4 x^3, \nonumber \\
\hat{\matB}(x) &= \tilde{\matB}_1 + \tilde{\matB}_2 x^4 + \tilde{\matB}_3 x^8 + \tilde{\matB}_4 x^{12}.
\end{align*}

Worker $\nd$ is then tasked to compute $\Stask=4$ sequentially-ordered subtasks: $\{\hat{\matA}(4\nd+i)\hat{\matB}(4\nd+i) | i \in [4] \}$. Due to the use of polynomial codes, the completion of any $\RT_{\text{\BICC}}=16$ subtasks enables the recovery of the $\matA\matB$ product. 

\textbf{Example~5 (Multilevel polynomial coding):} In this example each worker is provided $\LY = 4$ sequentially-ordered subtasks  The master first divides the cuboid into $\LY =4$ \taskboxes of dimensions $(\Xdimi{\ly},\Zdimi{\ly},\Ydimi{\ly}) \in \{(\Aa,\AB,\frac{\Bb}{2})$, $(\Aa,\AB,\frac{\Bb}{4})$, $(\frac{3\Aa}{4},\AB,\frac{\Bb}{4})$, $ (\frac{\Aa}{4},\AB,\frac{\Bb}{4}) \}$. We note that our selection here is for illustrative purpose and is not necessarily unique. Later in Sec.~\ref{CHAPTER:THEORY}, we show how to optimize the designing parameters in order to minimize the expected finishing time. The master then subdivides each \taskbox into $\dm{\ly} \in \{8,4,3,1\}$ \infoblocks each of which contains ${\Aa\AB\Bb}/{16}$ \MAC operations. In Fig.~\ref{FIG:HL4} the decomposition of the $\matA\matB$ product into \levels of computation is depicted by the solid (blue) lines. The partitioning of \taskboxes into \infoblocks is indicated by the dashed lines. In Fig.~\ref{FIG:HL4} the projection of the \infoblocks and \taskboxes onto the $\Xdir\Ydir$ plane represents the decomposition of $\matA\matB$ into the computations shown in (\ref{eq.basicEx4}). For instance, the sub-computations of the first \taskbox are the product of \datachunks $\inR{{\Aii{1}{\divxi,\divzi}}}{\frac{\Aa}{4}}{\AB}$ and $\inR{{\Bii{1}{\divzi,\divyi}}}{\AB}{\frac{\Bb}{4}}$, where $\divxi \in [4], \divzi \in [1]$ and $\divyi \in [2]$. 
\begin{align}	\label{eq.basicEx4}
	\matA\matB=\left[
	\begin{array}{cccc}
		\Aii{1}{1,1} \Bii{1}{1,1} & \Aii{1}{1,1} \Bii{1}{1,2} & \Aii{2}{1,1} \Bii{2}{1,1} & \Aii{3}{1,1} \Bii{3}{1,1}\\
		\Aii{1}{2,1} \Bii{1}{1,1} & \Aii{1}{2,1} \Bii{1}{1,2} & \Aii{2}{2,1} \Bii{2}{1,1} & \Aii{3}{2,1} \Bii{3}{1,1}\\
		\Aii{1}{3,1} \Bii{1}{1,1} & \Aii{1}{3,1} \Bii{1}{1,2} & \Aii{2}{3,1} \Bii{2}{1,1} & \Aii{3}{3,1} \Bii{3}{1,1}\\
		\Aii{1}{4,1} \Bii{1}{1,1} & \Aii{1}{4,1} \Bii{1}{1,2} & \Aii{2}{4,1} \Bii{2}{1,1} & \Aii{4}{1,1} \Bii{4}{1,1}
	\end{array}
	\right].
	\begin{tikzpicture}[remember picture,overlay]
\draw [dashed] ($(-3.7,-1)$) to ($(-3.7,1.2)$) ;
\draw [dashed] ($(-2.05,-1)$) to ($(-2.05,1.2)$);
\draw [dashed] ($(-0.4,-0.4)$) to ($(-2,-0.4)$);
\end{tikzpicture}
	                   \vspace*{-8ex}
\end{align}

The master encodes the data relevant to the $\ly$th \taskbox by applying a pair of polynomial codes (separately) to the \datachunks involved in that \levelno. In the above example, the polynomials used to encode are 

\begin{gather}
\begin{array}{l}
\cA{1}{x} = \Aii{1}{1,1} + \Aii{1}{2,1} x + \Aii{1}{3,1} x^2 + \Aii{1}{4,1} x^3 ,  \\ \vspace{2ex}
\cB{1}{x} = \Bii{1}{1,1} + \Bii{1}{1,2} x^4, \\ 
\cA{2}{x} = \Aii{2}{1,1} + \Aii{2}{2,1} x + \Aii{2}{3,1} x^2 + \Aii{2}{4,1} x^3, \\ \vspace{2ex}
\cB{2}{x} = \Bii{2}{1,1}, \\ 
\cA{3}{x} = \Aii{3}{1,1} + \Aii{3}{2,1} x + \Aii{3}{3,1} x^2 , \\  \vspace{2ex}
\cB{3}{x} = \Bii{3}{1,1},  \\ 
\cA{4}{x} = \Aii{4}{1,1},  \\
\cB{4}{x} = \Bii{4}{1,1}. 
\end{array}
\end{gather}

Worker $\nd \in [\ND]$ gets $\LY=4$ pairs of \encdatachunks $(\cA{\ly}{\nd}, \cB{\ly}{\nd})$, where $\ly \in [4]$. Each worker then computes the \enproducts $\inR{{\cA{\ly}{\nd}\cB{\ly}{\nd}}}{{\frac{\Aa}{4}}}{{\frac{\Bb}{4}}}$, working through the $\LY$ levels sequentially from $1$ to $\LY$. Each result is transmitted to the master as it is completed. The master can recover the first \taskboxno, \ie $\Aii{1}{\divxi,\divzi}\Bii{1}{\divzi,\divyi}$, where $\divxi \in [4], \divzi \in [1]$ and $\divyi \in [2]$, when it receives $\rt{1}=8$ \enproducts $\cA{1}{\nd} \cB{1}{\nd}$ from any $8$ of the $\ND$ workers (naturally, $\ND$ can be much larger than $8$). Similarly the master can recover the second \taskbox when it receives $\rt{2}=4$ \enproducts $\cA{2}{\nd}\cB{2}{\nd}$ from any of the $\ND$ workers, and so forth.

To compare MLCC and BICC in Examples~4 and~5, we recall that each worker is tasked with the same number and size of subtasks. In those examples one can observe that BICC has a more flexible recovery rule than MLCC. While for BICC, the $\matA\matB$ product can be recovered from any $\RT_{\text{\BICC}}$ completed subtasks, in MLCC the completed subtasks must follow a specific profile $\{\rt{\ly}\}_{\ly=1}^{4}$. On the other hand, from a decoding perspective, MLCC is much less complex than BICC. In BICC the master needs to deal with decoding $\Aa\Bb/16$ polynomials of degree $15$ when using polynomial codes. On the other hand, in MLCC the master is required to decode $\LY=4$ sets of polynomials of (in the example) degrees $7,3,2$ and $0$, each set consisting of $\Aa\Bb/16$ polynomials. As was discussed in Sec.~\ref{SEC:MLCC}, in MLCC the master can perform such decoding either in a serial, a parallel, or a streaming manner across levels. Parallel and streaming decoding are not possible for BICC. In the numerical results of Sec~\ref{sec.sim}, we observe that even serial decoding of MLCC takes less time than decoding BICC. As would be guessed, the parallel decoding time of MLCC is much less than the decoding time of BICC. This is due to the fact that in the decoding phase of multilevel polynomial codes, in a worst-case scenario, the master needs to deal with decoding a polynomial code of rate $8/\ND$.  This is much less computationally intensive than the decoding of the rate $16/\ND$ polynomial code used in bit-interleaved polynomial codes. Streaming decoding takes the least time of all. Furthermore, in the next section we design MLCC in such a way that it also outperforms BICC in terms of communication time. Compared to BICC, the difference of MLCC follows from the distinct rates applied across the levels, $8/\ND$, $4/\ND$, $2/\ND$, and $1/\ND$ in the multilevel polynomial coding example and ${\rt{\ly}}/{\ND}$ in general. In BICC the sub-computations are encoded jointly as a part of a single code with the code rate ${\rt{\text{\bicc}}}/{(\ND\Stask)}$.  

\textbf{Example~6 (HHCC ($\LY_{\text{HHCC}}=2$, $\RT_{\ly,\text{HHCC}} \in \{12,4\}$)):} The master first uses a two-level multilevel polynomial code with profile $(12,4)$ to partition the cuboid into task blocks. To accomplish this it partitions the cuboid into $\LY_{\text{HHCC}}=2$ task blocks of dimensions $(\Xdimi{\ly},\Zdimi{\ly},\Ydimi{\ly}) \in \{(\Aa,\AB,\frac{3\Bb}{4}),(\Aa,\AB,\frac{\Bb}{4})\}$, as is depicted by the solid (blue) lines in Fig.~\ref{FIG:HL2}. The master then subdivide each task block into $\DM_{\ly,\text{HHCC}}\in \{12,4\}$ information blocks. This is depicted by the dashed lines in Fig.~\ref{FIG:HL2}. The projection of such a partitioning onto the $\Xdir\Ydir$ plane represents the decomposition of the $\matA\matB$ product into
\begin{align*}	\label{eq.basicEx5}
	\left[
	\begin{array}{cccc}
		\Aii{1}{1,1} \Bii{1}{1,1} & \Aii{1}{1,1} \Bii{1}{1,2} & \Aii{1}{1,1} \Bii{1}{1,3} & \Aii{2}{1,1} \Bii{2}{1,1}\\
		\Aii{1}{2,1} \Bii{1}{1,1} & \Aii{1}{2,1} \Bii{1}{1,2} & \Aii{1}{2,1} \Bii{1}{1,3} & \Aii{2}{2,1} \Bii{2}{1,1}\\
		\Aii{1}{3,1} \Bii{1}{1,1} & \Aii{1}{3,1} \Bii{1}{1,2} & \Aii{1}{3,1} \Bii{1}{1,3} & \Aii{2}{3,1} \Bii{2}{1,1}\\
		\Aii{1}{4,1} \Bii{1}{1,1} & \Aii{1}{4,1} \Bii{1}{1,2} & \Aii{1}{4,1} \Bii{1}{1,3} & \Aii{2}{4,1} \Bii{2}{1,1}
	\end{array}
	\right]
	\begin{tikzpicture}[remember picture,overlay]
\draw [dashed] ($(-2.05,-1)$) to ($(-2.05,1.2)$);
\end{tikzpicture}
\end{align*}    

The master next encodes the computation of two task blocks using two independent bit-interleaved polynomial codes. For instance, the master can use two bit-interleaved polynomial codes, one with parameter set $(\RT_{\text{\BICC},1}=12,\Stask_{1,\text{HHCC}}=3)$ for the first level and a bit-interleaved polynomial code with $(\RT_{\text{\BICC},2}=4,\Stask_{2,\text{HHCC}}=1)$ for the second level. Therefore, the polynomials used to encode are
\begin{gather}
\begin{array}{l}
\cA{1}{x} = \Aii{1}{1,1} + \Aii{1}{2,1} x + \Aii{1}{3,1} x^2 + \Aii{1}{4,1} x^3 ,  \\ \vspace{2ex}
\cB{1}{x} = \Bii{1}{1,1} + \Bii{1}{1,2} x^4 + \Bii{1}{1,3} x^8, \\ 
\cA{2}{x} = \Aii{2}{1,1} + \Aii{2}{2,1} x + \Aii{2}{3,1} x^2 + \Aii{2}{4,1} x^3, \\ \vspace{2ex}
\cB{2}{x} = \Bii{2}{1,1}. 
\end{array}
\end{gather}

Worker $\nd \in [\ND]$ receives $\Stask=4$ pairs of encoded matrices $\{(\cA{1}{3\nd+1}, \cB{1}{3\nd+1}),(\cA{1}{3\nd+2}, \cB{1}{3\nd+2}),(\cA{1}{3\nd+3}, \cB{1}{3\nd+3}),(\cA{2}{3\nd+1}, \cB{2}{3\nd+1})\}$ and sequentially multiplies the matrices in each pair. The master is able to recover the $\matA\matB$ product when it receives $12$ completed subtasks from the set $\{\cA{1}{\nd}, \cB{1}{\nd}|\nd\in[3\ND]\}$ and $4$ from the the set $\{\cA{2}{\nd}, \cB{2}{\nd}|\nd\in[\ND]\}$. 

\textbf{Example~7 (HHCC ($\LY_{\text{HHCC}}=3,\RT_{\ly,\text{HHCC}} \in \{8,6,2\}$)):} In this example the master first uses a three-level multilevel polynomial code with profile $(8,6,2)$. It divides the cuboid into $\LY_{\text{HHCC}}=3$ task blocks and then subdivides the $\ly$th task blocks into $\DM_{\ly,\text{HHCC}}\in \{8,6,2\}$ information blocks. The partitioning used in this example is depicted in Fig.~\ref{FIG:HL3}. Through this partitioning, the matrix product $\matA\matB$ is partitioned into $16$ equal-sized subcomputations
 \begin{align*}	\label{eq.basicEx6}
	\left[
	\begin{array}{cccc}
		\Aii{1}{1,1} \Bii{1}{1,1} & \Aii{1}{1,1} \Bii{1}{1,2} & \Aii{2}{1,1} \Bii{2}{1,1} & \Aii{2}{1,1} \Bii{2}{1,2}\\
		\Aii{1}{2,1} \Bii{1}{1,1} & \Aii{1}{2,1} \Bii{1}{1,2} & \Aii{2}{2,1} \Bii{2}{1,1} & \Aii{2}{2,1} \Bii{2}{1,2}\\
		\Aii{1}{3,1} \Bii{1}{1,1} & \Aii{1}{3,1} \Bii{1}{1,2} & \Aii{2}{3,1} \Bii{2}{1,1} & \Aii{2}{3,1} \Bii{2}{1,2}\\
		\Aii{1}{4,1} \Bii{1}{1,1} & \Aii{1}{4,1} \Bii{1}{1,2} & \Aii{3}{1,1} \Bii{3}{1,1} & \Aii{3}{1,1} \Bii{3}{1,2}
	\end{array}
	\right]
	\begin{tikzpicture}[remember picture,overlay]
\draw [dashed] ($(-3.7,-1)$) to ($(-3.7,1.2)$) ;
\draw [dashed] ($(-0.4,-0.4)$) to ($(-3.6,-0.4)$);
\end{tikzpicture}
\end{align*}

The master next encodes the computation of the three task blocks using three independent bit-interleaved polynomial codes using following parameters: for the first, $(\RT_{\text{\BICC},1}=8,\Stask_{1,\text{HHCC}}=2)$, for the second $(\RT_{\text{\BICC},2}=6,\Stask_{2,\text{HHCC}}=1)$, and for the third, $(\RT_{\text{\BICC},3}=2,\Stask_{3,\text{HHCC}}=1)$. To generate encoded matrices, the master uses the polynomials 
\begin{gather}
\begin{array}{l}
\cA{1}{x} = \Aii{1}{1,1} + \Aii{1}{2,1} x + \Aii{1}{3,1} x^2 + \Aii{1}{4,1} x^3 ,  \\ \vspace{2ex}
\cB{1}{x} = \Bii{1}{1,1} + \Bii{1}{1,2} x^4, \\ 
\cA{2}{x} = \Aii{2}{1,1} + \Aii{2}{2,1} x + \Aii{2}{3,1} x^2, \\ \vspace{2ex}
\cB{2}{x} = \Bii{2}{1,1}+\Bii{2}{1,1} x^3, \\ 
\cA{3}{x} = \Aii{3}{1,1}, \\  \vspace{2ex}
\cB{3}{x} = \Bii{2}{1,1}+\Bii{2}{1,1} x. 
\end{array}
\end{gather}

Worker $\nd \in [\ND]$ is tasked with a sequence of $\Stask=4$ matrix  products $\cA{1}{2n+1}\cB{1}{2n+1},\cA{1}{2n+2}\cB{1}{2n+2},\cA{2}{n}\cB{2}{n},$ and $\cA{3}{n}\cB{3}{n}$. The master is able to recover the $\matA\matB$ product if it receives $12$ completed computations from the set $\{\cA{1}{n}\cB{1}{n}| n\in[2N]\}$, $6$ completed computations from the set $\{\cA{2}{n}\cB{2}{n}| n\in[N]\}$, and $2$ completed computations from the set $\{\cA{3}{n}\cB{3}{n}| n\in[N]\}$.

In Examples~6 and~7 we consider HHCC, as a generalization and unification of BICC and MLCC. HHCC provides a tradeoff between the time to compute (recovery flexibility rule) and the times to communicate and to decode. It falls between two extreme cases of hierarchical coding: BICC and MLCC. This means that it has lower communication and decoding times when compared to BICC, and larger communication and decoding times when compared to MLCC. On the other hand, the recovery conditions of HHCC are much more flexible than MLCC, while the recovery conditions of BICC are more flexible still. 

\begin{remark}\label{remark_exampels}
In general we can use different coding schemes across different levels of computations. For instance, we can use polynomial codes~\cite{POLY:NIPS17} to encode data in the first level, and use MatDot codes~\cite{MATDOT:2018} for the second level. The flexibility to use different types of code at different levels of hierarchical coding can be considered for future work. In this paper we focus on single-type hierarchical coding. We also note that the idea of hierarchical coding is not limited to polynomial codes. For instance, one can apply hierarchical coding to product codes and repetition codes by leveraging their corresponding cuboid partitioning structures as is discussed in Sec.~\ref{SEC:3dcoded}. In Sec.~\ref{sec.sim} we apply hierarchical coding for other types of codes. 
\end{remark}

\section{Theoretical Analysis}
\label{CHAPTER:THEORY}
In this section we use the performance measures introduced in
Sec.~\ref{SEC:model} to determine the finishing times of hierarchical
coding, and compare it to non-hierarchical coding. To accomplish this we assert probabilistic models on worker computation and communication times. We use shifted exponential models for analytic
tractability. We then compute the expected finishing times and optimize
the parameters of the schemes to minimize that time.

\subsection{Finishing Time Formulation}
\label{fin}
We term the time it takes the system to compute the $\matA\matB$
product the \emph{finishing time}. A number of cumulative effects contribute to the finishing time: (\rom{1}) the time to encode the data, (\rom{2}) the time to distribute the encoded data, (\rom{3}) the computation time required by the workers, (\rom{4}) the time for the master to aggregate completed tasks, and (\rom{5}) the time to decode. As encoding time is generally negligible when compared to
computation time, we ignore (\rom{1}) in the following.  We denote the
finishing time of the non-hierarchical, BICC, MLCC, and HHCC as $\fin^{\text{Non-h}}$, $\fin^{\text{\bicc}}$, $\fin^{\text{\mlcc}}$, and $\fin^{\text{HHCC}}$, respectively.

\textbf{Non-hierarchical scheme:} Recall from Sec.~\ref{SEC:model} that the time to distribute to the $\nd$th worker the encoded data is $\commi{\nd}\tcommi{\nd}$ sec. The time for that worker to complete its task is ${\comp{\nd}\tcompi{\nd}}$ sec. And the time for that worker to transmit its results to the master is $\commo{\nd}\tcommi{\nd}$ sec. Therefore, the finishing time of worker $\nd$ is the random variable
	\begin{align}
	 \tcom_{\nd} = \commi{\nd}\tcommi{\nd} + {\comp{\nd}\tcompi{\nd}} + \commo{\nd}\tcommi{\nd}. 
	\label{eq:perworker_fin_nonh}
	\end{align}
	
In non-hierarchical schemes, the master is required to collect (at
least) $\RT$ completed tasks. The time required is
$\tcom_{(\RT:\ND)}$, where $(.)_{(\RT:\ND)}$ is an order-statistic operation that selects the $\RT$th
element of the (sorted) $\ND$-element sequence
($\{\tcom_{\nd}\}_{\nd\in[\ND]}$). Let $\{i_1,\ldots,i_{\ND}|\,1\leq i_j\leq\ND, i_j\neq i_k\,\text{if}\, j\neq k\}$ denote the indices of the sorted sequence $\tcom_{i_1}\leq\tcom_{i_2}\leq\ldots\leq\tcom_{i_{\ND}}$. The outcome of $\tcom_{(\RT:\ND)}$ is derived at the index $\RT^*=i_{\RT}$, \ie $\tcom_{(\RT:\ND)}=\tcom_{\RT^*}$.    The master then spends
$\tdec{\text{non-h}}$ sec to decode yielding a total finishing time of
	\begin{align}\label{eq:fin_nonh}
	\mathbb \fin^{\text{non-h}} = \tcom_{\RT^*} + \tdec{\text{non-h}}.
	\end{align}

We further expand~(\ref{eq:perworker_fin_nonh}) by substituting in
expressions for $\commi{\nd}$, $\comp{\nd}$ and $\commo{\nd}$. The
encoded matrices are of dimensions
$\inR{\Ac{\nd}}{\frac{\Aa}{\Divx}}{\frac{\AB}{\Divz}}$ and $
\inR{\Bc{\nd}}{\frac{\AB}{\Divz}}{\frac{\Bb}{\Divy}}$ where
$\DM=\Divx\Divz\Divy$ so
\begin{align} \label{eq:deterministic_nonh_1}
     \commi{\nd} &= \frac{\Aa\AB}{\Divx\Divz} + \frac{\Bb\AB}{\Divy\Divz}.
\end{align}
The computation of
$\Ac{\nd}\Bc{\nd}$ requires
${\Aa\AB\Bb}/{(\Divx\Divy\Divz)}$ \MAC operations so
\begin{align} \label{eq:deterministic_nonh_2}
        \comp{\nd} &= \frac{\Aa\AB\Bb}{\Divx\Divy\Divz}.
   \end{align}
The (encoded) completed result is of size
$\frac{\Aa}{\Divx} \times \frac{\Bb}{\Divy}$, so
\begin{align} \label{eq:deterministic_nonh_3}
     \commo{\nd}&= \frac{\Aa\Bb}{\Divx \Divy}.
\end{align}

Using~(\ref{eq:deterministic_nonh_1})--(\ref{eq:deterministic_nonh_3})
and (\ref{eq:perworker_fin_nonh}) in (\ref{eq:fin_nonh}), an
expression for the finishing time of non-hierarchical schemes can be obtained as in~(\ref{eq:detailed_fin_nonh}). This is an order statistic amenable to analysis. Note that in~(\ref{eq:detailed_fin_nonh}) the $(.)_{(\RT:\ND)}$ operation is applied into a sequence of weighted sum of $\tcomp_{\nd}$ and $\tcomm_{\nd}, {\nd \in [\ND]}$. It is important to note that this is \emph{not} necessarily equal to the weighted sum of  $\tcomp_{(\RT:\ND)}$ and $\tcomm_{(\RT:\ND)}$, where $\tcomp_{(\RT:\ND)}$ and $\tcomm_{(\RT:\ND)}$ denote, respectively, the $\RT$th element of the sorted $\ND$-element sequences $\{\tcomp_{\nd}\}_{\nd \in [\ND]}$ and $\{\tcomm_{\nd}\}_{\nd \in [\ND]}$. 

\begin{align}\label{eq:detailed_fin_nonh}
	\mathbb \fin^{\text{non-h}} \begin{aligned}[t] &= 
	  \left(\frac{\Aa\AB}{\Divx\Divz} + \frac{\Aa\Bb}{\Divx \Divy} + \frac{\Bb\AB}{\Divy\Divz}\right) \tcomm_{\RT^*}
	\\ &+ \left(  \frac{\Aa\AB\Bb}{\Divx\Divy\Divz}\right) \tcomp_{\RT^*} + \tdec{\text{non-h}}. 
	\end{aligned}
\end{align}

\textbf{Bit-interleaved coding scheme:} Recall that in BICC each worker is assigned $\Stask$ subtasks. Further, these subtasks are tackled in order.  We use $\tcombicc_{(\nd-1)\Stask + \stask}$ to denote the time worker $\nd$
takes to finish subtask $\stask \in [\Stask]$.  As different workers will complete different numbers of jobs we need a per-worker count of the number of subtasks provided to the master.  The variable $p$ will play that role in the ensuing discussion. The master first spends $\commi{\nd}\tcommi{\nd}$ sec to distribute the data pertinent to all $\Stask$ (encoded) subtasks to worker $\nd$. That worker then spends $\sum_{i\in[\stask]}\comp{\nd,i}\tcompi{\nd}$ sec to finish its first $\stask \in [\Stask]$ subtasks, where $\comp{\nd,i}$ denotes the computation load of the $i$th subtask of worker $\nd$. Each subtask is transmitted to the master upon completion. The $\stask$th such transmission takes $\commo{\nd,\stask}\tcommi{\nd}$ seconds where $\commo{\nd,\stask}$ is the output communication load of the $\stask$th subtask of worker $\nd$. In aggregate, to complete its $\stask$th subtask, worker $\nd$ requires 

	\begin{align}
	 \tcombicc_{(\nd-1)\Stask + \stask} = \commi{\nd}\tcommi{\nd} +  \commo{\nd,\stask}\tcommi{\nd} + \sum_{i\in[\stask]}\comp{\nd,i}\tcompi{\nd} \text{ sec}.
          \label{eq:perworker_fin_sumRT}
	\end{align}
Note that to derive~(\ref{eq:perworker_fin_sumRT}) we do not consider the effect of multiple incoming completed results from a single worker which could lead to queuing delays. Instead, we consider a parallel communication model, wherein each worker can set up up to $\Stask$ parallel communication channels with the master, at most one channel for each of $\Stask$ subtasks. We also note that due to possibly lower $\commo{\nd,\stask}$ of later subtasks, in general the per-worker finishing times are not necessarily in order for all $\Stask$ subtasks. In BICC the master is able to recover $\matA\matB$ when it has received a total of at least $\RT_{\text{\bicc}}$ subtasks. The
finishing time of BICC can therefore be written as
	\begin{align}
	\mathbb \fin^{\text{\bicc}} = \tcombicc_{\RT_{\text{\bicc}}^*} + \tdec{\text{\bicc}}
\label{eq:fin_sumRT}
	\end{align}
where $ \tcombicc_{\RT_{\text{\bicc}}^*}$ is equal to $\tcombicc_{(\RT_{\text{\bicc}}:\Stask\ND)}$ and denotes the $\RT_{\text{\bicc}}$th element of the sorted $\Stask\ND$-element sequence $\{\tcombicc_{(n-1)\Stask+\stask}\}_{n\in[\ND],\stask\in[\Stask]}$. As before, we now make the expressions for $\commi{\nd}$, $\commo{\nd,i}$ and $\comp{\nd,j}$ explicit. In BICC the master  conveys $2\Stask$ distinct encoded submatrices $\inR{\Ac{(\nd-1)\Stask+\stask}}{\frac{\Aa}{\Divxi{,\text{\bicc}}}}{\frac{\AB}{\Divzi{,\text{\bicc}}}}$ and $\inR{\Bc{(\nd-1)\Stask+\stask}}{\frac{\AB}{\Divzi{,\text{\bicc}}}}{\frac{\Bb}{\Divyi{,\text{\bicc}}}}$ where $\stask \in [\Stask]$ to worker $\nd$. Thus,
\begin{equation}\label{bicc_comm_in}
\commi{\nd} = \frac{\Stask \Aa\AB}{\Divxi{,\text{\bicc}}\Divzi{,\text{\bicc}}} + \frac{\Stask \AB\Bb}{\Divzi{,\text{\bicc}}\Divyi{,\text{\bicc}}}. 
\end{equation}

The worker is tasked with $\Stask$ subtasks, each consisting of
\begin{equation}\label{comp_load_bicc_perw_pert}
\comp{\nd,\stask} = \frac{{\Aa\AB\Bb}}{{\Divxi{,\text{\bicc}}\Divzi{,\text{\bicc}}\Divyi{,\text{\bicc}}}} = \frac{\Aa\AB\Bb}{\Stask\Divx\Divz\Divy}
\end{equation}
basic multiply-and-accumulate operations. Note the last step follows because $\DM_{\text{\bicc}}=\Stask\DM$ (cf.~Remark~\ref{remark_bicc} in Sec.~\ref{BICC}). Finally, 
the $\stask$th worker-to-master transmission contains
\begin{equation}\label{comm_load_bicc_perw_pert}
\commo{\nd,\stask}= \frac{\Aa\Bb}{\Divxi{,\text{\bicc}}\Divyi{,\text{\bicc}}}
\end{equation}
entries. From~(\ref{comp_load_bicc_perw_pert}), we can conclude that per-worker computation time is linear in the number of subtasks the worker completes. In other words, worker $\nd$ spends $\sum_{i\in[\stask]}\comp{\nd,i}\tcompi{\nd}={\tcompi{\nd}\stask\Aa\AB\Bb}/{(\Stask\Divx\Divz\Divy)}$ sec to complete its first $\stask$ subtasks. Also, note that the right-hand-side of~(\ref{comm_load_bicc_perw_pert}) is independent of $n$ and $\stask$, thus the $\nd$th worker's finishing times ($\{\tcombicc_{(\nd-1)\Stask + \stask} \}_{\stask \in [\Stask]}$) are in ascending order, where $\nd\in [\ND]$.    
Thus, merging $\commi{\nd}, \comp{\nd,\stask},$ and $\commo{\nd,\stask}$ into~(\ref{eq:perworker_fin_sumRT}) and~(\ref{eq:fin_sumRT}) yields 
\begin{align}\label{BICC_FIN}
	\begin{aligned}[t] & \fin^{\text{\bicc}} =   \left( \frac{\Stask \Aa\AB}{\Divxi{,\text{\bicc}}\Divzi{,\text{\bicc}}}+\frac{\Stask \AB\Bb}{\Divzi{,\text{\bicc}}\Divyi{,\text{\bicc}}} \right. \\  &+ \left. \frac{\Aa\Bb}{\Divxi{,\text{\bicc}}\Divyi{,\text{\bicc}}}\right) \tcomm_{\nd^*} +   \frac{\stask^*\Aa\AB\Bb}{\Stask\Divx\Divz\Divy}\tcomp_{\nd^*} + \tdec{\text{\bicc}},
	 \end{aligned}
	\end{align}	
where $\nd^*$ and $\stask^*$ correspond to the $\RT_{\text{\bicc}}$th order statistics of $\{\tcombicc_{(n-1)\Stask+\stask}\}_{n\in[\ND],\stask\in[\Stask]}$, such that $(\nd^*-1)\Stask+\stask^*=\RT_{\text{\bicc}}^*$, $\nd^*\in [\ND]$, and $\stask^*\in[\Stask]$. As before, note that in~$(\ref{BICC_FIN})$, we do \emph{not} apply the $(.)_{(\RT_{\text{\bicc}:\ND\Stask})}$ operation to individual communication and computation terms. 

\textbf{Multilevel coding scheme:} In MLCC the overall job of computing $\matA\matB$ completes when each of the $\LY$ levels completes. For level $\ly$ to complete, at least $\RT_{\ly}$ workers must finish their $\ly$th subtask. We denote the finishing time of level $\ly$ by worker $\nd$ as $\tcommlcc_{\nd}^{\ly}$, which can be written as 
		\begin{align}\label{eq:perworker_fin_hier}
	\begin{aligned}[t]
	 \tcommlcc_{\nd}^{\ly} = \commi{\nd}\tcommi{\nd} + \commo{\nd,\ly}\tcommi{\nd} +  {\sum_{i\in[\ly]}\comp{\nd,i}\tcompi{\nd}} .
	  \end{aligned}
	\end{align}
In above expression $ \comp{\nd,\ly}$ and $\commo{\nd,\ly}$ are the level-$\ly$ computation load and output communication load of worker $\nd$, similar to BICC. Note that~(\ref{eq:perworker_fin_hier}) is similar to~(\ref{eq:perworker_fin_sumRT}), except that in BICC  when we define the per-subtask finishing time of all workers we use a single sequence, $\{\tcombicc_{(\nd-1)\Stask + \stask}\}_{\nd \in [\ND], \stask \in [\Stask]}$. In contrast, in MLCC we use $\LY$ distinct sequences, $( \{\tcommlcc_{\nd}^{1}\}_{\nd\in[\ND]}, \ldots,\{\tcommlcc_{\nd}^{\LY}\}_{\nd\in[\ND]})$. 
 We compute the finishing time of MLCC as
	\begin{align}\label{eq:fin_hier} 
	\mathbb \fin^{\text{\mlcc}}  = \max_{\ly \in [\LY]}\left[\tcommlcc_{\rt{\ly}^*}^{\ly}\right] + \tdec{\text{\mlcc}},
	\end{align}
where $\tcommlcc_{\rt{\ly}^*}^{\ly}$ is equal to $\tcommlcc_{(\rt{\ly}:\ND)}^{\ly}$ and denotes the $\rt{\ly}$th order statistics of the sequence $\{\tcommlcc_{\nd}^{\ly} \}_{\nd \in [\ND]}$. As before, we now make the expressions for $\commi{\nd}$, $\commo{\nd,i}$ and $\comp{\nd,j}$ explicit. Worker $\nd$ requires all encoded matrices $\inR{\cA{\ly}{\nd}}{\frac{\Xdimi{\ly}}{\divx}}{\frac{\Zdimi{\ly}}{\divz}}$ and $\inR{\cB{\ly}{\nd}}{\frac{\Zdimi{\ly}}{\divz}}{\frac{\Ydimi{\ly}}{\divy}}$, $\ly \in [\LY]$. The worst case is when all $\{\cA{\ly}{\nd} | \ly \in [\LY]\}$ (and $\{\cB{\ly}{\nd} | \ly \in [\LY]\}$) are distinct. In this case, the master conveys all $\{(\cA{\ly}{\nd},\cB{\ly}{\nd}) | \ly \in [\LY]\}$ to worker $\nd$. However, in Sec.~\ref{MLCC_EFIN} we make a particular choice of parameters $\{\Xdimi{\ly},\Zdimi{\ly},\Ydimi{\ly},\divx,\divz,\divy\}_{\ly \in [\LY]}$ which yields a set of encoded matrices in which many elements are equal. In that situation the master is able to convey encoded matrices to workers by sending only a subset of matrices as representative of all elements. Through such a design, the total input communication load of worker $\nd$ can be reduced. In general, an upper bound is 
\begin{align}
\commi{\nd} \leq \sum_{\ly=1}^{\LY} \left(\frac{\Xdimi{\ly}\Zdimi{\ly}}{\divx\divz}+\frac{\Ydimi{\ly}\Zdimi{\ly}}{\divy\divz}\right).
\end{align}

Each worker multiplies the encoded matrices $\inR{\cA{\ly}{\nd}}{\frac{\Xdimi{\ly}}{\divx}}{\frac{\Zdimi{\ly}}{\divz}}$ and $\inR{\cB{\ly}{\nd}}{\frac{\Zdimi{\ly}}{\divz}}{\frac{\Ydimi{\ly}}{\divy}}$ as its $\ly$th level subtask. This requires ${\Xdimi{\ly}\Zdimi{\ly}\Ydimi{\ly}}/{(\divx\divz\divy)}$ basic operations. Therefore,
\begin{align}\label{comp_mlcc1}
\comp{\nd,\ly} = \frac{\Xdimi{\ly}\Zdimi{\ly}\Ydimi{\ly}}{\divx\divz\divy}=\frac{\Aa\AB\Bb}{\dmsum}.
\end{align}
 The last step follows from the assumption of fixed per-level computation load (\ref{assumption}). Each worker has a computation load of at most ${\LY\Aa\AB\Bb}/{\dmsum}$. In Sec.~\ref{MLCC_EFIN} we design a MLCC scheme so that the computation load of each worker is at most equal to that of the non-hierarchical scheme, i.e., ${\LY\Aa\AB\Bb}/{\dmsum}= {\Aa\AB\Bb}/{\Divx\Divy\Divz}$. Solving for $\dmsum$ we get
\begin{align}\label{fixed_comp}
\dmsum = \LY \Divx\Divy\Divz.
\end{align}
Note that $\comp{\nd,\ly}$ is independent of $\nd$ and $\ly$. Therefore, similar to BICC, the per-worker computation time of MLCC is linear in the number of subtasks each worker completes, \ie $\sum_{i\in[\ly]}\comp{\nd,i}\tcompi{\nd}={\tcompi{\nd}\ly\Aa\AB\Bb}/{(\LY\Divx\Divz\Divy)}$. 
If the $\nd$th worker finishes the $\ly$th level task, the result $\cA{\ly}{\nd}\cB{\ly}{\nd}$ which is of dimension of $\frac{\Xdimi{\ly}}{\divx}\times \frac{\Ydimi{\ly}}{\divy}$ is sent to the master; hence, 
\begin{align}\label{commout_perl_perw_mlcc}
\commo{\nd,\ly}=\frac{\Xdimi{\ly}\Ydimi{\ly}}{\divx\divy}.
\end{align}
Incorporating $\commi{\nd}, \comp{\nd,\ly},$ and $\commo{\nd,\ly}$ into (\ref{eq:perworker_fin_hier}) and (\ref{eq:fin_hier}) yields a bound on the finishing time of MLCC:
	\begin{align}\label{eq:detailed_fin_hier_first}
	 \fin^{\text{\mlcc}}\begin{aligned}[t] & \leq  \max_{\ly\in [\LY]}  \left[  \left(\sum_{i=1}^{\LY} \left(\frac{\Xdimi{i}\Zdimi{i}}{\Divxi{i}\Divzi{i}}+\frac{\Ydimi{i}\Zdimi{i}}{\Divyi{i}\Divzi{i}}\right)\right)\tcomm_{\rt{\ly}^*} \right. \\  &+ \left.  \frac{\Xdimi{\ly}\Ydimi{\ly}}{\Divxi{\ly}\Divyi{\ly}}\tcomm_{\rt{\ly}^*} + \frac{\ly\Aa\AB\Bb}{\dmsum}\tcomp_{\rt{\ly}^*}\right] + \tdec{\text{\mlcc}}.
	 \end{aligned}
	\end{align}
As before, note that the $(.)_{(\RT_{\ly}:\ND)}$ operation is {\em not} applied to individual communication and computation terms. Instead, $\RT_{\ly}^*$ index corresponds to the index of $\RT_{\ly}$th order statistics of a sequence of weighted sum of computation and communication terms. 
	
\textbf{Hybrid hierarchical coding scheme:} In HHCC the master can recover the overall $\matA\matB$ product when all $\LY_{\text{HHCC}}$ levels complete. To complete level $\ly \in [\LY_{\text{HHCC}}]$, the master needs to receive at least $\RT_{\ly,\text{HHCC}}$ subtasks that are completed by workers in level $\ly$. For worker $\nd \in [\ND]$, we denote the finishing time of subtask $\stask$ in level $\ly$ as $\hat{T}_{(\nd-1)\Stask_{\ly,\text{HHCC}}+\stask}^{\ly}$, which can be written as 
\begin{align}\label{eq:perworker_fin_hhcc}
	\begin{aligned}[t]
	 \hat{T}_{(\nd-1)\Stask_{\ly,\text{HHCC}}+\stask}^{\ly} &= \commi{\nd}\tcommi{\nd} + \commo{\nd,\ly,\stask}\tcommi{\nd} \\ &+  {\sum_{i\in[\ly],j \in [\stask]}\comp{\nd,i,j}\tcompi{\nd}} ,
	  \end{aligned}
\end{align}
where $\comp{\nd,i,j}$ and $\commo{\nd,i,j}$ are, respectively, the subtask-$j$ computation load and output communication load of worker $\nd$ pertinent to level $i$. Note that the per-subtask (or per-level) finishing time of worker $\nd$ in BICC (or MLCC) is a special case for $\hat{T}_{(\nd-1)\Stask_{\ly,\text{HHCC}}+\stask}^{\ly}$. We then compute the finishing time of HHCC as
\begin{align}\label{fin_HHCC}
\fin^{\text{HHCC}}=\max_{\ly \in [\LY_{\text{HHCC}}]} \hat{T}_{\RT_{\ly,\text{HHCC}}^*}^{\ly} + \tdec{\text{HHCC}}, 
\end{align} 
where $\hat{T}_{\RT_{\ly,\text{HHCC}}^*}^{\ly}$ is equal to $\hat{T}_{(\RT_{\ly,\text{HHCC}}:\Stask_{\ly,\text{HHCC}}\ND)}^{\ly}$. As before, we now find the explicit expression for $\commi{\nd}$, $\comp{\nd,i,j}$, and $\commo{\nd,i,j}$ in~(\ref{eq:perworker_fin_hhcc}). Worker $\nd \in [\ND]$ requires all encoded matrices $\inR{\cA{\ly}{(\nd-1)\Stask_{\ly,\text{HHCC}}+\stask}}{\Xdimi{\ly,\text{HHCC}}/\Divxi{\ly,\text{HHCC}}}{\Zdimi{\ly,\text{HHCC}}/\Divzi{\ly,\text{HHCC}}}$ and $\inR{\cB{\ly}{(\nd-1)\Stask_{\ly,\text{HHCC}}+\stask}}{\Zdimi{\ly,\text{HHCC}}/\Divzi{\ly,\text{HHCC}}}{\Ydimi{\ly,\text{HHCC}}/\Divyi{\ly,\text{HHCC}}}$, for all $\ly \in [\LY_{\text{HHCC}}]$ and $\stask \in [\Stask_{\ly,\text{HHCC}}]$. The worst case is when all these encoded matrices are distinct and the master requires to convey $\sum_{\ly \in [\LY_{\text{HHCC}}]} \Stask_{\ly,\text{HHCC}}$ distinct data to each worker. This upper bounds the input communication load as
\begin{align}
\commi{\nd} &\leq \sum_{\ly \in [\LY_{\text{HHCC}}]} \left(\frac{\Stask_{\ly,\text{HHCC}}\Xdimi{\ly,\text{HHCC}}\Zdimi{\ly,\text{HHCC}}}{\Divxi{\ly,\text{HHCC}}\Divzi{\ly,\text{HHCC}}} \right. \nonumber \\ &+ \left. \frac{\Stask_{\ly,\text{HHCC}}\Ydimi{\ly,\text{HHCC}}\Zdimi{\ly,\text{HHCC}}}{\Divyi{\ly,\text{HHCC}}\Divzi{\ly,\text{HHCC}}} \right).
\end{align} 
Worker $\nd$ multiplies the encoded matrices $\cA{\ly}{(\nd-1)\Stask_{\ly,\text{HHCC}}+\stask}$ and $\cB{\ly}{(\nd-1)\Stask_{\ly,\text{HHCC}}+\stask}$ as its $\stask$th subtask in level $\ly$. This multiplication consists of 
\begin{align}
\comp{\nd,\ly,\stask}=\frac{\Xdimi{\ly,\text{HHCC}}\Zdimi{\ly,\text{HHCC}}\Ydimi{\ly,\text{HHCC}}}{\Divxi{\ly,\text{HHCC}}\Divzi{\ly,\text{HHCC}}\Divyi{\ly,\text{HHCC}}}
\end{align}
basic multiply-and-accumulate operations. After completing the $\cA{\ly}{(\nd-1)\Stask_{\ly,\text{HHCC}}+\stask}\cB{\ly}{(\nd-1)\Stask_{\ly,\text{HHCC}}+\stask}$ matrix product, the worker sends its result, which is a ${\Xdimi{\ly,\text{HHCC}}/\Divxi{\ly,\text{HHCC}}}\times{\Ydimi{\ly,\text{HHCC}}/\Divyi{\ly,\text{HHCC}}}$ matrix, to the master. Thus, the $\nd$th worker's output communication load pertinent to the subtask $\stask$ in level $\ly$ is
\begin{align}
\commo{\nd,\ly,\stask} = \frac{\Xdimi{\ly,\text{HHCC}}\Ydimi{\ly,\text{HHCC}}}{\Divxi{\ly,\text{HHCC}}\Divyi{\ly,\text{HHCC}}}.
\end{align} 
Using $\commi{\nd}$, $\comp{\nd,\ly,\stask}$, and $\commo{\nd,\ly,\stask}$ in~(\ref{eq:perworker_fin_hhcc}) and~(\ref{fin_HHCC}) yields the following upper bound on the finishing time of HHCC.
\begin{align}\label{HHCC_fin}
 \begin{aligned}[t] &\fin^{\text{HHCC}} \leq \max_{\ly \in [\LY_{\text{HHCC}}]} \left[ \tcommi{\nd_{\ly}^*} \sum_{i \in [\LY_{\text{HHCC}}]} \left(\frac{\Stask_{i,\text{HHCC}}\Xdimi{i,\text{HHCC}}\Zdimi{i,\text{HHCC}}}{\Divxi{i,\text{HHCC}}\Divzi{i,\text{HHCC}}} \right.\right. \\ &+ \left. \left. \frac{\Stask_{i,\text{HHCC}}\Ydimi{i,\text{HHCC}}\Zdimi{i,\text{HHCC}}}{\Divyi{i,\text{HHCC}}\Divzi{i,\text{HHCC}}}  \right) + \frac{\Xdimi{\ly,\text{HHCC}}\Ydimi{\ly,\text{HHCC}}}{\Divxi{\ly,\text{HHCC}}\Divyi{\ly,\text{HHCC}}} \tcommi{\nd_{\ly}^*}  \right. \\
  &+ \left. \frac{\stask_{\ly}^*\Xdimi{\ly,\text{HHCC}}\Zdimi{\ly,\text{HHCC}}\Ydimi{\ly,\text{HHCC}}}{\Divxi{\ly,\text{HHCC}}\Divzi{\ly,\text{HHCC}}\Divyi{\ly,\text{HHCC}}}\tcompi{\nd_{\ly}^*} \right]  + \tdec{\text{HHCC}}, \end{aligned}
\end{align}
where $\nd_{\ly}^*$ and $\stask_{\ly}^*$ correspond to the $\RT_{\ly,\text{HHCC}}$th order statistics of $\{ \hat{T}_{(\nd-1)\Stask_{\ly,\text{HHCC}}+\stask}^{\ly} \}_{\nd \in [\ND], \stask \in [\Stask_{\ly,\text{HHCC}}]}$, such that $(\nd_{\ly}^*-1)\Stask_{\ly,\text{HHCC}}+\stask_{\ly}^*=\RT_{\ly,\text{HHCC}}^*$, $\nd_{\ly}^* \in [\ND]$, and $\stask_{\ly}^* \in [\Stask_{\ly,\text{HHCC}}]$. Note that both~(\ref{BICC_FIN}) and~(\ref{eq:detailed_fin_hier_first}) are special cases for the above upper bound. In BICC the above upper bound is achieved as in BICC $\LY_{\text{HHCC}}=1$ and for that single level the master is required to send all $\Stask_{\ly,\text{HHCC}}=\Stask$ distinct encoded data, where $\Xdimi{\ly,\text{HHCC}}=\Aa,\Ydimi{\ly,\text{HHCC}}=\Bb,$ and $\Zdimi{\ly,\text{HHCC}}=\AB$. In MLCC we have the same upper bound to (\ref{HHCC_fin}) if we set $\Stask_{\ly,\text{HHCC}}=1$ for all $\LY_{\text{HHCC}}=\LY$ levels and consider the assumption of fixed per-level computation load (\ref{assumption}). While in Sec.\ref{SEC:EFIN} we will compute the expected finishing time of BICC and MLCC based on the probabilistic model we consider in Sec.~\ref{SEC:Prob}, generalizing this computation to expected finishing time of HHCC is significantly more complicated and can be considered for future work. 

\subsection{Probabilistic Model}
\label{SEC:Prob}
We now assign a random distribution model to the computation time $\tcompi{\nd}$ and the communication time $\tcommi{\nd}$. We denote $\Fcompi{s} (t)$ as the probability that a worker is able to finish $s$ basic operations by time $t$, i.e., $\Fcompi{s} (t)=\mathbb{P}(s\tcompi{\nd}\leq t)$. Similarly, $\tcommi{\nd}$ is a random variable with distribution $\Fcommi{s}(t) = \mathbb{P}(s\tcommi{\nd}\leq t)$. $\Fcompi{s} (t)$ and $\Fcommi{s}(t)$ are assumed not to be a function of $\nd$, \ie all workers are identical and therefore share independent yet identical distributed computation and communication times. 

\textbf{Computation model:} In our analysis, we assume that workers complete tasks according to a shifted exponential distribution. The shifted exponential model is a widely used (e.g., in~\cite{ SPEEDUP:TIT17, SHORT:NIPS16}) model of computation time. Importantly, it provides design guidance in the choice of per-level parameters such as the recovery profile. The shifted exponential distribution is parameterized by a scale parameter $\mu$ and a shift parameter $\alpha$. We use a shifted exponential distribution with the parameter set $(\scalep, \shiftp)$ to define the computation time $\tcompi{\nd}$. $\tcompi{\nd}$ is assumed to be conditionally deterministic. This means that the probability of a worker is able to finish $s$ basic operations by time $t$ satisfies 
\begin{align}
\Fcompi{s}(t) = 1-e^{-\frac{1}{\scalep}\left(\frac{t}{s} -\shiftp\right)}, \text{ for } t\geq s\shiftp,
\end{align}
which is a  shifted exponential model. In other words, we assumed that workers make linear progress conditioned on the time it takes to compute one basic operation.

\textbf{Communication model:} Similar to the computation model, we use a shifted exponential distribution with parameter set $(\scalem,\shiftm)$ to describe $\tcommi{\nd}$, the time worker $\nd$ takes to communicate an element of a matrix with the master node. The shifted exponential distribution we assert on communication time is motivated by the experiments we conducted on Amazon EC2. As demonstrated in App.~\ref{ec2_appendix}, a shifted exponential distribution provides a good fit to the distribution of communication time. As noted before, we make a conditionally deterministic assumption: the probability that the $\nd$th worker is able to communicate a matrix of $s$ entries to the master by time $t$ is
\begin{align}
\Fcommi{s}(t) = 1-e^{-\frac{1}{\scalem}\left(\frac{t}{s} -\shiftm\right)}, \text{ for } t\geq s\shiftm.
\end{align}

Regarding the above two probabilistic models, we make two further assumptions. First, we assume there are $\ND$ distinct and independent routes between the master and the workers. This means that either the master or any worker is able to send its resulting matrices as soon as they are completed. Second, all workers are assumed to be statistically identical, having independent yet identical distributed computation and communication times. This models a homogeneous computation fabric. Such a homogeneous assumption is easily relaxed to a heterogeneous one by assigning each worker specific $\scalep, \shiftp, \scalem$, and $\shiftm$ parameters. 

\subsection{Expected Finishing Time}
\label{SEC:EFIN}
 In this section we ignore decoding time, assuming it is negligible when compared to computation and communication times. We note that we do consider decoding time in our results conducted on Amazon EC2 in Sec.~\ref{sec.sim}. To calculate the expected finishing time, we consider two regimes: the {\em  fast-network} and the {\em fast-worker} regimes. Each regime is determined by how the computation and communication times influence the total finishing time. The fast-network regime corresponds to a master-worker model where the network is fast, but workers are slow. This means that computation time plays a much more substantial role than does communication time. In the fast-worker regime the network is slow and the workers are fast. We next calculate the expected finishing time of non-hierarchical, BICC, and MLCC schemes.
 
\textbf{Non-hierahrchical coding scheme:} To compute the expected finishing time of the non-hierarchical scheme, we take the expectation of~(\ref{eq:detailed_fin_nonh}). The expected finishing time is
	
\begin{align}\label{eq:nonh_E1_2}
	\mathbb E[\fin^{\text{Non-h}}] \begin{aligned}[t] 
	&=
	 \left(\frac{\Aa\AB}{\Divx\Divz} + \frac{\Aa\Bb}{\Divx \Divy} + \frac{\Bb\AB}{\Divy\Divz}\right)\mathbb E[\tcomm_{\RT^*}]  
	\\& + \left(  \frac{\Aa\AB\Bb}{\Divx\Divy\Divz}\right)\mathbb E[\tcomp_{\RT^*}].  \end{aligned}
	\end{align}	
	
In the fast-network regime we ignore the communication term when we compute $\mathbb E[\fin^{\text{Non-h}}]$. Therefore, the $\RT$th order statistics of the sequence $\{\tcomp_{n}\}_{n\in [\ND]}$ is obtained at index $\RT^*$. From App.~\ref{exp_order_Exponen}, we approximate $\mathbb E[\fin^{\text{Non-h}}]$ as 
	\begin{align}\label{eq:nonh_E1_2}
	\mathbb E[\fin^{\text{Non-h}}] \begin{aligned}[t] 
	\approx \left( \shiftp + \scalep  \log \left(\frac{\ND}{\ND-\RT}\right) \right)  \left( \frac{\Aa\AB\Bb }{ \Divx \Divy \Divz}\right)	
 \end{aligned}.
	\end{align}

In the fast-worker regime the computation time is negligible when compared to communication time. Therefore, the $\RT$th order statistics of the sequence $\{\tcomm_{n}\}_{n\in [\ND]}$ is obtained for index $\RT^*$ in this regime. Using App.~\ref{exp_order_Exponen}, we have
	
		\begin{align}\label{eq:nonh_E1_2}
	\mathbb E[\fin^{\text{Non-h}}] \begin{aligned}[t] 
	&\approx \left( \shiftm + \scalem  \log \left(\frac{\ND}{\ND-\RT}\right) \right)\\ &\times \left(\frac{\Aa\AB}{\Divx\Divz} + \frac{\Aa\Bb}{\Divx \Divy} + \frac{\Bb\AB}{\Divy\Divz}\right).	
 \end{aligned} 
	\end{align}

\textbf{Bit-interleaved coding scheme:} To calculate the expected finishing time of BICC, we take the expectation of~(\ref{BICC_FIN}).
	
 \begin{align}\label{BICC_E0}
	\begin{aligned}[t] &\mathbb E[  \fin^{\text{\bicc}}] =    \left( \frac{\Stask \Aa\AB}{\Divxi{,\text{\bicc}}\Divzi{,\text{\bicc}}}+\frac{\Stask \AB\Bb}{\Divzi{,\text{\bicc}}\Divyi{,\text{\bicc}}} + \right. \\ &  \left.  \frac{\Aa\Bb}{\Divxi{,\text{\bicc}}\Divyi{,\text{\bicc}}}\right)\mathbb E[ \tcomm_{\nd^*}] +  \left(\frac{\Aa\AB\Bb}{\Divx\Divz\Divy}\right) \mathbb E\left[  \tcomp_{\nd^*}\frac{\stask^*}{\Stask}\right] .
	 \end{aligned}
	\end{align}

Recall that $\nd^*$ and $\stask^*$ corresponded to the $\RT_{\text{\bicc}}$th order statistic of $\{\tcombicc_{(\nd-1)\Stask+\stask}\}_{\nd\in[\ND],\stask\in[\Stask]}$, such that $(\nd^*-1)\Stask+\stask^*=\RT_{\text{\bicc}}^*$, $\nd^*\in[\ND]$, and $\stask^*\in[\Stask]$. To compare (\ref{BICC_E0}) with $\mathbb E[ \fin^{\text{Non-h}}]$, we first provide an effective choice of parameters $\Divxi{,\text{\bicc}},\Divzi{,\text{\bicc}},$ and $\Divyi{,\text{\bicc}}$, such that they match (\ref{comp_load_bicc_perw_pert}). To do this, we use the following particular choice of parameters: $\Divxi{,\text{\bicc}}=\Divx,\Divzi{,\text{\bicc}}=\Stask\Divz$, and $\Divyi{,\text{\bicc}}=\Divy$. Incorporating these parameters into (\ref{BICC_E0}), yields
	
	 \begin{align}\label{BICC_Ez}
	\mathbb E[ \fin^{\text{\bicc}}]\begin{aligned}[t] &=  \left( \frac{\Aa\AB}{\Divx\Divz}+\frac{\AB\Bb}{\Divz\Divy} + \frac{\Aa\Bb}{\Divx\Divy}\right)\mathbb E[ \tcomm_{\nd^*}] \\&+ \left(\frac{\Aa\AB\Bb}{\Divx\Divz\Divy}\right)\mathbb E\left[ \tcomp_{\nd^*}\frac{\stask^*}{\Stask}\right].
	 \end{aligned}
	\end{align}

In App.~\ref{proof_ebicc_enonh_fn} we prove that $\mathbb E[ \fin^{\text{\bicc}}]\leq\mathbb E[ \fin^{\text{Non-h}}]$. 

\textbf{Multilevel coding scheme:}
\label{MLCC_EFIN}
To compute the expected finishing time of MLCC, we first cluster non-hierarchical coding to three different categories. Each category is characterized by whether $\Divx$, $\Divy$, or $\Divz$ {\em dominates} the partitioning structure of the non-hierarchical scheme. 

A non-hierarchical coding scheme of parameters $(\Divx,\Divz,\Divy)$ is called to be {\em $\Divx$-dominated} if $\Divx$ is the maximum element of the set $\{\Divx,\Divz,\Divy \}$. Analogous definitions hold for {\em $\Divy$-dominated} and {\em $\Divz$-dominated}. For example, in polynomial codes we have $\Divz=1$; hence, polynomial codes are either $\Divx$- or $\Divy$-dominated. In contrast, for MatDot codes $\Divy=\Divz=1$. Therefore, MatDot codes are $\Divz$-dominated. We now introduce a particular choice of parameters for MLCC in each of the above categories. In each category we first specify a subset of parameters with the objective of reducing the input and output communication loads. We then optimize the remaining parameters to minimize the upper and lower bounds of the expected finishing time. 

\textit{Communication load reduction:} To reduce the input communication load, it is required to encode data in such a way that many encoded matrices of the set $\{\cA{\ly}{\nd} | \ly \in [\LY] \}$ or $\{\cB{\ly}{\nd} | \ly \in [\LY] \}$ become equal. As a result, the master is able to amortize the input communication load by sending only a subset of these two sets as representatives of all elements. For instance, if we use an $\Divx$-dominated scheme as a baseline, we set $\divy=\Divy,\divz=\Divz,\Ydimi{\ly}=\Bb,$ and $\Zdimi{\ly}=\AB $. These equalize encoded submatrices $\cB{\ly}{\nd}$ across all levels, $\ly \in [\LY]$. Therefore, the master distributes only a single $\frac{\AB}{\Divz} \times \frac{\Bb}{\Divy}$ encoded submatrix as a representative of the set $\{\cB{\ly}{\nd} | \ly \in [\LY] \}$ to each worker. To convey the set $\{\cA{\ly}{\nd} | \ly \in [\LY] \}$ to each worker, the master needs to transmit $\sum_{\ly=1}^{\LY} {\Xdimi{\ly}\Zdimi{\ly}}/{(\divx\divz)}={\Aa\AB}/{\Divx\Divz}$ real numbers (the proof is provided in App.~\ref{app_proof_reduce_comm}). The input communication load of MLCC thus satisfies 
\begin{align}\label{commin_perl_perw_mlcc1}
\commi{\nd} = \frac{\Aa\AB}{\Divx\Divz} + \frac{\Bb\AB}{\Divy\Divz}.
\end{align} 

To calculate the output communication load, we incorporate the above choice of parameters into~(\ref{commout_perl_perw_mlcc}). Therefore,
\begin{align}\label{commout_perl_perw_mlcc1}
 \commo{\nd,\ly}=\frac{\Aa\Bb}{\LY\Divx\Divy}.
\end{align} 

If we use an $\Divy$-dominated scheme as a baseline, we set $\divx=\Divx,\divz=\Divz,\Xdimi{\ly}=\Aa,$ and $\Zdimi{\ly}=\AB $. With this set of parameters, all encoded submatrices $\cA{\ly}{\nd}$, $\ly \in [\LY]$, assigned to each worker are equal, and we achieve the same input and output communication load as~(\ref{commin_perl_perw_mlcc1}) and~(\ref{commout_perl_perw_mlcc1}). 
If we use an $\Divz$-dominated scheme as a baseline, we set $\divx=\Divx,\divy=\Divy, \Xdimi{\ly}=\Aa,$ and $\Ydimi{\ly}=\Bb$. Merging these parameters with (\ref{fixed_volume}),~(\ref{fixed_comp}), and~(\ref{assumption}) results in $\Zdimi{\ly}={\AB\divz}/{(\LY\Divz)}$ and $\sum_{\ly=1}^{\LY} \Zdimi{\ly} = \AB$. Through these results, we again achieve the same input communication load as~(\ref{commin_perl_perw_mlcc1}). However, the output communication load is
\begin{align}\label{commout_perl_perw_mlcc2}
\commo{\nd,\ly}=\frac{\Aa\Bb}{\Divx\Divy}.
\end{align}
Comparing~(\ref{commin_perl_perw_mlcc1})-(\ref{commout_perl_perw_mlcc2}) in MLCC with (\ref{eq:deterministic_nonh_1}) and~(\ref{eq:deterministic_nonh_3}) in non-hierarchical coding shows that MLCC has $\commi{\nd}$ and $\commo{\nd,\ly}$ no larger than those of the non-hierarchical coded scheme. 

\textit{Expected finishing time reduction:} To obtain $\mathbb E[ \fin^{\text{\mlcc}}]$, we merge the parameter selections made in the previous part into~(\ref{eq:perworker_fin_hier}) and~(\ref{eq:fin_hier}). In either the $\Divx$- or $\Divy$-dominated situations

\begin{align}\label{eq:reg1_fin_1_2}
	 \mathbb E[ \fin^{\text{\mlcc}}] \begin{aligned}[t] &=   \mathbb E \left[\max_{\ly\in [\LY]} \left[  \left(\frac{\Aa\AB}{\Divx\Divz}+\frac{\Bb\AB}{\Divy\Divz} + \frac{\Aa\Bb}{\LY \Divx\Divy}\right) \right. \right. \\ &\times \left. \left. \tcomm_{\rt{\ly}^*}  +  \left(\frac{\ly\Aa\AB\Bb}{\LY\Divx\Divz\Divy}\right)  \tcomp_{\rt{\ly}^*}\right]\right], 
	\end{aligned}
	\end{align}
and in the $\divz$-dominated situation
		
\begin{align}\label{eq:reg1_fin_3}
	 \mathbb E[\fin^{\text{\mlcc}}] \begin{aligned}[t] &=  \mathbb E \left[ \max_{\ly\in [\LY]}\left[  \left(\frac{\Aa\AB}{\Divx\Divz}+\frac{\Bb\AB}{\Divy\Divz} + \frac{\Aa\Bb}{\Divx\Divy}\right)\tcomm_{\rt{\ly}^*} \right.\right. \\ &+ \left. \left. \left(\frac{\ly\Aa\AB\Bb}{\LY\Divx\Divz\Divy} \right)\tcomp_{\rt{\ly}^*} \right]\right] .\end{aligned}
		\end{align}
		
In the fast-network regime, communication time is negligible in comparison to computation time. In the following optimization problem and theorem we optimize the choice of $\{\rt{\ly}\}_{\ly\in\LY}$ to minimize both upper and lower bounds on $\mathbb E[\fin^{\text{\mlcc}}]$ in the fast-network regime. The optimization problem for the fast-worker regime is provided in App.~\ref{proof_opt1}. Given the result of these optimization problems, plus the parameter selections made in the previous part, we are able to optimize the choice of $\{\divx\}_{\ly \in [\LY]}$ in the $\Divx$-dominated situation, $\{\divz\}_{\ly \in [\LY]}$ in the $\Divz$-dominated situation, and $\{\divy\}_{\ly \in [\LY]}$ in the $\Divy$-dominated situation.
\begin{optproblem}
	The solution to the following convex optimization programs yields the optimal set $\{\rt{\ly} \}$ that minimizes both the upper and lower bounds on $\mathbb E[\fin^{\text{\mlcc}}]$ in the fast-network regime.
	\begin{mini}
	{z,\{\rt{\ly}\}}{z}{\label{thm.opt_Fnetwork}}{}
		\addConstraint{\begin{aligned}[t] \left( \shiftp + \scalep \log \left(\frac{\ND}{\ND-\rt{\ly}}\right)\right)\ly \leq z , \; \forall \ly \in [\LY]  \end{aligned}}
			\addConstraint{\rt{\ly} \leq \rt{\ly-1} \leq \ND , \; \forall \ly \in [\LY]}
		\addConstraint{\sum_{\ly=1}^{\LY}\rt{\ly}=\LY\RT}.	
\end{mini}
\end{optproblem} 
The above optimization problem is justified in App.~\ref{proof_opt1}. In the following theorem and its corollaries, we provide the explicit solution to the optimization problem~1.

\begin{theorem}\label{theorem}
Let $\{\bar{\RT}_{\ly}\}_{\ly=1}^{\LY}$ be the set of optimization variables that solve the optimization problem~(\ref{thm.opt_Fnetwork}). Then  $\forall i \neq j \in [\LY]$ 
\begin{align}\label{statement1}
& \left( \shiftp + \scalep \log \left(\frac{\ND}{\ND-\bar{\RT}_i}\right)\right)i = \nonumber \\ & \left( \shiftp + \scalep \log \left(\frac{\ND}{\ND-\bar{\RT}_j}\right)\right)j. 
\end{align}
\textit{The proof is provided in App.~\ref{explicit_solution}.}
\end{theorem}

\begin{corollary}
If we define $z_i = \log \left(\ND/(\ND-\bar{\RT}_i) \right)$ for all $i \in [\LY]$, from~\eqref{statement1} we can rewrite $z_i$ as a function of $\gamma=z_1$:
\begin{align*}
z_i = \frac{1}{i}\left(\frac{\shiftp}{\scalep} + \gamma \right) - \frac{\shiftp}{\scalep}
\end{align*} 
Inserting $\{z_i\}_{i=1}^{\LY}$ into $\sum_{\ly=1}^{\LY} \bar{\RT}_{\ly}=\LY\RT$ yields
\begin{align}\label{z1_condition}
\sum_{i=1}^{\LY} e^{-\frac{1}{i}\left(\frac{\shiftp}{\scalep}+\gamma \right)+\frac{\shiftp}{\scalep}} = \LY-\frac{\RT\LY}{\ND}. 
\end{align}
Therefore, the optimal choice of optimization parameters in Thm.~1 can be written as  
\begin{align}\label{rt_sol}
\bar{\RT}_{\ly} = \ND\left(1-e^{ -\frac{1}{\ly}\left(\frac{\shiftp}{\scalep}+\gamma \right)+\frac{\shiftp}{\scalep}}\right), 
\end{align}
where $\gamma$ is the unique positive solution to~\eqref{z1_condition}. \end{corollary}

\begin{corollary} Replacing the optimal recovery thresholds $\{\bar{\RT}_{\ly}\}_{{\ly}=1}^{\LY}$~\eqref{rt_sol} in the upper and lower bound of $\mathbb{E}(\fin^{\text{\mlcc}})$, which we express in (\ref{up_fn}) and (\ref{low_fn}) in App.~\ref{proof_opt1}, yields 
\begin{align}\label{lower_upper_mlcc}
\mathbb{E}(\fin^{\text{\mlcc}}) &\geq (\shiftp+\scalep\gamma)\frac{\Aa\AB\Bb}{\LY\Divx\Divz\Divy}, \nonumber \\ 
\mathbb{E}(\fin^{\text{\mlcc}}) &\leq \left((\shiftp+\scalep\gamma) + \sqrt{\frac{L-1}{L}\sum_{j=1}^{L}j^2\sum_{i=1}^{N}\frac{1}{i^2}}\right) \nonumber \\ &\times \frac{\Aa\AB\Bb}{\LY\Divx\Divz\Divy}.
\end{align}
\end{corollary}

In our results in Sec.~\ref{sec.sim} we show that using the optimal recovery profile obtained from~(\ref{thm.opt_Fnetwork}) yields $\mathbb E[\fin^{\text{\mlcc}}] \leq \mathbb E[\fin^{\text{Non-h}}]$.   

\subsection{Algorithm Implementation}
\label{SEC:IMP}
In Sec.~\ref{MLCC_EFIN} we develop an approach for choosing the parameter sets $\{\divx,\divz,\divy,\Xdimi{\ly},\Ydimi{\ly},\Zdimi{\ly},\rt{\ly}| \, \ly \in [\LY] \}$. Given these parameters, in Alg.~\ref{Alg:partition}, we describe one method of selecting \taskboxes and \infoblocks of a multilevel polynomial code. Algorithm~\ref{Alg:partition} works iteratively level-by-level, placing cuboids of volume $\volm{\ly}=\Xdimi{\ly}\Zdimi{\ly}\Ydimi{\ly}$. This yields the needed partitioning of the overall cuboid and avoids overlaps with cuboids placed in previous levels. The cuboid in the $\ly$th level is then partitioned into $\dm{\ly}$ equally sized \infoblocks according to $\{ \divx,\divz,\divy\}$. We comment that rounding errors in Alg.~\ref{Alg:partition} result in at most $\Aa\AB(\Bb-\sum_\ly ( \Bb/\dmsum-1 )\dm{\ly})=\dmsum\Aa\AB$ extra unit cubes that require additional  $\Aa\AB\dmsum$ basic operations to multiply $\matA$ and the last $\dmsum$ columns of $\matB$. We assign this negligible computation, $\Aa\AB\dmsum \ll \Aa\AB\Bb$, to the master.   

\begin{figure}[ht]
  \centering
  \begin{minipage}{0.92\linewidth}
\removelatexerror
\begin{algorithm}[H]
\caption{Partition a $\Aa \times \AB \times \Bb$ cuboid into $\LY$ \taskboxes given $\Xdimi{\ly},\Zdimi{\ly},\Ydimi{\ly}$ for all $\ly \in [\LY]$ and then partition the $\ly$th \taskbox into \infoblocks given $\dm{\ly}$, $\rt{\ly}$, $\divx$, $\divz$, $\divy$, and $\dmsum$, where ${\ly \in [\LY]}$, $\dmsum=\sum_{\ly} \dm{\ly}, $ and $\dm{\ly}=\divx\divz\divy$.}\label{Alg:partition}
\Input{$\LY,\dmsum,$ $\{\dm{\ly}, \rt{\ly},\Xdimi{\ly},\Zdimi{\ly},\Ydimi{\ly}, \divx,\divz,\divy\}_{\ly \in [\LY]}$}
\begin{algorithmic}[1]
\State  \textbf{\emph{for}} $\ly \in [\LY]$:
\State \multiline {Slice the $\ly$th \taskbox from the remaining un-allocated, unit cubes such that the $\ly$th \taskbox contains all \MAC operations indexed by $(\Aai,\ABi,\Bbi)$, where $ \Aai \in [\Aa], \ABi \in [\AB], \Bbi \in [\floor{\frac{\Bb}{\dmsum}}\dm{\ly}]+\floor{\frac{\Bb}{\dmsum}}\sum_{i=1}^{{\ly-1}}\dm{i}$.}
\State  \multiline {Given $\{\divx,\divz,\divy \}$, decompose the $\ly$th \taskbox into $\dm{\ly}$ equally sized \infoblocksno.}
\State  \textbf{\emph{end for}}
\end{algorithmic}
\end{algorithm}
  \end{minipage}
\end{figure}
\section{Evaluation}
\label{sec.sim}
In this section we evaluate the performance of our scheme. We present our results both for numerical simulations and for experiments that were conducted on EC2. We compare BICC and MLCC with non-hierarchical schemes. The latter include uncoded computation, polynomial~\cite{POLY:NIPS17}, MatDot~\cite{MATDOT:2018}, polyDot~\cite{MATDOT:2018}, and entangled polynomial~\cite{ENTGL:ISIT18} coded schemes. These results demonstrate that BICC has the least computation time, and MLCC also outperforms non-hierarchical coding in terms of computation time. On the other hand, the decoding time of MLCC can be smaller than that of BICC. 

\subsection{Numerical Simulations}
\label{sec:num_sim}
We first explain the experimental setup for the shifted exponential model. In each trial we generate $\ND$ pairs of independent shifted exponential random variables $(\tcompi{\nd}, \tcommi{\nd})$, $\nd \in [\ND]$, one pair per worker. $\tcompi{\nd}$ and $ \tcommi{\nd}$ are shifted exponential distributions with parameters $(\scalep,\shiftp)$ and $(\scalem,\shiftm)$, respectively. We recall from the discussion in Sec.~\ref{Sec:metrics} that the realization of $\tcompi{\nd}$ and $\tcommi{\nd}$ set the speed of computation and communication of the $\nd$th worker. Once these two speeds are set, the processor is modeled as progressing through the equal-sized jobs in a (conditionally) deterministic fashion. 

\textbf{Effect of $\LY$:} Figures~\ref{FIG:fin_vs_l_fn1} and~\ref{FIG:fin_vs_l_fn2} plot expected finishing time vs. number of levels $\LY$ based on the shifted exponential distribution model corresponding to $(\scalep,\shiftp,\scalem,\shiftm)=(10^{-6},10^{-7},10^{-8},10^{-9})$. In Fig.~\ref{FIG:fin_vs_l_fn1} a polynomial code corresponding to $\ND=300$ and $(\Divx,\Divz,\Divy)=(42,1,1)$ (an $\Divx$-dominated scenario) achieves an expected finishing time of $5.98$ msec. For the same per-worker computation load ($\DM_{\text{sum}}/\LY=42$), we plot (the solid line) the performance of MLCC using polynomial coding, for different choices of $\LY$. The decrease in finishing time as $\LY$ is increased illustrates that the division of the job into smaller information blocks (larger $\LY$) results in a reduction in the completion time of $\matA\matB$. In particular, when compared to non-hierarchical polynomial coding, MLCC observes a $35\%$ improvement in expected finishing time for $\LY=96$. We also plot (the dashdotted line) the expected finishing time of the randomized-order RMLCC (cf. Remark~\ref{remark_mlcc1} in Sec.~\ref{SEC:MLCC}) using polynomial codes, where $\rt{\ly}=42$ for all $\ly \in [\LY]$. For the same number of workers ($\ND=300$), RMLCC outperforms MLCC for small values of $\LY<32$. For small $\LY$, if we shuffle the order in which workers work through levels, on average each level gets the same attention. However, as $\LY$ increases (e.g., $\LY>32$), the chance of completing all levels at the same time decreases. Thus, the design of an optimal profile for MLCC is necessary for $\LY>32$. Finally, we plot (the dotted line) the expected finishing time of BICC using polynomial coding for different values of $\LY$. It can be observed that the performance of BICC lower bounds MLCC and RMLCC in this simulation. BICC attains a $66\%$ improvement in expected finishing time for $\LY=32$ when compared to non-hierarchical polynomial coding. 
	\begin{figure*}[h]
\centering 
	\subfloat[$(\ND,\DM_{\text{Poly}})=(300,42)$]{\tikzset{every mark/.append style={scale=0.8}}
		\begin{tikzpicture}[scale=0.82]
	\begin{axis}[
	height=10cm,
	width=10cm,
	grid=major,
	xlabel={\large Number of levels, $\LY$},
	ylabel={\large Expected finishing time (msec)},
legend style={at={(0.65,0.84)},anchor=west,nodes=right},
y tick label style={font=\large},
    x tick label style={font=\large},	
	axis on top,xmin=1, xmax=96, ymin=2, ymax=6.1]
	
	\addlegendentry{Poly}
	\addplot [line width=0.5mm, color=black, dashed, every mark/.append style={solid, fill=black}] coordinates {
		(1, 5.981555573953)
		(2,	5.981555573953)
		(4,	5.981555573953)
		(8,	5.981555573953)
		(16,5.981555573953)
		(32,5.981555573953)
		(48,5.981555573953)
		(64,5.981555573953)
		(96,5.981555573953)
 	};
 	
 	\addlegendentry{RML-Poly}
		\addplot [line width=0.5mm, color=olive, dashdotted, every mark/.append style={solid, fill=olive},mark=diamond*] coordinates {
		(1,5.981555573953)
		(2,4.229617925382)
		(4,3.274100601579)
		(8,2.792983140728)
		(16,2.614185878927)
		(32,2.747639309786)
		(48,3.01133022366)
		(64,3.289085972891)
		(96,3.865597383125)
	};

	\addlegendentry{ML-Poly}
		\addplot [line width=0.5mm, color=blue, solid, every mark/.append style={solid, fill=blue},mark=square*] coordinates {
		(1,	5.981555573953)
		(2,	4.998175032436)
		(4,	3.301403294153)
		(8, 2.868696460263)
		(16,2.729322134277)
		(32,2.726554598637)
		(48,2.719748398231)
		(64,2.713281997338)
		(96,2.697044179944)
	}; 	
		\addlegendentry{BI-Poly}
		\addplot [line width=0.5mm, color=red, dotted, every mark/.append style={solid, fill=red},mark=otimes*] coordinates {
		(1,	5.981555573953)
		(2,	4.069473238777)
		(4, 3.030886186562)
		(8,	2.458458925296)
		(16,2.142633050548)
		(32,2.028936367439)
		(48,2.044479964254)
		(64,2.084716961838)
		(96,2.190392937998)
	}; 	
 	
	\end{axis}
	\end{tikzpicture} 
 \label{FIG:fin_vs_l_fn1}}
	\quad%
		\subfloat[$(\ND,\DM_{\text{Poly}})=(20,4)$]{\tikzset{every mark/.append style={scale=0.8}}
	   		\begin{tikzpicture}[scale=0.82]
	\begin{axis}[
	height=10cm,
	width=10cm,
	grid=major,
	xlabel={\large Number of levels, $\LY$},
	ylabel={\large Expected finishing time (msec)},
legend style={at={(0.65,0.84)},anchor=west,nodes=right},
y tick label style={font=\large},
    x tick label style={font=\large},	
	axis on top,xmin=1, xmax=96, ymin=20, ymax=80.5]
	
	\addlegendentry{Poly}
	\addplot [line width=0.5mm, color=black, dashed, every mark/.append style={solid, fill=black}] coordinates {
		(1, 79.058463596022)
		(2,	79.058463596022)
		(4,	79.058463596022)
		(8,	79.058463596022)
		(16,79.058463596022)
		(32,79.058463596022)
		(48,79.058463596022)
		(64,79.058463596022)
		(96,79.058463596022)
 	};
 	
 	\addlegendentry{RML-Poly}
		\addplot [line width=0.5mm, color=olive, dashdotted, every mark/.append style={solid, fill=olive},mark=diamond*] coordinates {
		(1,79.058463596022)
		(2,61.665055500612)
		(4,52.028908321162)
		(8,48.371424639715)
		(16,48.280653618093)
		(32,49.937939236748)
		(48,51.177200969253)
		(64,52.267122670146)
		(96,54.757577530965)
	};

	\addlegendentry{ML-Poly}
		\addplot [line width=0.5mm, color=blue, solid, every mark/.append style={solid, fill=blue},mark=square*] coordinates {
		(1,	79.058463596022)
		(2,	56.743538450107)
		(4,	46.725642452914)
		(8, 43.171117392199)
		(16,42.679535505562)
		(32,42.51647220053)
		(48,42.402080627627)
		(64,42.357889812564)
		(96,42.261167910957)
	}; 	
		\addlegendentry{BI-Poly}
		\addplot [line width=0.5mm, color=red, dotted, every mark/.append style={solid, fill=red},mark=otimes*] coordinates {
		(1,	79.058463596022)
		(2,	55.726526728792)
		(4, 41.981880875908)
		(8,	34.485063258054)
		(16,30.200867527649)
		(32,28.417605565456)
		(48,27.648003844306)
		(64,27.449683555659)
		(96,27.497966409342)
	}; 	
 	
	\end{axis}
	\end{tikzpicture} 
 \label{FIG:fin_vs_l_fn2}}
	\quad
	\setlength{\belowcaptionskip}{-12pt}
\caption{The expected finishing time vs. the number of levels for the shifted exponential distribution (a) where $\DM_{\text{Poly}}=\DM_{\text{\bicc}}/\Stask=\DM_{\text{sum}}/\LY=42$, and $N=300$, (b) where $\DM_{\text{Poly}}=\DM_{\text{\bicc}}/\Stask=\DM_{\text{sum}}/\LY=4$, and $N=20$. In both sub-figures $(\Aa,\AB,\Bb)=(1000,1000,1000)$ and $(\scalep,\shiftp,\scalem,\shiftm)=(10^{-6},10^{-7},10^{-8},10^{-9})$.}
\label{fin_vs_L}
\end{figure*}

Figure~\ref{FIG:fin_vs_l_fn2} shows that the expected finishing time of the RMLCC is $48$ msec for $(\ND, \rt{\ly},\LY)=(20,4,16)$, where as for the polynomial coding of $(\Divx,\Divz,\Divy)=(4,1,1)$ it is $79$ msec. This is a $39\%$ improvement in expected finishing time. For the same per-worker computation load ($\DM_{\text{sum}}/\LY=4$) and for the same number of workers ($\ND=20$) MLCC outperforms RMLCC and attains a $46\%$ improvement compared to non-hierarchical polynomial coding. In this simulation BICC lower bounds MLCC and RMLCC. BICC achieves a $65\%$ improvement in expected finishing time when compared to non-hierarchical polynomial coding for $\LY=96$.  

Figure~\ref{FIG:rate_vs_lidx} graphs the optimal per-level recovery thresholds, $\{\rt{\ly}\}_{\ly \in [8]}$ for different values of $(\scalem,\shiftm)$, where $(\scalep,\shiftp)=(10^{-6},10^{-7})$. For the line with parameters $(\scalep,\shiftp,\scalem,\shiftm)=(10^{-6},10^{-7},10^{-8},10^{-9})$, the network is fast and workers are slow. This yields the most diverse profile. In the case that $(\scalep,\shiftp,\scalem,\shiftm)=(10^{-6},10^{-7},10^{-1},10^{-2})$ the network is slow and workers are fast. In this case the recovery thresholds are all set equal to the recovery threshold of the polynomial code (\ie $\RT=4$). This is due to the fact that the same group of workers needs to complete all their levels. For the parameter setting corresponding to fast workers, $(\scalep,\shiftp,\scalem,\shiftm)=(10^{-6},10^{-7},10^{-2},10^{-3})$, the optimal profile changes from the diverse profile of the fast-network case to a constant one. We note that all profiles approximately have the same average recovery threshold, equal to $\RT_{\text{Poly}}=4$.
	\begin{figure}[h]
\centering 
\tikzset{every mark/.append style={scale=0.8}}
	   		\begin{tikzpicture}[scale=0.8]
	\begin{axis}[
	height=10cm,
	width=10cm,
	grid=major,
	xlabel={\large $\ly$, index of layer},
	ylabel={\large Recovery profile, $\rt{\ly}$},
	legend style={nodes=right},
	y tick label style={font=\large},
    x tick label style={font=\large},
	axis on top,xmin=1, xmax=8, ymin=1, ymax=10]
	\addlegendentry{\large $(\scalem,\shiftm)=(10^{-8},10^{-9})$}
	\addplot [line width=0.5mm, color=blue, solid, every mark/.append style={solid, fill=blue},mark=square*] coordinates {
		(1,	9)
		(2,	6)
		(3,	4)
		(4,	3)
		(5,	2)
		(6,	2)
		(7,	1)
		(8,	1)		
	};

	\addlegendentry{\large $(\scalem,\shiftm)=(10^{-2},10^{-3})$}
	\addplot [line width=0.5mm, color=blue, dashed, every mark/.append style={solid, fill=blue},mark=otimes*] coordinates {
		(1,	5)
		(2,	4)
		(3,	4)
		(4,	4)
		(5,	3)
		(6,	3)
		(7,	3)
		(8,	2)
	};	
	
	\addlegendentry{\large $(\scalem,\shiftm)=(10^{-1},10^{-2})$}
	\addplot [line width=0.5mm, color=blue, dotted, every mark/.append style={solid, fill=blue},mark=triangle*] coordinates {
		(1,	4)
		(2,	4)
		(3,	4)
		(4,	4)
		(5,	4)
		(6,	4)
		(7,	4)
		(8,	4)
	};




	\end{axis}
	\end{tikzpicture} 
	\setlength{\belowcaptionskip}{-12pt}
\caption{Optimal recovery profile of MLCC for different values of $(\scalem,\shiftm)$, where $\LY=8, \ND=20, (\Aa,\AB,\Bb)=(1000,1000,1000), \DM_{\text{Poly}}=\DM_{\text{\bicc}}/\Stask=\DM_{\text{sum}}/\LY=4$ and $(\scalep,\shiftp)=(10^{-6},10^{-7})$.}
 \label{FIG:rate_vs_lidx}
\end{figure}

\textbf{Tradeoffs:} Fig.~\ref{FIG:comp_vs_r} illustrates the tradeoff between the total computation load and the recovery threshold of various non-hierarchical schemes. The total computation load is the maximum number of basic operations that each worker is assigned to complete. This is equal to $\comp{\nd}$ in non-hierarchical schemes and $\sum_{\ly\in[\LY]} \comp{\nd,\ly}$ in MLCC. Both are equal to ${\Aa\AB\Bb}/{(\Divx\Divz\Divy)}$.

In Fig.~\ref{FIG:comp_vs_r} we first plot (dashed lines) the tradeoff between total computation load and recovery threshold for polyDot and entangled polynomial codes. Entangled polynomial codes outperform polyDot codes at the same recovery threshold. Entangled polynomial codes have a smaller total computation load than do polyDot codes. If the total computation load is fixed, entangled polynomial codes have a lower recovery threshold than do polyDot codes. For instance, as is illustrated in Fig.~\ref{FIG:comp_vs_r}, entangled polynomial and polyDot codes with a parameter set $(\Divx,\Divz,\Divy)=(6,6,6)$ have, respectively, recovery thresholds of $\rt{\text{Ent}}=221$ and $\rt{\text{Pdot}}=396$ (cf. (\ref{EQ:rthr_ent}) and (\ref{EQ:rthr_pdot})). The total computation load of each of these codes is $4.6e6$. 
    	\begin{figure}[h]
\centering 
\tikzset{every mark/.append style={scale=0.8}}
			\begin{tikzpicture}[scale=0.82]
	\begin{axis}[
	height=10cm,
	width=10cm,
	grid=major,
	xlabel={\large Recovery threshold},
	ylabel={\large Total computation load},
	legend style={nodes=right},
	legend style={cells={anchor=north east}, at={(0.99,0.99)}},
	y tick label style={font=\large},
    x tick label style={font=\large},
	axis on top,xmin=0, xmax=1400, ymin=0, ymax=3E7]

		\addlegendentry{\large PolyDot}
	\addplot [line width=0.5mm, color=red, dashed, every mark/.append style={solid, fill=red},mark=otimes*] coordinates {
 (1296.0, 771604.938271605) 
 (972.0, 1543209.87654321)
 (720.0, 2314814.814814815)
 (567.0, 3086419.75308642)
 (396.0, 4629629.62962963) 
 (272.0, 6944444.444444444)
 (207.0, 9259259.25925926)
 (140.0, 1.388888888888889E7)
 (71.0,  2.777777777777778E7) 
	};

		\addlegendentry{\large EntPoly}
	\addplot [line width=0.5mm, color=blue, dashed, every mark/.append style={solid, fill=blue},mark=square*] coordinates {
 (1296.0, 771604.938271605) 
 (649.0, 1543209.87654321)
 (434.0, 2314814.814814815)
 (327.0, 3086419.75308642)
 (221.0, 4629629.62962963) 
 (152.0, 6944444.444444444)
 (119.0, 9259259.25925926)
 (89.0, 1.388888888888889E7)
 (71.0,  2.777777777777778E7) 
	};
	\addlegendentry{\large ML-PolyDot}
	\addplot [line width=0.5mm, color=red,solid, every mark/.append style={solid, fill=red},mark=otimes*] coordinates {
(1377.0,578703.7037037037)
(721.0,1157407.407407407)
(420.0,1736111.111111111)
(253.0,2314814.814814815)
(154.0,2893518.518518519)
(99.0,3472222.222222222)
(74.0,4050925.925925926)
(66.0,4629629.62962963)
	};

	\addlegendentry{\large ML-EntPoly}
	\addplot [line width=0.5mm, color=blue,solid, every mark/.append style={solid, fill=blue},mark=square*] coordinates {
(864.0,578703.7037037037)
(372.0,1157407.407407407)
(193.0,1736111.111111111)
(115.0,2314814.814814815)
(77.0,2893518.518518519)
(57.0,3472222.222222222)
(46.0,4050925.925925926)
(41.0,4629629.62962963)
	};


		\addplot [line width=0.5mm, color=red, dashed, every mark/.append style={solid, fill=red},mark size = 3pt, mark=otimes*] coordinates {
 (396.0, 4629629.62962963) 
	};	
			\addplot [line width=0.5mm, color=blue, dashed, every mark/.append style={solid, fill=blue},mark size = 3pt, mark=square*] coordinates {
 (221.0, 4629629.62962963) 
	};	
		\addplot [line width=0.5mm, color=red, dashed, every mark/.append style={solid, fill=red},mark size = 3pt, mark=otimes*] coordinates {
 (1296.0, 771604.938271605) 
	};	
			\addplot [line width=0.5mm, color=blue, dashed, every mark/.append style={solid, fill=blue},mark size = 3pt, mark=square*] coordinates {
 (1296.0, 771604.938271605) 
	};	
		\addplot [line width=0.5mm, color=red, dashed, every mark/.append style={solid, fill=red},mark size = 3pt, mark=otimes*] coordinates {
 (71.0,  2.7777777777778E7) 
	};	
			\addplot [line width=0.5mm, color=blue, dashed, every mark/.append style={solid, fill=blue},mark size = 3pt, mark=square*] coordinates {
 (71.0,  2.7777777777778E7) 
	};		
	
		\addplot [line width=0.5mm, color=blue, dashed, every mark/.append style={solid, fill=blue},mark size = 3pt, mark=otimes*] coordinates {
 (71.0,  2.7777777777778E7) 
	};	
    \node at (axis cs: 210,2.8E7) [] {\large MatDot};
    \node at (axis cs: 1270, 0.2E7) [] {\large Poly};
    \node at (axis cs: 730, 0.48E7) [] {\large $\Divx=\Divz=\Divy=6$};

	\addplot [line width=0.5mm, color=black, dotted, every mark/.append style={solid, fill=blue}] coordinates {
 (0,0.462962962963E7) 
 (396.0, 0.462962962963E7) 
	};	
	\end{axis}
	\end{tikzpicture}  
	\setlength{\belowcaptionskip}{-12pt}
\caption{Total computation load vs. recovery threshold, when $(\Aa,\AB,\Bb)=(1000,1000,1000)$, $\Divx\Divz=36,$ and $ \Divx=\Divy$. In the hierarchical schemes $\LY=8$ and $\Divx=\Divy=\Divz=6$.} \label{FIG:comp_vs_r}
\end{figure}

We next plot (solid lines) the tradeoff between the total computation load and the recovery thresholds used across levels for two MLCC approaches. The first is a multilevel entangled polynomial code. The second is a multilevel polyDot code. We apply MLCC with $\LY=8$ to entangled polynomial and polyDot codes with $(\Divx,\Divz,\Divy)=(6,6,6)$. The per-worker computation load in MLCC is at most equal to the computation load of the corresponding non-hierarchical schemes. In other words, workers that complete all their levels have a computation load of $4.6e6$; other workers have lower computation loads. The recovery threshold of the first level, $\rt{1}$, is equal to $\rt{1,\text{M-Ent}}=41$ in the multilevel entangled polynomial code and is equal to $\rt{1,\text{M-Pdot}}=66$ in the multilevel polyDot code. Both of these recovery thresholds are lower than those of the corresponding non-hierarchical schemes. I.e., $\rt{1,\text{M-Ent}}=41<\rt{\text{Ent}}=221$ and $\rt{1,\text{M-Pdot}}=66 < \rt{\text{Pdot}}=396$. However, the recovery threshold of the last level, $\rt{8}$, is equal to $\rt{8,\text{M-Ent}}=864$ in the multilevel entangled polynomial code, and is equal to $\rt{8,\text{M-Pdot}}=1377$ in the multilevel polyDot code. Both of these recovery thresholds are larger than those of the corresponding non-hierarchical schemes. I.e., $\rt{8,\text{M-Ent}}=864>\rt{\text{Ent}}=221$ and $\rt{8,\text{M-Pdot}}=1377 > \rt{\text{Pdot}}=396$. 

We now analyze BICC with $\LY=8$ using polyDot and entangled polynomial codes with $(\Divx,\Divz,\Divy)=(6,6,6)$. To apply BICC using polyDot and entangled polynomial codes, we use $(\Divxi{,\text{BICC}},\Divzi{,\text{BICC}},\Divyi{,\text{BICC}})=(6,48,6)$. This yields BICC with recovery thresholds $\RT_{\text{B-Pdot}}=3420$ and $\RT_{\text{B-Ent}}=1775$. Both of these recovery thresholds are approximately $\LY=8$ times larger than those of their corresponding non-hierarchical codes. I.e., $\RT_{\text{B-Pdot}}=3420 \approx 8 \rt{\text{Pdot}}$ and $\RT_{\text{B-Ent}}=1775=8\rt{\text{Ent}}$. However, the per-subtask computation load of these BICC approaches are $1/8$ of their corresponding non-hierarchical codes. For illustrative reasons, the two points corresponding to bit-interleaved polyDot code and bit-interleaved entangled polynomial code are not shown in Fig.~\ref{FIG:comp_vs_r}.
    
\subsection{Experiments on Amazon EC2}
 From this section onwards we use Python to implement large matrix multiplications on a cluster of $\ND+1$ ``t2.micro'' instances ($\ND$ workers and a master). We use the function ``numpy.matmul'' linked to the library ``openBlas'' to multiply matrices with entries of type ``float32''. We use the package ``mpi4py'' for the message passing interface between instances. 

We first note that to illustrate the strengths and weaknesses of various coded computing schemes on EC2 we need to artificially inject delays into computation (see App.~\ref{ec2_appendix_artificial}). In this \emph{artificial-straggler} scenario we assign workers to be stragglers independently with probability $\prob$. Workers that are designated stragglers are assigned one more extra computation per-level than non-stragglers (i.e., stragglers are half as fast as non-stragglers). In this artificial-straggler scenario the probability of realizing $\success$ stragglers among $\ND$ workers is ${\ND \choose \success}\prob^{\success} (1-\prob)^{(1-\success)}$. 

\begin{figure*}[h]
\centering 
	\subfloat[]{\tikzset{every mark/.append style={scale=0.8}}
			\begin{tikzpicture}[scale=0.8]
	\begin{axis}[
	height=10cm,
	width=10cm,
	grid=major,
	xlabel={{\large $L$}},
	ylabel={{\large Avg. decoding time (sec)}},
legend style={at={(0.05,0.85)},anchor=west,nodes=right},	
		y tick label style={font=\large},
    x tick label style={font=\large},
	axis on top,xmin=1, xmax=32, ymin=0, ymax=1]

		\addlegendentry{{\large BI-Poly}}
		\addplot [line width=0.5mm, color=red, dotted, every mark/.append style={solid, fill=red},mark=otimes*] coordinates {
 	(1, 0.1018)
 	(2, 0.1382)
 	(4, 0.1639)
 	(8, 0.2532)
 	(16,0.5177)
 	(32,0.8088)
	};

	\addlegendentry{{\large ML-Poly, serial decoding}}
		\addplot [line width=0.5mm, color=blue, solid, every mark/.append style={solid, fill=blue},mark=*] coordinates {
 	(1, 0.1018)
 	(2, 0.1050)
 	(4, 0.1066)
 	(8, 0.1090)
 	(16,0.1153)
 	(32,0.1138)
	};

	\addlegendentry{{\large Poly}}
	\addplot [line width=0.5mm, color=black, dashed, every mark/.append style={solid, fill=black}] coordinates {
 	(1, 0.1018)
 	(2, 0.1018)
 	(4, 0.1018)
 	(8, 0.1018)
 	(16,0.1018)
 	(32,0.1018)
 	};

	\addlegendentry{{\large ML-Poly, parallel decoding}}
		\addplot [line width=0.5mm, color=blue, solid, every mark/.append style={solid, fill=blue},mark=square*] coordinates {
 	(1, 0.1018)
 	(2, 0.0840)
 	(4, 0.0502)
 	(8, 0.0211)
 	(16,0.0111)
 	(32,0.0055)
	};


	\end{axis}
	\end{tikzpicture} 
 \label{FIG:dec_vs_L}}
	\quad
\subfloat[]{\tikzset{every mark/.append style={scale=0.8}}
			\begin{tikzpicture}[scale=0.8]
	\begin{axis}[
	height=10cm,
	width=10cm,
	grid=major,
	xlabel={{\large $L$}},
	ylabel={{\large Avg. computation time (sec)}},
legend style={at={(0.66,0.88)},anchor=west,nodes=right},	
		y tick label style={font=\large},
    x tick label style={font=\large},
	axis on top,xmin=1, xmax=32, ymin=2.4, ymax=4.5]

	\addlegendentry{{\large Poly}}
	\addplot [line width=0.5mm, color=black, dashed, every mark/.append style={solid, fill=black}] coordinates {
 	(1, 4.144)
 	(2, 4.144)
 	(4, 4.144)
 	(8, 4.144)
 	(16,4.144)
 	(32,4.144)
 	};

	\addlegendentry{{\large ML-Poly}}
		\addplot [line width=0.5mm, color=blue, solid, every mark/.append style={solid, fill=blue},mark=*] coordinates {
 	(1, 4.144)
 	(2, 3.132)
 	(4, 3.178)
 	(8, 3.178)
 	(16,3.249)
 	(32,3.192) 	
	};

		\addlegendentry{{\large BI-Poly}}
		\addplot [line width=0.5mm, color=red, dotted, every mark/.append style={solid, fill=red},mark=otimes*] coordinates {
 	(1, 4.144)
 	(2, 3.092)
 	(4, 2.931)
 	(8, 2.68)
 	(16,2.567) 
 	(32,2.513) 	   
	};



	\end{axis}
	\end{tikzpicture} 
 \label{FIG:comp_vs_L}}
	\quad
	\setlength{\belowcaptionskip}{-12pt}
\caption{(a) The average decoding time vs. the number of levels; (b) The average computation time vs. the number of levels. Both (a) and (b) plot results from Amazon EC2 where $(\Aa,\AB,\Bb)=(8192,8192,8192),\DM_{\text{Poly}}=\DM_{\text{\bicc}}/\Stask=\DM_{\text{sum}}/\LY=8,$ and $\ND=12$.}
\end{figure*}
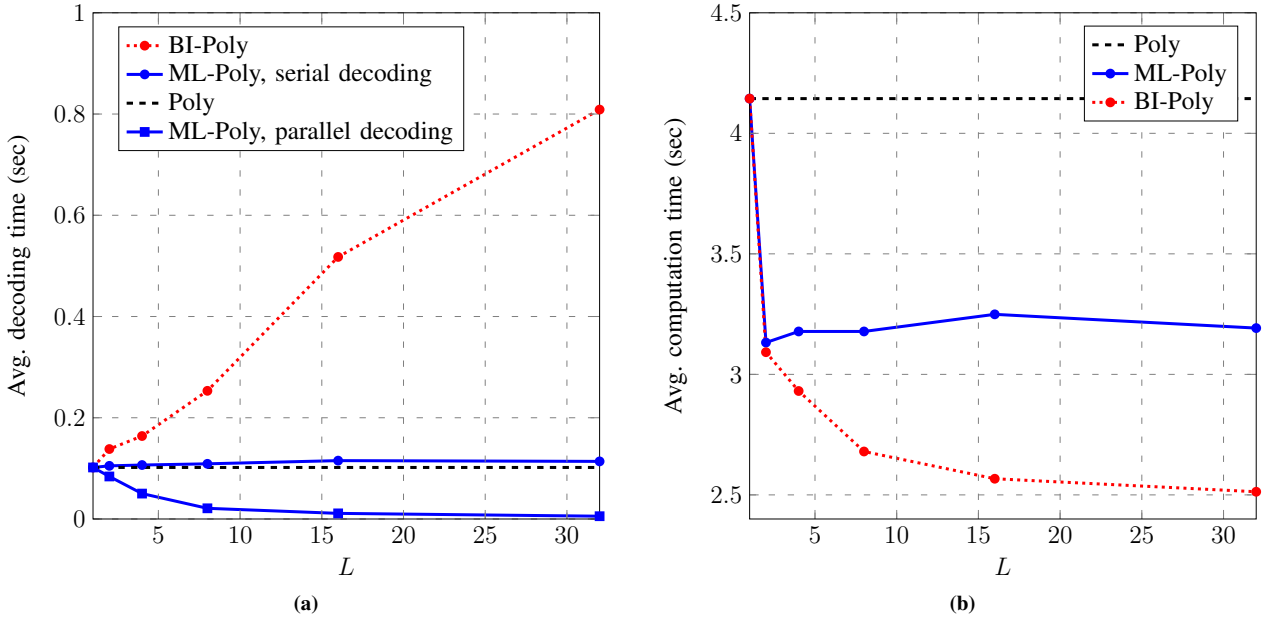

\textbf{Effect of $\LY$ and $\Stask$:} We now discuss the result of implementing coded computing on EC2. The decoder we implemented in the master solves a system of linear equations which involves a Vandermonde matrix. Both $\matA$ and $\matB$ are $8192 \times 8192$ matrices and the average recovery threshold per level (respectively, $\DM_{\text{Poly}}, \DM_{\text{B-poly}}/\Stask, $ and $\DM_{\text{sum}}/\LY$, where $\Stask=\LY$) is set to $8$. In Fig.~\ref{FIG:dec_vs_L} we plot decoding time versus the number of levels. We plot (solid lines) the serial and parallel decoding times for MLCC. Each data point on these lines corresponds to a different number of levels, where $\LY \in \{1,2,4,8,16,32\}$. One can observe that when the decoding process of each level is carried out serially the decoding time of MLCC is very small. It is close to that of polynomial codes. The decoding time of MLCC can be further reduced when the decoding of levels is conducted in parallel. The decoding time of BICC is the largest and increases dramatically as $\LY$ is increased. 

Figure~\ref{FIG:comp_vs_L} plots the average computation time vs. the number of levels, where $(\Aa,\AB,\Bb)= (8192, 8192,$ $ 8192)$. We consider the artificial-straggler scenario with $\prob=0.33$. We first plot (dashed line) the average computation time of polynomial coding where $\DM_{\text{Poly}}=8$. In bit-interleaved polynomial coding (and multilevel polynomial coding) we implement a matrix multiplication of dimensions $\Aa\times \AB \times \frac{\Bb}{8\Stask}$ (and $\Aa\times \AB \times \frac{\Bb}{8\LY}$) on each of $\ND=12$ workers per subtask (and per level). We assume a $\Divy$-dominated scheme, where $\Divx=\Divz=1$ and $\Divy=8$. As in the shifted exponential model, the bit-interleaved polynomial coding (the dotted line) has the smallest average computation time. Compared to polynomial codes, we observe an improvement of $23\%$ in multilevel polynomial coding where $\LY=32$. This improvement increases to $39\%$ in bit-interleaved polynomial codes with $\Stask=32$. 

\begin{figure}[h]
\centering 
	\tikzset{every mark/.append style={scale=1}}
		\begin{tikzpicture}[scale=0.82]
	\begin{axis}[
	height=10cm,
	width=10cm,
	grid=major,
	xlabel={{\large Avg. computation time (sec)}},
	ylabel={{\large Avg. decoding time (sec)}},
	legend style={nodes=right},
	legend style={cells={anchor=north east}, at={(0.99,0.99)}},
			y tick label style={font=\large},
    x tick label style={font=\large},
	axis on top,xmin=2.5, xmax=6.5, ymin=0, ymax=0.4]

		\addlegendentry{{\large Serial}}
	\addplot [line width=0.5mm, color=blue, dashed] coordinates {
 (2.7775, 0.3718)
 (2.9611, 0.2643)
 (3.3037, 0.1823)
 (3.5056, 0.1595)
 	};

		\addlegendentry{{\large Parallel}}
	\addplot [line width=0.5mm, color=red, solid ] coordinates {
  (2.7775, 0.3718)
 (2.9611, 0.2080)
 (3.3037, 0.0564)
 (3.5056, 0.0237)
	};

			\addlegendentry{{\large Uncoded}}			
	\addplot [only marks, mark size = 3pt, mark=*, ] coordinates {
 (6.2872,0)
	};
	
				\addlegendentry{{\large Poly}}			
	\addplot [only marks, mark size = 3pt, mark=star] coordinates {
(4.3113,0.1370)
	};
	
		\addlegendentry{{\large BI-Poly}}
		\addplot [only marks,mark size = 3pt, mark=otimes*,mark options={scale=0.6}] coordinates {
 (2.7775, 0.3718)
	};

			\addlegendentry{{\large HH-Poly($\LY_{\text{HHCC}}=2,\Stask_{\ly,\text{HHCC}}=5$)}}
			\addplot [only marks,mark size = 3pt, mark=square*] coordinates {
 (2.9611, 0.2643)
  (2.9611, 0.2080)
	};	
			\addlegendentry{{\large HH-Poly($\LY_{\text{HHCC}}=5,\Stask_{\ly,\text{HHCC}}=2$)}}	
		\addplot [only marks,mark size = 3pt, mark=triangle*] coordinates {
 (3.3037, 0.0564)
 (3.3037, 0.1823)
	};
			\addlegendentry{{\large ML-Poly}}			
			\addplot [only marks,mark size = 3pt, mark=diamond*] coordinates {
 (3.5056, 0.0237)
 (3.5056, 0.1595)
	};

	\end{axis}
	\end{tikzpicture} 
	\quad
	\setlength{\belowcaptionskip}{-12pt}
\caption{A tradeoff between the average decoding time and the average computation time for different hierarchical and non-hierarchical schemes, where $(\Aa,\AB,\Bb)=(10^4,10^4,10^4)$, $\DM_{\text{Poly}}=\DM_{\text{\bicc}}/\Stask=\DM_{\text{sum}}/\LY=8, \ND=12, \prob=0.33$ and $ \Stask=\LY=10$.}	\label{FIG:hybrid_tradeoff}
\end{figure}
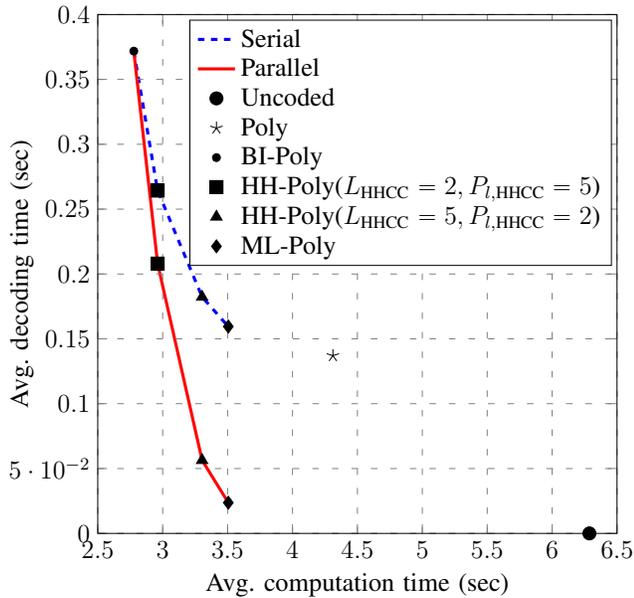

\textbf{Tradeoffs:} Figure~\ref{FIG:hybrid_tradeoff} plots a tradeoff between average decoding time and average computation time. For these results, we set $(\Aa,\AB,\Bb)=(10^4,10^4,10^4)$, $\DM_{\text{Poly}}=8, \ND=12$, and $\LY=\Stask=10$. We again consider an artificial straggler scenario with $p=0.33$ and a $\Divy$-dominated scheme with $\Divx=\Divz=1$ and $\Divy=8$. The dashed blue line corresponds to hierarchical polynomial coding when we use serial decoding. For the solid red line we decode each level in parallel. These two lines intersect at a small circle marker symbol which corresponds to bit-interleaved polynomial coding with parameters $\Stask=10,\Divxi{,\text{\bicc}}=\Divzi{,\text{\bicc}}=10,\Divyi{,\text{\bicc}}=8,$ and $\DM_{\text{\bicc}}=80$. They meet due to the fact that BICC consists of a single code. Hence, both serial and parallel decoding take the same amount of time, in this case $0.37$ sec. The diamond marker symbols on each line correspond to multilevel polynomial coding with parameters $\LY=10,\DM_{\text{sum}}=80,\divx=\divz=1,$ for all $\ly\in [\LY]$ and $\divy \in \{12,12,12,11,10,9,7,5,2,0 \}$. The triangle marker symbols correspond to hybrid polynomial coding with $\LY_{\text{HHCC}}=5$ levels and $\Stask_{\ly,\text{HHCC}}=2$ subtasks per level, $\ly \in [\LY]$. The square marker symbols correspond to hybrid polynomial coding with $\LY_{\text{HHCC}}=2$ levels and $\Stask_{\ly,\text{HHCC}}=5$ subtasks per level. Figure~\ref{FIG:hybrid_tradeoff} demonstrates that hierarchical coding schemes with a lower number of levels ($\LY$) have a less constrained recovery condition, and thus a lower average computation time when compared to hierarchical coding schemes with a larger $\LY$. However, the average decoding time in schemes with a larger $\LY$ is lower than that of schemes with smaller $\LY$, especially when the decoding process is conducted in parallel. The average computation time of all hierarchical coding approaches outperforms that of the (non-hierarchical) polynomial coding, which is depicted by the star marker. Polynomial coding has a lower average decoding time when compared to the serial decoder in hierarchical coding. However, its decoding time is larger than that of multilevel polynomial coding and hybrid polynomial coding, where $\LY_{\text{HHCC}}=5$ and when decoding is conducted in parallel. Figure~\ref{FIG:hybrid_tradeoff} also illustrates that, while the uncoded method has zero decoding time, it has the largest average computation time. 

\begin{figure*}[h]
\centering
\subfloat[($\Aa,\AB,\Bb$)=($2^{13},2^{13},2^{13}$)]{\tikzset{every mark/.append style={scale=0.8}}
	\begin{tikzpicture}[scale=0.55]
\begin{axis}[
height=12cm,
	width=6cm,
	grid=major,
	     ymin=1,
    ymax=6.5,
    ybar stacked,
  bar width=25pt,
  nodes near coords,
    enlargelimits=0.15,
    legend style={at={(0.5,-0.15)},
      anchor=north,legend columns=-1},
    ylabel={{\Large Avg. computation + decoding times (sec)}},
    symbolic x coords={Uncoded, Poly, BICC, MLCC},
        every node near coord/.append style={font=\small},
    xtick=data,
    y tick label style={font=\Large},
    x tick label style={font=\Large, rotate=45,anchor=east},
    ]
\addplot+[ybar,blue,pattern color = blue, pattern = north west lines] plot coordinates {(Uncoded,6.42) (Poly,4.144) 
  (BICC,2.931) (MLCC,3.178)};
\addplot+[ybar,red,pattern color = red, pattern = north east lines] plot coordinates {(Uncoded,0) (Poly,0.1102) 
  (BICC,0.1639) (MLCC,0.0502)};
\legend{ \strut \large comp, \strut \large dec}
\end{axis}
\end{tikzpicture}
 \label{FIG:compdec_square}}
	\quad
	\subfloat[($\Aa,\AB,\Bb$)=($2^{14},2^{11},2^{14}$)]{\tikzset{every mark/.append style={scale=0.8}}
	\begin{tikzpicture}[scale=0.55]
\begin{axis}[
height=12cm,
	width=6cm,
	grid=major,
    ybar stacked,
  bar width=25pt,
    	     ymin=1,
    ymax=6.5,
  nodes near coords,
    enlargelimits=0.15,
    legend style={at={(0.5,-0.15)},
      anchor=north,legend columns=-1},
    ylabel={{\Large Avg. computation + decoding times (sec)}},
    symbolic x coords={Uncoded, Poly, BICC, MLCC},
        every node near coord/.append style={font=\small},
    xtick=data,
    y tick label style={font=\Large},
    x tick label style={font=\Large, rotate=45,anchor=east},
    ]
\addplot+[ybar,blue,pattern color = blue, pattern = north west lines] plot coordinates {(Uncoded,4.16) (Poly,2.81) 
  (BICC,1.98) (MLCC,2.26)};
\addplot+[ybar,red,pattern color = red, pattern = north east lines] plot coordinates {(Uncoded,0) (Poly,0.4401) 
  (BICC,0.6541) (MLCC,0.1995)}; 
\legend{ \strut \large comp, \strut \large dec}
\end{axis}
\end{tikzpicture}
 \label{FIG:compdec_smallNx}}
	\quad
\subfloat[($\Aa,\AB,\Bb$)=($2^{13},2^{13},2^{13}$)]{\tikzset{every mark/.append style={scale=0.8}}
	\begin{tikzpicture}[scale=0.55]
\begin{axis}[
height=12cm,
	width=6cm,
	grid=major,
    ybar stacked,
  bar width=25pt,
      ymin=1.5,
    ymax=10.5,
  nodes near coords,
    enlargelimits=0.15,
    legend style={at={(0.5,-0.15)},
  anchor=north,legend columns=-1},
    ylabel={{\Large Avg. finishing time (sec)}},
    symbolic x coords={Uncoded, Poly, BICC, MLCC},
        every node near coord/.append style={font=\small},
    xtick=data,
    y tick label style={font=\Large},
    x tick label style={font=\Large, rotate=45,anchor=east},
    ]
\addplot+[ybar,olive,pattern color = olive,pattern = crosshatch] plot coordinates {(Uncoded,3.41) (Poly,3.41) 
  (BICC,3.41) (MLCC,3.38)}; 
\addplot+[ybar,blue,pattern color = blue, pattern = north west lines] plot coordinates {(Uncoded,6.42) (Poly,4.144) 
  (BICC,2.931) (MLCC,3.178)}; 
\addplot+[ybar,black] plot coordinates {(Uncoded,0.28) (Poly,0.28) 
  (BICC,0.25) (MLCC,0.16)}; 
\addplot+[ybar,red,pattern color = red, pattern = north east lines] plot coordinates {(Uncoded,0) (Poly,0.1102) 
  (BICC,0.1639) (MLCC,0.0502)};
\legend{\strut \large input comm, \strut \large comp, \strut \large output comm, \strut \large dec}
\end{axis}
\end{tikzpicture}
 \label{FIG:fin_square}}
	\quad
	\subfloat[($\Aa,\AB,\Bb$)=($2^{14},2^{11},2^{14}$)]{\tikzset{every mark/.append style={scale=0.8}}
	\begin{tikzpicture}[scale=0.55]
\begin{axis}[
height=12cm,
	width=6cm,
	grid=major,
    ybar stacked,
       ymin=1.5,
    ymax=10.5,
  bar width=25pt,
  nodes near coords,
    enlargelimits=0.15,
    legend style={at={(0.5,-0.15)},
   anchor=north,legend columns=-1},
    ylabel={{\Large Avg. finishing time (sec)}},
    symbolic x coords={Uncoded, Poly, BICC, MLCC},
        every node near coord/.append style={font=\small},
    xtick=data,
    y tick label style={font=\Large},
    x tick label style={font=\Large, rotate=45,anchor=east},
    ]
\addplot+[ybar,olive,pattern color = olive,pattern = crosshatch] plot coordinates {(Uncoded,1.56) (Poly,1.56) 
  (BICC,1.56) (MLCC,1.53)}; 
\addplot+[ybar,blue,pattern color = blue, pattern = north west lines] plot coordinates {(Uncoded,4.16) (Poly,2.81) 
  (BICC,1.98) (MLCC,2.26)}; 
\addplot+[ybar,black] plot coordinates {(Uncoded,1.29) (Poly,1.29) 
  (BICC,1.29) (MLCC,0.33)}; 
  \addplot+[ybar,red,pattern color = red, pattern = north east lines] plot coordinates {(Uncoded,0) (Poly,0.4401) 
  (BICC,0.6541) (MLCC,0.1995)}; 
\legend{\strut \large input comm, \strut \large comp, \strut \large output comm, \strut \large dec}
\end{axis}
\end{tikzpicture}
 \label{FIG:fin_smallNx}}
 \quad
	\setlength{\belowcaptionskip}{-12pt}
\caption{(a),(b) The average computation plus decoding times of multiplication for matrices of different dimensions; (c),(d) The average finishing times of multiplication for matrices of different dimensions In all sub-figures $\DM_{\text{Poly}}=\DM_{\text{B-poly}}/\Stask=\DM_{\text{sum}}/\LY=8, \ND = 12, \prob=0.33$ and $ \LY = \Stask =4$.}\label{FIG:bar_fin}
\end{figure*}
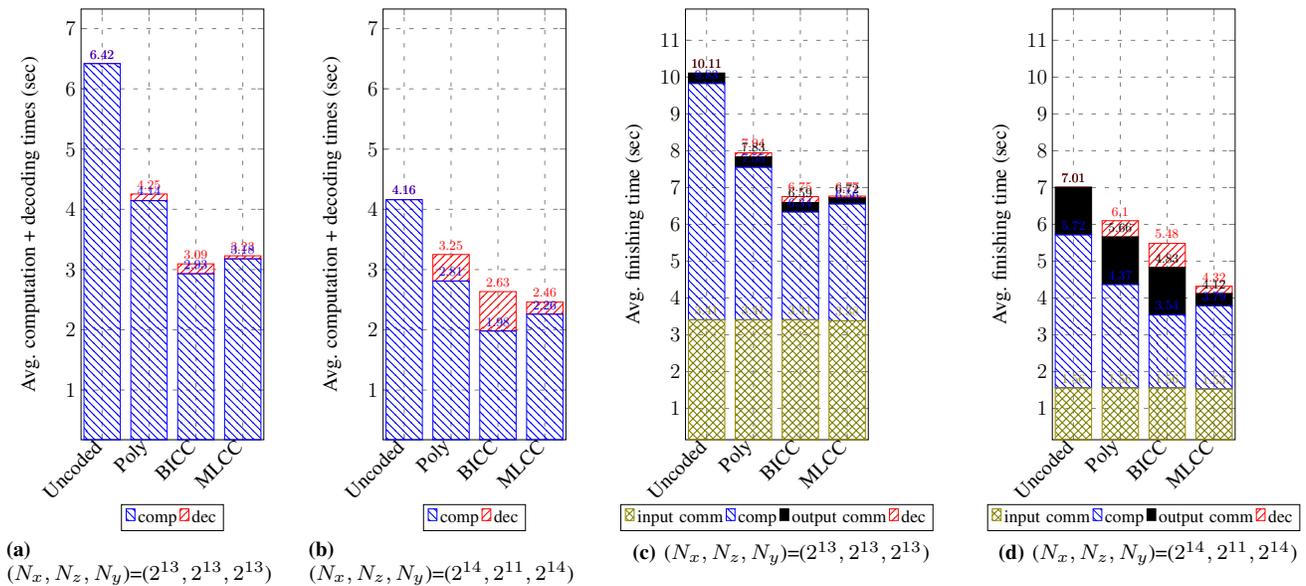

\textbf{Different size matrices:} In Figure~\ref{FIG:compdec_square} and~\ref{FIG:compdec_smallNx} we plot the sum of the average computation and decoding times for different matrix sizes. To keep the sub-figures comparable we assign the same computation load to each worker. To do this we set $\Aa\AB\Bb=2^{39}$ and $\DM_{\text{Poly}}=\DM_{\text{\bicc}}/\Stask=\DM_{\text{sum}}/\LY=8$ in all sub-figures. With this setting, each worker in each scheme is tasked with completing $2^{36}$ basic operations. We also set $ \ND = 12$ and $p=0.33$ and assume a $\Divy$-dominated scheme with $\Divx=\Divz=1$ and $\Divy=8$. In each sub-figure we compare BICC and MLCC with uncoded and polynomial coded schemes. In the hierarchical coding schemes we set $\Stask=\LY=4$. BICC achieves the lowest computation time. Compared to the uncoded scheme, BICC and MLCC, respectively, achieve computation time reductions of, on average, $53\%$ and $48\%$. Their average reductions are $29\%$ and $21\%$ when compared to polynomial coding.

We now consider the decoding time. The uncoded scheme has zero decoding time; no encoding nor decoding occurs. In polynomial coding, the master works with a Vandermonde matrix of dimension $\DM_{\text{Poly}} \times \DM_{\text{Poly}}$ ($8 \times 8$). The Vandermonde matrices in the decoding phases of BICC and MLCC are, respectively, of dimensions $\DM_{\text{B-poly}}\times \DM_{\text{B-poly}}$ ($32 \times 32$) and $\dm{\ly}\times \dm{\ly}$. As before $\sum_{\ly \in [4]} \dm{\ly}=\dmsum=32$. Figure~\ref{FIG:bar_fin} illustrates that MLCC outperforms polynomial coding in terms of decoding time, while BICC has the largest decoding time. In the decoding phase of BICC the master performs $1024\Aa\Bb$ basic operations to multiply the inverse of a Vandermonde matrix of dimension $32\times32$ by a $32 \times {\Aa\Bb}/{32}$ matrix, while in the computation phase workers perform $\Aa\AB\Bb/\DM_{\text{B-poly}}=\Aa\AB\Bb/32$ basic operations. Since in these subfigures $\AB > \DM_{\text{B-poly}}^2=2^{10}$, decoding time is smaller than computation time.

We now consider the sum of computation and decoding times. In Fig.~\ref{FIG:compdec_square} we consider multiplication of square matrices. This figure shows that the sum of the average computation and decoding times of BICC and MLCC are both approximately equal to $3$ seconds. For polynomial code and uncoded it is, respectively, $4$ and $6$ seconds. When compared to polynomial coding and uncoded approach, hierarchical coding, respectively, achieves a $27\%$ and a $52\%$ improvement in the sum of average computation and decoding times. In Fig.~\ref{FIG:compdec_smallNx} $(\Aa,\AB,\Bb)=(2^{14},2^{11},2^{14})$. In this figure MLCC has the lowest sum of average computation and decoding times. MLCC is the best choice. 

In Figs.~\ref{FIG:fin_square} and~\ref{FIG:fin_smallNx} we plot the average finishing time for four different schemes: uncoded, polynomial coded, BICC, and MLCC. As before, we use matrix dimensions $(\Aa,\AB,\Bb)\in\{(2^{13},2^{13},2^{13}),(2^{14},2^{11},2^{14})\}$. For each of these matrix multiplication problems, we separately measure average input communication time, computation time, output communication time, and decoding time as required by each scheme. We sum these to approximate the average finishing time. Fig.~\ref{FIG:fin_square} illustrates that BICC is the best choice and achieves, respectively, $33\%$ and $15\% $ improvements in comparison to uncoded and polynomial coded schemes. We note that in this figure the output communication time is smaller than the input communication time. This can be explained as follows. In BICC since $\Divxi{,\text{\bicc}}=1,\Divzi{,\text{\bicc}}=4,$ and $ \Divyi{,\text{\bicc}}=8$ the input communication load of the $\nd$th worker is $\commi{\nd}=\Stask\Aa\AB/4+\Stask\Bb\AB/8 \approx 2^{26}$ while the output communication load is smaller and equal to $\commo{\nd}=\Aa\Bb/8 = 2^{23}$ (cf.~(\ref{bicc_comm_in}) and~(\ref{comm_load_bicc_perw_pert})). Similarly in other coding schemes, due to the choice of parameter sets $\{\Aa,\AB,\Bb, \Divx,\Divz,\Divy,\LY,\Stask\}$ the input communication times dominate the output communication times.
On the other hand in Fig.~\ref{FIG:fin_smallNx} the input and output communication times are both on the same order of $2^{25}$ for the uncoded, polynomial coded and BICC schemes. However for MLCC, the input and output communication loads are respectively $\commi{\nd}=\Aa\AB+\Bb\AB/8 \approx 2^{25}$ and $\commo{\nd}=\Aa\Bb/(8\LY) = 2^{23}$ (cf.~(\ref{commin_perl_perw_mlcc1}) and~(\ref{commout_perl_perw_mlcc1})). Thus, the output communication time is smaller than the input communication time. Note that in Fig.~\ref{FIG:fin_smallNx} MLCC of profile $(12,11,7,2)$ outperforms the other approaches.

\section{Conclusion}
\label{sec.conclusion}
In this paper we introduce hierarchical coded computing to accelerate
distributed matrix multiplication. Through our hierarchical design, we can exploit the work completed by stragglers (and by leaders) while, at the same time, providing robustness to stragglers. To apply hierarchical coding to matrix multiplication, we connect the task allocation problem that underlies coded matrix multiplication to a geometric question of cuboid partitioning. We then develop three hierarchical approaches each with its particular strengths and ideal regime of operation. Due to parallelism with coded modulation, we term our approaches bit-interleaved coded computation (BICC), multilevel coded computing (MLCC), and hybrid hierarchical coded computing (HHCC). Our proposed schemes allow us to reap significant performance improvement, in terms of computation, decoding, and communication times. We analytically study our scheme under a probabilistic model of computation and communication time. This study is useful in developing design guidelines. Under this model, we numerically show that our method realizes a $66\%$ improvement in the expected finishing time. We also implemented our scheme in Amazon EC2 and measured a $27\%$ improvement in finishing time when compared to state-of-the-art approaches. 

A direct extension of this work is to apply hierarchical coding idea to more general computational problems beyond matrix-matrix multiplication, e.g., to tensor multiplication or non-linear computation. An extension can be made to an arbitrary $n$ nested loop computational problem by first viewing the problem as an $n$-dimensional cuboid and then generalizing the idea of hierarchical coding to the partitioned cuboid. Another extension is to use hierarchical coding in problems of large-scale machine learning, where matrix multiplication is a building block. Also extending hierarchical coding to coded MapReduce~\cite{li2015coded} or gradient coding~\cite{GC:ICML17} can be potential future research. In terms of analysis, we aim to extend our 
order-statistic analysis to characterize the performance of the hierarchical coded computing scheme, HHCC.
\appendix

\section{Appendix}
\subsection{Statistics Observed on Amazon EC2}
\label{ec2_appendix}
In this appendix we compare the empirical statistics observed on Amazon EC2 with the shifted exponential model of Sec.~\ref{SEC:Prob}. This comparison validates the model we assumed in Sec.~\ref{SEC:Prob}.

\begin{figure*}[h]
\centering 
	\subfloat[Histogram]{\tikzset{every mark/.append style={scale=0.8}}
		 			\begin{tikzpicture}[scale=0.82]
\begin{axis}[ ybar interval,
	height=8cm,
	width=10cm,
	grid=major,
	xlabel={{\large Communication time (sec)}},
	ylabel={{\large Frequency}},
	x label style={at={(axis description cs:0.5,-0.04)}},
minor xtick={1.01,1.06,1.12,1.18,1.23,1.29,1.34,1.4,1.45,1.51,1.56,1.62,1.67,1.73,1.79,1.84,1.9,1.95,2.01,2.06},	
	x tick label style={rotate=45,anchor=east,/pgf/number format/.cd,fixed,fixed zerofill,precision=2},
 axis on top,ymax=45,ymin=0, 
  tick label style={font=\small}]
\addplot coordinates { 
(1.01, 33) 
(1.06, 40) 
(1.12, 20) 
(1.18, 16) 
(1.23, 13) 
(1.29, 14)
(1.34,13)
(1.4, 5)
(1.45,6)
(1.51,3)
(1.56,4) 
(1.62,2)
(1.67,4)
(1.73,2)
(1.79,2)
(1.84,0)
(1.9,1)
(1.95,1)
(2.01,1)
(2.06,1)
};
\end{axis}
\end{tikzpicture}
 \label{FIG:hist_comm}}
	\quad
\subfloat[CCDF]{\tikzset{every mark/.append style={scale=0.8}}
		\begin{tikzpicture}[scale=0.82]
\begin{axis}[
	height=8cm,
	width=10cm,
	grid=major,
	xlabel={{\large Communication time in sec ($t$)}},
	ylabel={{ \large $P(\tcomm_{\nd}>t)$}},
		x label style={at={(axis description cs:0.5,-0.04)}},
	legend style={at={(0.32,0.85)},anchor=west,nodes=right},	
	axis on top,xmin=0.5, xmax=3, ymin=0, ymax=1]
		\addlegendentry{{Experimental data}}
		\addplot [line width=0.5mm, color=red, solid, every mark/.append style={solid, fill=red},mark=otimes*] coordinates {
(0.5,1)
(0.75,1)
(1,0.938888888888889)
(1.25,0.327777777777778)
(1.5,0.105555555555556)
(1.75,0.038888888888889)
(2,0.005555555555556)
(2.25,0) 	
(2.5,0)
(2.75,0)
(3,0)  
	}; 	
	
		\addlegendentry{{ShiftedExp ($\scalem,\shiftm$)}}
		\addplot [line width=0.5mm, color=blue, dotted, every mark/.append style={solid, fill=blue},mark=square*] coordinates {
(0.5,1)
(0.75,1)
(1,0.950718512212226)
(1.25,0.304228242246917)
(1.5,0.097352499390469)
(1.75,0.031152627604768)
(2,0.009968785729772)
(2.25,0.003189993800423) 	
(2.5,0.001020792373573)
(2.75,3.266517539336501E-4)
(3,1.04527983466807E-4)  
	}; 	
	
		\addlegendentry{{ShiftedExp using MLE}}
		\addplot [line width=0.5mm, color=black, dashed, every mark/.append style={solid, fill=black},mark=triangle*] coordinates {	
(0.5,1)
(0.75,1)
(1,0.918729663398827)
(1.25,0.306318829348813)
(1.5,0.102131485410518)
(1.75,0.034052233531752)
(2,0.011353546889484)
(2.25,0.003785449986754) 	
(2.5,0.001262128191454)
(2.75,4.208132658568509E-4)
(3,1.403057200687813E-4)  	   
	}; 

	\end{axis}
	\end{tikzpicture}  \label{FIG:ccdf_comm}}
	\quad
	\setlength{\belowcaptionskip}{-12pt}
\caption{(a) The histogram of master-to-worker communication times, where $\ND=12$ workers multiply matrices of dimensions $(\Aa,\AB,\Bb)=(5000,5000,5000)$ on Amazon EC2. (b) The Complementary of CDF of master-to-worker communication times both for the experiment on Amazon EC2 and for the shifted exponential models. The latter consists of a shifted exponential model with the parameter set $\{\scalem =0.22, \shiftm=0.99\}$ and a shifted exponential model using MLE. In both sub-figures straggling is natural.}
\label{EC2_comm} 
\end{figure*}
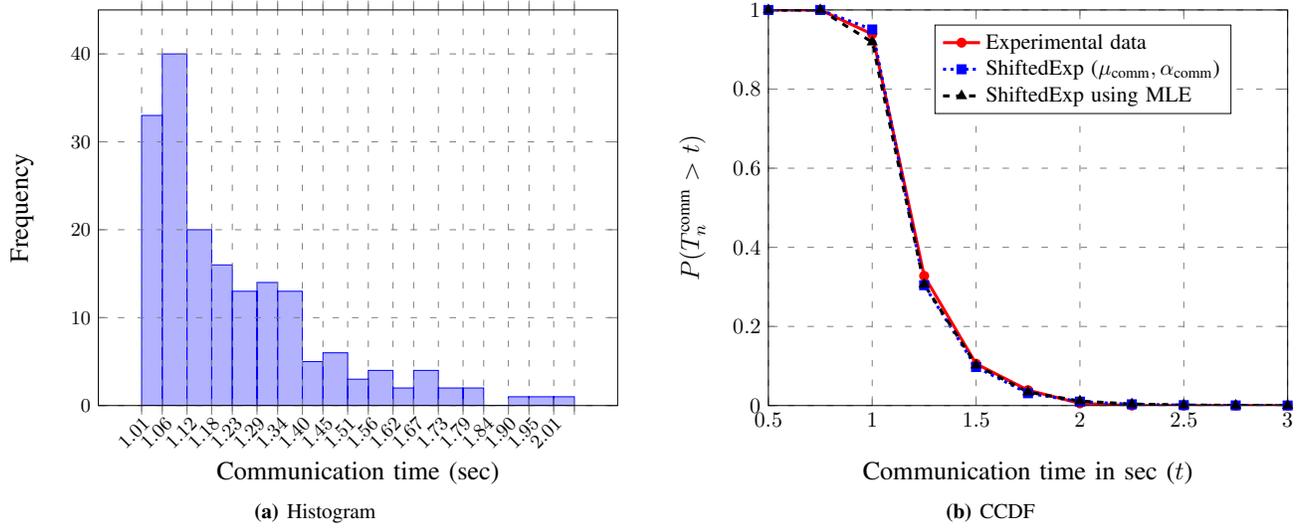

\textbf{Communication model:} Figure~\ref{FIG:hist_comm} shows the histogram of communication times observed between the master and $\ND=12$ workers on Amazon EC2. The master generates two random matrices of dimensions $5000 \times 5000$ for each worker. It then sends each matrix to the respective worker. This communication occurs in a serial manner from the master to all $12$ workers and is repeated for $15$ trials. We log the communication times for all workers. We then plot (solid line) the complementary CDF (the CCDF) of master-to-workers communication times derived from the Amazon EC2 empirical data in Fig.~\ref{FIG:ccdf_comm}. We also plot the CCDF of two shifted exponential distribution. The dotted line corresponds to the shifted exponential distribution that has the same scale and shift parameter to that of the empirical data ($\scalem =0.22,\shiftm=0.99$). The dashed line corresponds to the shifted exponential distribution the parameters of which correspond to the maximum likelihood estimation (MLE). It can be observed that the shifted exponential distribution is a good match to the empirical communication time data.  

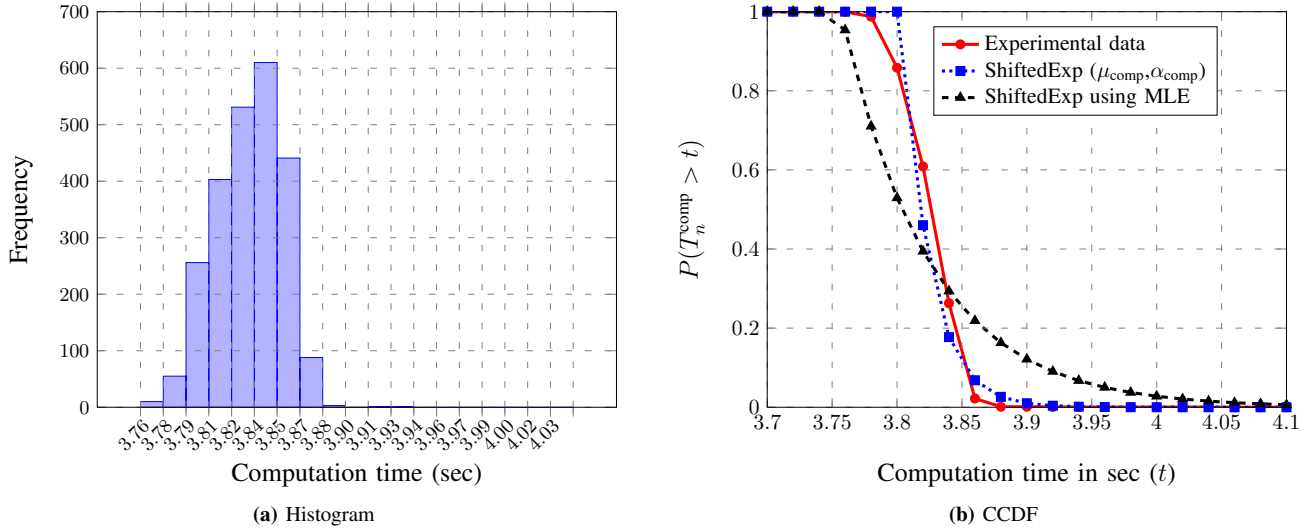
\begin{figure*}[h]
\centering 
	\subfloat[Histogram]{\tikzset{every mark/.append style={scale=0.8}}
	\begin{tikzpicture}[scale=0.82]
\begin{axis}[ ybar interval,
	height=8cm,
	width=10cm,
	grid=major,
	xlabel={{\large Computation time (sec)}},
	ylabel={{\large Frequency}},
	x label style={at={(axis description cs:0.5,-0.04)}},
	minor xtick={3.76,3.775,3.79,3.805,3.82,3.835,3.85,3.865,3.88,3.895,3.91,3.925,3.94,3.955,3.97,3.985,4.00,4.015,4.03,4.045},
	x tick label style={rotate=45,anchor=east,/pgf/number format/.cd,fixed,fixed zerofill,precision=2},
 axis on top,ymax=700,ymin=0,                                       tick label style={font=\small}]
\addplot coordinates { 
(3.76, 10) 
(3.775, 55) 
(3.79, 256) 
(3.805, 403) 
(3.82, 531) 
(3.835, 610)
(3.85,441)
(3.865,88)
(3.88,3)
(3.895,0)
(3.91,1) 
(3.925,1)
(3.94,0)
(3.955,0)
(3.97,0)
(3.985,0)
(4.00,0)
(4.015,0)
(4.03,0)
(4.045,1)
};
\end{axis}
\end{tikzpicture}
 \label{FIG:hist_comp}}
	\quad
\subfloat[CCDF]{\tikzset{every mark/.append style={scale=0.8}}
	 	\begin{tikzpicture}[scale=0.82]
	\begin{axis}[
	height=8cm,
	width=10cm,
	grid=major,
	xlabel={{\large Computation time in sec ($t$)}},
	ylabel={{\large $P(\tcomp_{\nd}>t)$}},
		x label style={at={(axis description cs:0.5,-0.04)}},
	legend style={at={(0.32,0.85)},anchor=west,nodes=right},
	axis on top,xmin=3.7, xmax=4.1, ymin=0, ymax=1]

		\addlegendentry{{Experimental data}}
		\addplot [line width=0.5mm, color=red, solid, every mark/.append style={solid, fill=red},mark=otimes*] coordinates {
(3.7,1)
(3.72,1)
(3.74,1)
(3.76,0.999583333333333)
(3.78,0.9875)
(3.8,0.858333333333333)
(3.82,0.60875)
(3.84,0.262916666666667)
(3.86,0.022083333333333)
(3.88,0.00125)
(3.9,0.00125)
(3.92,8.333333333333334E-4)
(3.94,4.166666666666667E-4)
(3.96,4.166666666666667E-4)
(3.98,4.166666666666667E-4)
(4,4.166666666666667E-4)
(4.02,4.166666666666667E-4)
(4.04,4.166666666666667E-4)
(4.06,0)
(4.08,0)
(4.1,0) 
	}; 	
	
		\addlegendentry{{ShiftedExp ($\scalep$,$\shiftp$)}}
		\addplot [line width=0.5mm, color=blue, dotted, every mark/.append style={solid, fill=blue},mark=square*] coordinates {
(3.7,1)
(3.72,1)
(3.74,1)
(3.76,1)
(3.78,1)
(3.8,1)
(3.82,0.460276310243068)
(3.84,0.177244283556605)
(3.86,0.06825364537424)
(3.88,0.026283274209996)
(3.9,0.010121225018973)
(3.92,0.003897505123076)
(3.94,0.00150086043497)
(3.96,5.779548644905478E-4)
(3.98,2.2256021786258E-4)
(4,8.570401188456349E-5)
(4.02,3.300310236775862E-5)
(4.04,1.270891224279913E-5)
(4.06,4.893977802310039E-6)
(4.08,1.884584476777155E-6)
(4.1,7.257202205602835E-7) 	   
	};

		\addlegendentry{{ShiftedExp using MLE}}
		\addplot [line width=0.5mm, color=black, dashed, every mark/.append style={solid, fill=black},mark=triangle*] coordinates {
(3.7,1)
(3.72,1)
(3.74,1)
(3.76,0.953601210488475)
(3.78,0.710376240971904)
(3.8,0.52918809056343)
(3.82,0.394213684302042)
(3.84,0.293665771513357)
(3.86,0.218763550816921)
(3.88,0.162965846919789)
(3.9,0.121399872890663)
(3.92,0.090435691995778)
(3.94,0.067369217051167)
(3.96,0.050186063775562)
(3.98,0.037385635569609)
(4,0.027850077128866)
(4.02,0.020746652671977)
(4.04,0.015455023521121)
(4.06,0.011513074220452)
(4.08,0.008576556213226)
(4.1,0.006389024779147) 
	}; 		

	\end{axis}
	\end{tikzpicture} 
	 \label{FIG:ccdf_comp}}
	\quad
	\setlength{\belowcaptionskip}{-12pt}
\caption{(a) The histogram of computation times, where $\ND=12$ workers multiply matrices of dimensions $(\Aa,\AB,\Bb)=(5000,5000,5000)$ on Amazon EC2. (b) The Complementary of CDF of computation times both for the experiment on Amazon EC2 and for the shifted exponential models. The latter consists of a shifted exponential model with the parameters $(\scalem =0.021, \shiftm=3.803)$ and a shifted exponential model using MLE. In both sub-figures straggling is natural.}
\label{EC2} 
\end{figure*}

\textbf{Worker computation model:} Similarly, the histogram of worker computation times is shown in Fig.~\ref{FIG:hist_comp}. The results are obtained from the time it takes to perform matrix multiplication across $\ND=12$ workers over $100$ trials\footnote{The reason that we use a different number of trials when gathering statistics about communication and computation is because of the speed variation inherent to burstable t2.micro instances on EC2~\cite{yang2019timely}. Burstable instances get access a short-term high level of resources in exchange for getting access to fewer resources most of the time. In the period of on-peak accessing resources the burstable instance operates fast, while in the off-peak period it operates much more slowly. The result is that on EC2, we experience a sudden slowdown in communicating $5000 \times 5000$ matrices after around 15 trials while slowing occurs after around 100 trials for multiplying the matrices of dimensions $5000 \times 5000$. Therefore, we select the number of trials to gather the most data while avoiding the slowdown period of t2.micro instances.}. In particular, each worker multiplies two $5000 \times 5000$ random matrices in each trial. Turning to the CCDF lines in Fig.~\ref{FIG:ccdf_comp}, it can be observed that the CCDF of the experimental computation time (solid line) closely matches the CCDF of the shifted exponential distribution that has the same scale and shift parameters as the empirical data ($\scalem =0.0210,\shiftm=3.8037$), plotted using the dotted line. The dashed line plots the CCDF of the shifted exponential distribution obtained from MLE. The difference between CCDF of empirical data and the shifted exponential distribution which is derived from MLE is due to the existence of outliers in the empirical data (cf. Fig.~\ref{FIG:hist_comp}). 

\subsection{Artificial Straggler}
\label{ec2_appendix_artificial}
\begin{figure*}[h]
\centering 
		\subfloat[Histogram]{\tikzset{every mark/.append style={scale=0.8}}
	   		\begin{tikzpicture}[scale=0.82]
\begin{axis}[ ybar interval,
	height=8cm,
	width=10cm,
	grid=major,
	xlabel={{\large Computation time (sec)}},
	ylabel={{\large Frequency}},
	x label style={at={(axis description cs:0.5,-0.04)}},
	x tick label style={rotate=70,anchor=east,/pgf/number format/.cd,fixed,fixed zerofill,precision=1},
 axis on top,ymax=1800,ymin=0,                                       tick label style={font=\footnotesize}]
\addplot coordinates { 
(3.76,1602)
(5.03,4)
(6.31,69)
(7.58,666)
(8.86,1)
(10.13,0)
(11.41,0)
(12.68,0)
(13.96,0)
(15.23,0)
(16.51,0)
(17.78,2)
(19.06,0)
(20.33,3)
(21.61,2)
(22.88,1)
(24.15,0)
(25.43,1)
(26.70,1)
(27.98,2)
(29.25,0)
(30.53,1)
(31.80,0)
(33.08,1)
(34.35,0)
(35.62,0)
(36.90,0)
(38.17,1)
(39.45,27)
(40.72,2)
(41.99,0)
};
\end{axis}
\end{tikzpicture} \label{FIG:hist_comp03}}
	\quad
\subfloat[CCDF]{\tikzset{every mark/.append style={scale=0.8}}
 	\begin{tikzpicture}[scale=0.82]
	\begin{axis}[
	height=8cm,
	width=10cm,
	grid=major,
	xlabel={{\large Computation time in sec ($t$)}},
	ylabel={{\large $P(\tcomp_{\nd}>t)$}},
		x label style={at={(axis description cs:0.5,-0.04)}},
	legend style={at={(0.32,0.85)},anchor=west,nodes=right},
	axis on top,xmin=0, xmax=48, ymin=0, ymax=1]

		\addlegendentry{{Experimental data}}
		\addplot [line width=0.5mm, color=red, solid, every mark/.append style={solid, fill=red},mark=otimes*] coordinates {
(0,1)
(3,1)
(6,0.33125)
(9,0.024583333333333)
(12,0.024166666666667)
(15,0.024166666666667)
(18,0.024166666666667)
(21,0.0225)
(24,0.020833333333333)
(27,0.020416666666667)
(30,0.019166666666667)
(33,0.01875)
(36,0.018333333333333)
(39,0.017916666666667)
(42,0.005833333333333)
(45,0.005416666666667)
(48,0.005416666666667)
	}; 	
	
		\addlegendentry{{ShiftedExp ($\scalep$,$\shiftp$)}}
		\addplot [line width=0.5mm, color=blue, dotted, every mark/.append style={solid, fill=blue},mark=square*] coordinates {
(0,0.879564317772919)
(3,0.567752530935344)
(6,0.366480233304222)
(9,0.236560392221329)
(12,0.15269805594523)
(15,0.098565512470224)
(18,0.063623339461523)
(21,0.041068414527434)
(24,0.026509370398848)
(27,0.017111610638728)
(30,0.0110454233445)
(33,0.007129742455868)
(36,0.004602198204772)
(39,0.002970686311197)
(42,0.001917556951455)
(45,0.00123776941652)
(48,7.989713824711619E-4)	   
	};

		\addlegendentry{{ShiftedExp using MLE}}
		\addplot [line width=0.5mm, color=black, dashed, every mark/.append style={solid, fill=black},mark=triangle*] coordinates {
(0,1)
(3,1)
(6,0.363562092177321)
(9,0.09367845301589)
(12,0.024137974635623)
(15,0.006219592667816)
(18,0.001602592327546)
(21,4.129373586793818E-4)
(24,1.064008976345227E-4)
(27,2.741614624949036E-5)
(30,7.064273816140875E-6)
(33,1.820239945296511E-6)
(36,4.690182663761823E-7)
(39,1.208511739141544E-7)
(42,3.113952543740599E-8)
(45,8.023670876011917E-9)
(48,2.067446225408021E-9) 
	}; 		

	\end{axis}
	\end{tikzpicture}  
\label{FIG:ccdf_comp03}}
	\quad
	\quad
	\setlength{\belowcaptionskip}{-12pt}
\caption{(a) The histogram of computation times; (b) The CCDF of computation times. In both figures straggling is artificially injected according to Bernoulli ($0.33$) and the empirical data refers to the experiment on Amazon EC2 where $\ND=12$ workers perform a matrix multiplication of dimension $(\Aa,\AB,\Bb)=(5000,5000,5000)$.}
\label{EC2_artificial} 
\end{figure*}
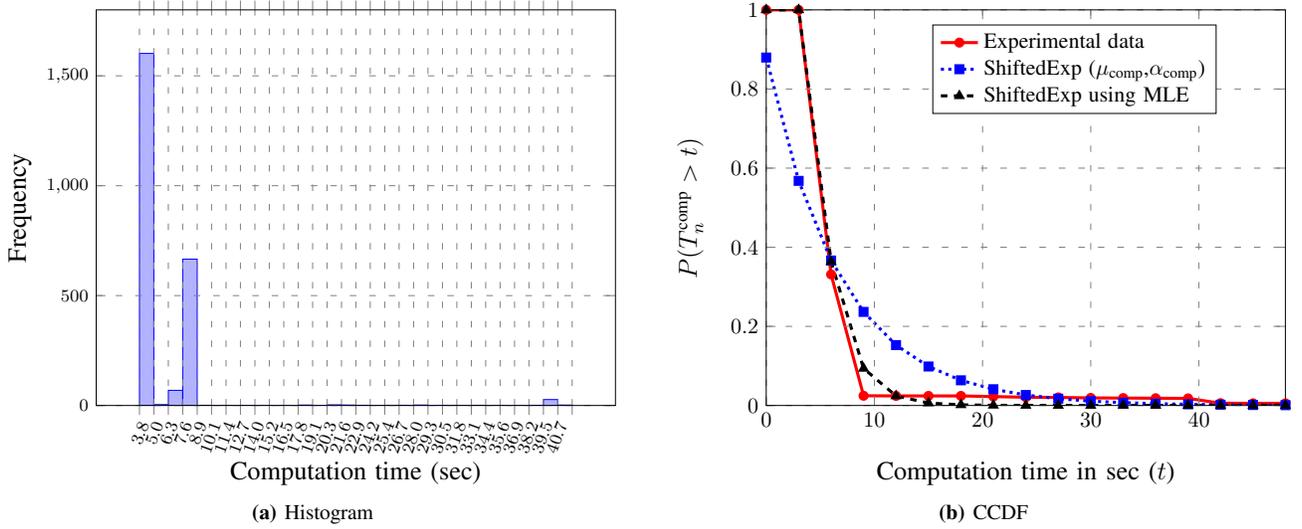

In Fig.~\ref{FIG:hist_comp03}, we plot the histogram of the artificial-straggler scenario in which we selected a subset of $12$ workers to be stragglers according to the Bernoulli(0.33) distribution, \ie $\prob=0.33$. Thus, roughly $33\%$ of workers are going to be stragglers. We plot the CCDF of this scenario in Fig.~\ref{FIG:ccdf_comp03}. Figure~\ref{FIG:ccdf_comp03} illustrates that the shifted exponential distribution derived from MLE while not able to capture the outliers in the straggler-scenario data, captures the majority of the probability mass and the variation about the mode of the distribution. The addition of artificial stragglers aids us in understanding how our techniques would work in the real-world overloaded computational system. The effect of natural stragglers seems to be reduced in many distributed computing platforms such as the on-demand instances on Amazon EC2 (cf.~\cite[Fig.~(2)]{GC:ICML17}). However, concerns about, and performance loss due to, stragglers remain in many platforms tailored to modern massive data analytics, such as in spot instances on Amazon EC2~\cite{yang2019coded} and in heterogeneous systems in federated settings~\cite{dhakal2019coded}. In this paper, we test our approaches in a deployed cloud computing system such as Amazon EC2, and due to the lack of an acute presence of straggling effects in EC2, we artificially inject stragglers to highlight the capabilities of our approaches. 

\subsection{Proof to Expected Value of Order Statistics (\ref{eq:nonh_E1_2})}
\label{exp_order_Exponen}
Given $\ND$ independent shifted exponential random variables $\tcom_{1},\ldots,\tcom_{\ND}$ with parameters $\scalee$ and $\shiftt$, let $\tcom_{\RT^*}$ denotes the $\RT$th order statistics ($\tcom_{[\RT:\ND]}$). Then according to~\cite{SPEEDUP:TIT17} the expected value of $\tcom_{\RT^*}$ is
\begin{align}\label{expected_order_exponential}
\mathbb E[\tcom_{\RT^*}] \begin{aligned}[t] = \shiftt + \sum_{\nd=\ND-\RT+1}^{\ND} \frac{\scalee}{\nd} \approx \shiftt + \scalee \log \left(\frac{\ND}{\ND-\RT}\right).
 \end{aligned}
\end{align}

\subsection{Proof to $\mathbb E[ \fin^{\text{BICC}}]\leq\mathbb E[ \fin^{\text{Non-h}}]$}
\label{proof_ebicc_enonh_fn}
The $\RT$th order statistics of the sequence 
\begin{align*}
&\left( \frac{\Aa\AB}{\Divx\Divz}+\frac{\AB\Bb}{\Divz\Divy} + \frac{\Aa\Bb}{\Divx\Divy}\right)\tcomm_{\nd}  \\ &+ \left(\frac{\Aa\AB\Bb}{\Divx\Divz\Divy}\right)\tcomp_{\nd},
\end{align*}
 where $n\in [\ND]$, is equal to or larger than $\RT-1$ elements of the sequence 
\begin{align*}
&\left( \frac{\Aa\AB}{\Divx\Divz}+\frac{\AB\Bb}{\Divz\Divy} + \frac{\Aa\Bb}{\Divx\Divy}\right)\tcomm_{\nd} \\ &+ \left(\frac{\Aa\AB\Bb}{\Divx\Divz\Divy}\right)\frac{\tcomp_{\nd}\Stask}{\Stask} ,
\end{align*}
 where $n\in [\ND]$. Let's $i_1,\ldots,i_{\RT-1}$ be the indices of these $\RT-1$ elements. That is, for all $j\in[\RT-1]$, 
\begin{align*}
& \left( \frac{\Aa\AB}{\Divx\Divz}+\frac{\AB\Bb}{\Divz\Divy} + \frac{\Aa\Bb}{\Divx\Divy}\right)\tcomm_{i_j} \\ &+ \left(\frac{\Aa\AB\Bb}{\Divx\Divz\Divy}\right)\frac{\tcomp_{i_j}\Stask}{\Stask}
\end{align*}
 is equal to or smaller than 
\begin{align*}
& \left( \frac{\Aa\AB}{\Divx\Divz}+\frac{\AB\Bb}{\Divz\Divy} + \frac{\Aa\Bb}{\Divx\Divy}\right)\tcomm_{\RT^*} \\ &+ \left(\frac{\Aa\AB\Bb}{\Divx\Divz\Divy}\right)\tcomp_{\RT^*}.
\end{align*} 
Furthermore, for $i_j\in \{i_1,\ldots,i_{\RT}\}$,
\begin{align*}
&\left( \frac{\Aa\AB}{\Divx\Divz}+\frac{\AB\Bb}{\Divz\Divy} + \frac{\Aa\Bb}{\Divx\Divy}\right)\tcomm_{i_j}  \\ &+ \left(\frac{\Aa\AB\Bb}{\Divx\Divz\Divy}\right)\frac{\tcomp_{i_j}\Stask}{\Stask}
\end{align*}
is equal to or larger than all $\Stask$ elements 
\begin{align*}
& \left( \frac{\Aa\AB}{\Divx\Divz}+\frac{\AB\Bb}{\Divz\Divy} + \frac{\Aa\Bb}{\Divx\Divy}\right)\tcomm_{i_j} \\ &+ \left(\frac{\Aa\AB\Bb}{\Divx\Divz\Divy}\right)\frac{\tcomp_{i_j}\stask}{\Stask},
\end{align*}
 where $\stask \in [\Stask]$. Therefore, the $\RT$th order statistics
\begin{align*}
& \left( \frac{\Aa\AB}{\Divx\Divz}+\frac{\AB\Bb}{\Divz\Divy} + \frac{\Aa\Bb}{\Divx\Divy}\right)\tcomm_{\RT^*} \\ &+ \left(\frac{\Aa\AB\Bb}{\Divx\Divz\Divy}\right)\tcomp_{\RT^*}
\end{align*}
is equal or larger than at least $\Stask\RT$ elements 
\begin{align*}
&\left( \frac{\Aa\AB}{\Divx\Divz}+\frac{\AB\Bb}{\Divz\Divy} + \frac{\Aa\Bb}{\Divx\Divy}\right)\tcomm_{i_j}  \\ &+ \left(\frac{\Aa\AB\Bb}{\Divx\Divz\Divy}\right)\frac{\tcomp_{i_j}\stask}{\Stask},
\end{align*}
 where $ j\in [\RT]$ and $ \stask\in [\Stask]$. Thus, the $\RT$th order statistics
\begin{align*}
& \left( \frac{\Aa\AB}{\Divx\Divz}+\frac{\AB\Bb}{\Divz\Divy} + \frac{\Aa\Bb}{\Divx\Divy}\right)\tcomm_{\RT^*} \\ &+ \left(\frac{\Aa\AB\Bb}{\Divx\Divz\Divy}\right)\tcomp_{\RT^*}
\end{align*}
is equal or larger than the $\RT_{\text{BICC}}$th order statistics

\begin{align*}
&\left( \frac{\Aa\AB}{\Divx\Divz}+\frac{\AB\Bb}{\Divz\Divy} + \frac{\Aa\Bb}{\Divx\Divy}\right)\tcomm_{\nd^*} \\ &+ 
 \left(\frac{\Aa\AB\Bb}{\Divx\Divz\Divy}\right) \frac{\tcomp_{\nd^*}\stask^*}{\Stask},
\end{align*}
 where $(\nd^*-1)\Stask+\stask^*=\RT_{\text{\bicc}}^*$. From this, we can conclude that $\mathbb E[ \fin^{\text{BICC}}]\leq\mathbb E[ \fin^{\text{Non-h}}]$.

\subsection{Proof to $\sum_{\ly=1}^{\LY} {\Xdimi{\ly}\Zdimi{\ly}}/{(\divx\divz)}={\Aa\AB}/{\Divx\Divz}$, Where $\divy=\Divy$, $\divz=\Divz$, $\Ydimi{\ly}=\Bb,$ and $\Zdimi{\ly}=\AB $}
\label{app_proof_reduce_comm}
Recall from (\ref{fixed_volume}) that $\sum_{l \in [\LY]} \Xdimi{\ly}\Zdimi{\ly}\Ydimi{\ly} = \Aa\AB\Bb.$ Incorporating the above choices of $\divy,\divz,\Ydimi{\ly},$ and $\Zdimi{\ly}$ into (\ref{fixed_volume}), yields 
\begin{align}\label{eq.app2}
\sum_{\ly=1}^{\LY}\Xdimi{\ly}=\Aa.
\end{align}

 As is shown in (\ref{fixed_comp}), we also have that $\dmsum = \sum_{\ly=1}^{\LY}\divx\divz\divy= \LY \Divx\Divy\Divz.$ Using the above choice of $\divy,\divz,\Ydimi{\ly},$ and $\Zdimi{\ly}$ in to (\ref{fixed_comp}), results in 
\begin{align}\label{eq.app3}
\sum_{\ly=1}^{\LY} \divx = \LY\Divx. 
\end{align}
Considering (\ref{eq.app2}) and (\ref{eq.app3}) together with the assumption (\ref{assumption}) that $\Xdimi{\ly}\Zdimi{\ly}\Ydimi{\ly}/(\divx\divz\divy) \approx \Aa\AB\Bb/\dmsum
$, yields 
\begin{align}
\Xdimi{\ly}={\Aa\divx}/{(\LY\Divx)}.
\end{align}

 Combining these three expressions proves that $\sum_{\ly=1}^{\LY} {\Xdimi{\ly}\Zdimi{\ly}}/{(\divx\divz)} = {\Aa\AB}/{(\Divx\Divz)}$. 

\subsection{Derivation of Optimization Problem~(\ref{thm.opt_Fnetwork}) and Extension to the Fast-Worker Regime}
\label{proof_opt1}
\subsubsection{Fast-network regime}
To obtain the optimization problem~(\ref{thm.opt_Fnetwork}), we first find the upper and lower bounds for $ \mathbb E[\fin^{\text{\mlcc}}]$. Since $ \mathbb E[\fin^{\text{\mlcc}}]$ is an expectation of max function (cf. (\ref{eq:reg1_fin_1_2}) and (\ref{eq:reg1_fin_3})), the upper bound is~\cite{upperbound}
\begin{align}\label{up_fn0}
\mathbb E[ \fin^{\text{\mlcc}}] \begin{aligned}[t]  &\leq  \left(\max_{\ly\in [\LY]} \left[\mathbb E\left[\tcomp_{\rt{\ly}^*}\ly\right]\right] \right. \\ &+ \left. \sqrt{\frac{\LY-1}{\LY}\sum_{\ly=1}^{\LY} \Var\left[\tcomp_{\rt{\ly}^*}\ly \right]}\right)\left(\frac{\Aa\AB\Bb}{\LY\Divx\Divz\Divy}\right) 
	\end{aligned}
	\end{align}	  
This upper bound (the right-hand-side of above inequality) can be further bounded and yield the new upper bound

\begin{align}\label{up_fn}
\mathbb E[ \fin^{\text{\mlcc}}] \begin{aligned}[t]
&\leq \left(\max_{\ly\in [\LY]} \left[\left(\shiftp + \scalep  \log \left(\frac{\ND}{\ND-\rt{\ly}^*}\right)\right)\ly\right]  
 \right. \\ &+ \left. \sqrt{\frac{(\LY-1)}{\LY}\sum_{j=1}^{\LY}{{j^2}}\sum_{i=1}^{\ND}{\frac{1}{i^2}}}\right)\left(\frac{\Aa\AB\Bb}{\LY\Divx\Divz\Divy}\right)
	\end{aligned}
\end{align}
Inequality~(\ref{up_fn}) follows App.~\ref{exp_order_Exponen} and~\cite{bookorder}, from which we know 
\begin{align}
\Var\left[\tcomp_{\rt{\ly}^*}\ly\right] = \ly^2\sum_{i=\ND-\rt{\ly}+1}^{\ND} \frac{1}{i^2} \leq \ly^2\sum_{i=1}^{\ND} \frac{1}{i^2}.
\end{align}

Following the convexity property of $\max$ function, the lower bound of $\mathbb E[ \fin^{\text{\mlcc}}]$ in a fast-network regime is

	\begin{align}\label{low_fn}
\mathbb E[ \fin^{\text{\mlcc}}] \begin{aligned}[t]  &\geq  \max_{\ly\in [\LY]} \left[ \left(\frac{\Aa\AB\Bb}{\LY\Divx\Divz\Divy}\right)\mathbb E\left[\tcomp_{\rt{\ly}^*}\ly\right]\right]
\\ &\approx \max_{\ly\in [\LY]} \left[\left(\shiftp + \scalep  \log \left(\frac{\ND}{\ND-\rt{\ly}^*}\right)\right)\ly \right. \\ &\times \left. 
 \left(\frac{\Aa\AB\Bb}{\LY\Divx\Divz\Divy}\right)\right].
	\end{aligned}
	\end{align} 
The second approximation in (\ref{low_fn}) follows App.~\ref{exp_order_Exponen}. 

To optimize the recovery profile, we first use the set $\{\rt{\ly}\}_{\ly\in[\LY]}$ that minimizes the upper-bound (\ref{up_fn}). To obtain this optimization problem,  we use (\ref{up_fn}) as the objective function of a minimization problem where $\{\rt{\ly}\}_{\ly \in [\LY]}$ are the variables. To solve this min-max problems, 
	
	\begin{align}\label{minmax1}
	\begin{aligned}[t] &\min_{\rt{\ly}}  \left[\left(\max_{\ly\in [\LY]} \left[\left(\shiftp + \scalep  \log \left(\frac{\ND}{\ND-\rt{\ly}^*}\right)\right)\ly\right] 
 \right. \right. \\ &+ \left. \left. \sqrt{\frac{(\LY-1)}{\LY}\sum_{j=1}^{\LY}{{j^2}}\sum_{i=1}^{\ND}{\frac{1}{i^2}}}\right)\left(\frac{\Aa\AB\Bb}{\LY\Divx\Divz\Divy}\right)\right],
 	\end{aligned}
	\end{align}	
we note that the term $\sqrt{{(\LY-1)}/{\LY}\sum_{j=1}^{\LY}{{j^2}}\sum_{i=1}^{\ND}{{1}/{i^2}}}$ and the coefficient $\frac{\Aa\AB\Bb}{\LY\Divx\Divz\Divy}$ in (\ref{minmax1}) are independent of $\rt{\ly}$, so can be ignored. We then introduce an auxiliary variable $z$ to recast the problem as the optimization problem~(\ref{thm.opt_Fnetwork}). The second constraint, $\rt{\ly} \leq \rt{\ly-1}$, is due to the sequential behavior of workers. This means that more workers finish their first subtasks than their second, etc. The third constraint, $\sum_{\ly=1}^{\LY}\rt{\ly}=\LY\RT$, is due to the assumption that the master requires the same total amount of completed computations in MLCC and non-hierarchical coding (cf. Remark.~\ref{remark_mlcc2} in Sec.~\ref{SEC:MLCC}). We note that while the per-level recovery thresholds are integer, we relax the integer optimization problem into the convex optimization problem~(\ref{thm.opt_Fnetwork}). 

The difference between the upper-bound (\ref{up_fn}) and the lower-bound (\ref{low_fn}) is 
\begin{align*}
\sqrt{\frac{(\LY-1)}{\LY}\sum_{j=1}^{\LY}{{j^2}}\sum_{i=1}^{\ND}{\frac{1}{i^2}}}\left(\frac{\Aa\AB\Bb}{\LY\Divx\Divz\Divy}\right),
\end{align*}
which is constant with respect to the recovery profile $\rt{\ly}$. Therefore, the optimization problem which is yielded from minimizing the lower-bound (\ref{low_fn}) is identical to the optimization problem~(\ref{thm.opt_Fnetwork}). This shows that the same recovery profile $\{\rt{\ly}\}_{\ly\in [\LY]}$ minimizes both the upper and lower bounds. 
\subsubsection{Fast-worker regime}
Similar to the previous regime, in a fast-worker regime the optimal recovery profile is obtained by minimizing the upper and lower bounds of $\mathbb E[\fin^{\text{\mlcc}}]$ in the following optimization problem.  

\begin{optproblem}\label{thm.opt_Fworker}
The solution set to the following optimization program yields the set $\{\rt{\ly}= \RT\}_{\ly \in [\LY]}$.
	\begin{mini}
	{z,\{\rt{\ly}\}}{z}{\label{eqn:optimization_expected_fw}}{}
		\addConstraint{\begin{aligned}[t] \left( \shiftm + \scalem \log \left(\frac{\ND}{\ND-\rt{\ly}}\right)\right) \leq z , \; \forall \ly \in [\LY]  \end{aligned}}
			\addConstraint{\rt{\ly} \leq \rt{\ly-1} \leq \ND , \; \forall \ly \in [\LY]}
		\addConstraint{\sum_{\ly=1}^{\LY}\rt{\ly}=\LY\RT}.	
\end{mini}
\end{optproblem} 
\textit{Proof}: Since $\shiftm$ and $\scalem$ are fixed variables that are independent of $\rt{\ly}$, we can assume without loss of generality that the objective of the above optimization problem is to minimize $\log \ND(\ND-\rt{\ly})$. This is equivalent to minimizing $\rt{\ly}$. We can therefore determine optimal recovery profile, by solving the optimization problem
	\begin{mini}
	{\{\rt{\ly}\}_{\ly \in [\LY]}}{{\rt{1}}}{\label{eqn:zopt4}}{}
	\addConstraint{\sum_{\ly=1}^{\LY}\rt{\ly}=\LY\RT, \;\;\forall \ly \in [\LY]}	
	\addConstraint{\rt{\ly} \leq \rt{\ly-1} \leq \ND , \;\; \forall r \in [R]}
	\addConstraint{\rt{j} \in \mathbb Z^+,\;\; \forall j \in [\LY]}.
\end{mini}
Relaxing the integer constraints on the $\{\rt{\ly}\}_{\ly \in [\LY]}$, we can reformulate the relaxed problem as the following linear program;
	\begin{mini}
	{\{\rt{\ly}\}_{\ly \in [\LY]}}{{\rt{1}}}{\label{eqn:zopt5}}{}
	\addConstraint{\sum_{\ly=1}^{\LY}\rt{\ly}=\LY\RT, \;\;\forall \ly \in [\LY]}	
	\addConstraint{\rt{\ly} \leq \rt{\ly-1} \leq \ND , \;\; \forall r \in [R]}.
\end{mini}

Since $\rt{1}$ is the maximum element of the sequence $\{\rt{\ly}\}_{\ly \in [\LY]}$, it is always larger or equal than the average, $\sum_{\ly=1}^{\LY}\rt{\ly}/\LY$. Therefore, merging this with the first constraint results in $\rt{1}\geq\RT$. The set $\{\rt{\ly}=\RT\}_{\ly \in [\LY]}$ is the only solution that satisfies the above constraints and minimizes the above optimization problem.

\subsection{Proof to Theorem~\ref{theorem}}\label{explicit_solution}
We prove by contradiction, assuming that~(\ref{statement1}) does not hold. First sort the sequence $\{ \left( \shiftp + \scalep \right.$ $ \left.  \log \left( \ND /(\ND-\bar{\RT}_\ly) \right) \right) \ly \}_{\ly=1}^{\LY}$ in ascending order. The index set of the sorted sequence is $\{i_1,i_2,\ldots,i_{\LY}\}$ such that
\begin{align*}
&\left( \shiftp + \scalep \log \left(\frac{\ND}{\ND-\bar{\RT}_{i_{j}}}\right)\right)i_{j} \leq \\ &\left( \shiftp + \scalep \log \left(\frac{\ND}{\ND-\bar{\RT}_{i_{j+1}}}\right)\right)i_{j+1}, \; \forall j \in [\LY-1].
\end{align*}
Since, by assumption~(\ref{statement1}) does not hold, there must exist at least a single value of $m$, $1 \leq m \leq \LY-1$ such that
\begin{align*}
&\left( \shiftp + \scalep \log \left(\frac{\ND}{\ND-\bar{\RT}_{i_{m}}}\right)\right)i_{m} < \\ &\left( \shiftp + \scalep \log \left(\frac{\ND}{\ND-\bar{\RT}_{i_{m+1}}}\right)\right)i_{m+1}. 
\end{align*}
Define $\epsilon>0$ as the difference between both side of above inequality,
\begin{align}
&\left( \shiftp + \scalep \log \left(\frac{\ND}{\ND-\bar{\RT}_{i_{m+1}}}\right)\right)i_{m+1} - \nonumber \\ &\left( \shiftp + \scalep \log \left(\frac{\ND}{\ND-\bar{\RT}_{i_{m}}}\right)\right)i_{m}=\epsilon. \label{proof2}
\end{align}
For such $\epsilon$, we alternatively define a new set of recovery thresholds $\{ \hat{\RT}_{i_l}\}_{l\in [\LY]}$ as

\begin{align}\label{new_optimal_solution}
\hat{\RT}_{i_l} =  \left\{
                \begin{array}{ll}
                  \bar{\RT}_{i_l} & l<m\\
                  \bar{\RT}_{i_l}+\theta & l=m\\
                   \bar{\RT}_{i_l}-\frac{\theta}{\LY-m} & l>m
                \end{array}
              \right.
\end{align}
where 
$0<\theta <
(1-\exp({-\frac{\epsilon}{2i_{m}\scalep}}))(\ND-\bar{\RT}_{i_m})$. The alternative defined set $\{ \hat{\RT}_l\}_{l\in [\LY]}$ provides a contradiction to the assumption that $\{\bar{\RT}_l\}_{l\in[\LY]}$ is the optimal solution to~(\ref{thm.opt_Fnetwork}). This follows because we can decrease the objective while keeping the equality constraint satisfied. To see we can decrease the objective, observe that since $\hat{\RT}_{i_l}<\bar{\RT}_{i_l}$ for $l>m$, we have 
\begin{align*}
&\left( \shiftp + \scalep \log \left(\frac{\ND}{\ND-\hat{\RT}_{i_{\ly}}}\right)\right)i_{\ly} < \\ & \left( \shiftp + \scalep \log \left(\frac{\ND}{\ND-\bar{\RT}_{i_{\ly}}}\right)\right)i_{\ly}.
\end{align*}
Also due to the above choice of $\{\hat{\RT}_{\ly}\}_{\ly\in [\LY]}$ in~\eqref{new_optimal_solution}, we have
\begin{align*}
&0<\left( \shiftp + \scalep \log \left(\frac{\ND}{\ND-\hat{\RT}_{i_{m}}}\right)\right)i_{m} -\\ & \left( \shiftp + \scalep \log \left(\frac{\ND}{\ND-\bar{\RT}_{i_{m}}}\right)\right)i_{m} <\frac{\epsilon}{2}.
\end{align*}
Then, applying~(\ref{proof2}) we find that 
\begin{align*}
&\left( \shiftp + \scalep \log \left(\frac{\ND}{\ND-\hat{\RT}_{i_{m}}}\right)\right)i_{m} < \\ &\left( \shiftp + \scalep \log \left(\frac{\ND}{\ND-\bar{\RT}_{i_{m+1}}}\right)\right)i_{m+1}
\end{align*}
Therefore, the objective of the optimization problem is decreased if we use $\{\hat{\RT}_{\ly}\}_{\ly=1}^{\LY}$ rather than $\{\bar{\RT}_{\ly}\}_{\ly=1}^{\LY}$.
Note that the set $\{\hat{\RT}_{\ly}\}_{\ly\in [\LY]}$ in~(\ref{new_optimal_solution}) satisfies $\sum_{\ly=1}^{\LY} \hat{\RT}_{\ly}=\LY\RT$ (i.e., the constraint of the optimization problem~1). $\blacksquare$

\bibliographystyle{IEEEtran} 
\bibliography{reference}


\begin{IEEEbiographynophoto}{Shahrzad Kianidehkordi} (GS'20) is currently pursuing a Ph.D. degree in Electrical and Computer Engineering at the University of Toronto (UofT), ON, Canada. She received her B.Sc. degree in Electrical Engineering and a minor degree in Economics from the Sharif University of Technology (SUT), Tehran, Iran, in 2017, and the M.A.Sc. degree in Electrical and Computer Engineering from the University of Toronto in 2019. During her B.Sc. degree, she conducted research on biomedical and sparse signal processing in the Image and Multimedia Processing laboratory at SUT and also spent a summer internship in the Department of Information Engineering at the Chinese University of Hong Kong (CUHK) in the field of network information theory. As part of her graduate studies, she was trained in the 2019 international high performance computing (HPC) summer school, in Kobe, Japan (granted with full scholarship). Her research interests include distributed computing, approximation algorithms, coding theory and statistical learning. 

Shahrzad won a gold medal in the 29th Iranian National Mathematical Olympiad, Iran, in 2011. She is also a recipient of the Ontario Graduate Scholarship (OGS) for two consecutive years, 2019-2020 and 2020-2021 and won the DiDi scholarship for 2020-2021.
\end{IEEEbiographynophoto}

\begin{IEEEbiographynophoto}{Nuwan Ferdinand} is a Senior Research Engineer at the Huawei Technologies, Canada. He worked as a Postdoctoral Fellow at the University of Toronto during 2016-2018 and held an NSERC Postdoctoral Fellowship. He received the Ph.D. degree from the University of Oulu, the M.Eng. degree from the Asian Institute of Technology, and the B.Sc. degree from the University of Moratuwa. His research interests are communication theory, information theory, coding theory, machine learning, and their applications to communication networks.
\end{IEEEbiographynophoto}

\begin{IEEEbiographynophoto}{Stark C. Draper} (S'99-M'03-SM'15) is a Professor of Electrical and Computer Engineering at the University of Toronto (UofT) and was an Associate Professor at the University of Wisconsin, Madison. As a research scientist he has worked at the Mitsubishi Electric Research Labs (MERL), Disney's Boston Research Lab, Arraycomm Inc., the C. S. Draper Laboratory, and Ktaadn Inc. He completed postdocs at the University of Toronto and at the University of California, Berkeley. He received the M.S. and Ph.D. degrees from the Massachusetts Institute of Technology (MIT), and the B.S. and B.A. degrees in Electrical Engineering and in History from Stanford University. His research interests include information theory, optimization, error-correction coding, security, and the application of tools and perspectives from these fields in communications, computing, and learning. 

Prof. Draper has received the NSERC Discovery Award, the NSF CAREER Award, the 2010 MERL President's Award, and teaching awards from the University of Toronto, the University of Wisconsin, and MIT. He received an Intel Graduate Fellowship, Stanford's Frederick E. Terman Engineering Scholastic Award, and a U.S. State Department Fulbright Fellowship. He spent the 2019-20 academic year on sabbatical at the Chinese University of Hong Kong, Shenzhen, and visiting the Canada-France-Hawaii Telescope (CFHT) in Hawaii, USA.  He chairs the Machine Intelligence major at UofT, is a member of the IEEE Information Theory Society Board of Governors, and serves as the Faculty of Applied Science and Engineering representative on the UofT Governing Council. 
\end{IEEEbiographynophoto}
\end{document}